\documentclass[12pt]{article}
\usepackage{amsfonts,amssymb,epsfig,amsmath}
\usepackage{color}

\usepackage[vcentermath]{youngtab}

\renewcommand{\baselinestretch}{1.2}

\def\det{{\rm det}}

\newcommand{\be}{\begin{eqnarray}}
\newcommand{\ee}{\end{eqnarray}}
\newcommand{\nn}{\nonumber}
\newcommand{\bn}{\begin{enumerate}}
\newcommand{\en}{\end{enumerate}}

\begin{document}

\makeatletter \@addtoreset{equation}{section} \makeatother
\renewcommand{\theequation}{\thesection.\arabic{equation}}
\renewcommand{\thefootnote}{\alph{footnote}}

\begin{titlepage}

\begin{center}
\hfill {\tt KIAS-P12070}\\
\hfill {\tt SNUTP12-004}\\

\vspace{2cm}

{\Large\bf Instantons on the 5-sphere and M5-branes}

\vspace{2cm}

\renewcommand{\thefootnote}{\alph{footnote}}

{\large Hee-Cheol Kim$^1$, Joonho Kim$^2$ and Seok Kim$^{2,3}$}

\vspace{1cm}

\textit{$^1$School of Physics, Korea Institute for Advanced Study,
Seoul 130-722, Korea.}

\textit{$^2$Department of Physics and Astronomy \& Center for
Theoretical Physics,\\
Seoul National University, Seoul 151-747, Korea.}

\textit{$^3$Perimeter Institute for Theoretical Physics,
Waterloo, Ontario, Canada N2L 2Y5.}


\vspace{0.7cm}

E-mails: {\tt heecheol1@gmail.com, joonho0@snu.ac.kr, skim@phya.snu.ac.kr}

\end{center}

\vspace{1.5cm}

\begin{abstract}

We calculate the partition functions of supersymmetric gauge theories on $S^5$, which
acquire non-perturbative contributions from instanton loops wrapping its Hopf fiber.
The instantons on the $\mathbb{CP}^2$ base equivariantly localize to 3 fixed points
of $SU(3)$ Cartans. Using our results, we study the index of the 6d $(2,0)$
theory. We first study the partition functions of maximal SYM with AD(E) gauge groups,
which agree with the gravity dual indices on $AdS_7$ in the large $N$ limits. We also
show that the most general partition function of the Abelian theory agrees with the
index for a free 6d tensor supermultiplet. We explain possible ambiguities in our 5d approach,
role of maximal SUSY, and the Wilson loops on $S^5$ dual to M2-brane
Wilson surfaces in $AdS_7$.

\end{abstract}

\end{titlepage}

\renewcommand{\thefootnote}{\arabic{footnote}}

\setcounter{footnote}{0}

\renewcommand{\baselinestretch}{1}

\tableofcontents

\renewcommand{\baselinestretch}{1.2}

\section{Introduction and discussions}

The discovery of M-theory as a non-perturbative type IIA string theory \cite{Hull:1994ys}
was perhaps one of the most dramatic events in the history of string theory. The quantum
mechanical emergence of a new (circle) direction of M-theory, often showing its Kaluza-Klein
states as quantum bound states of D0-branes, is  the essence of this phenomenon.
The type IIA string theory description and a more geometric M-theory description often
exhibit very different pictures of various phenomena at weak/strong couplings.

Such weak/strong-coupling dual descriptions of type IIA/M-theory have been studied
in various 10d/11d manifolds. The classic studies of \cite{Hull:1994ys} start
from the type IIA string theory on $\mathbb{R}^{9,1}$. Similar studies can be made on
various curved backgrounds, also including interesting Euclidean spaces. Among others,
studies by Gopakumar and Vafa \cite{Gopakumar:1998ii} suggest that the topological string
partition function on $\mathbb{R}^4\times$CY$_3$ with a gravi-photon deformation has a dual
interpretation at strong coupling as a Witten index counting 5d BPS states on
$\mathbb{R}^4\times S^1$, with the emerging M-theory circle at strong coupling interpreted
as a Euclidean time direction.

The idea that Euclidean type IIA partition functions of various sorts can count
states of M-theory was applied to the study of M5-branes from 5d gauge theories
living on D4-branes, the type IIA counterpart of M5's. More generally, there have been
studies on the 6d $(2,0)$ superconformal theories on curved backgrounds including a circle
factor, from the dimensionally reduced 5d gauge theories. It was often the Yang-Mills
instanton solitons in 5d which carried information on the quantum emergence of an extra
circle direction and the 6d theory. The earliest study of this sort is
\cite{Aharony:1997th}, with various other ingredients like DLCQ and so on.

At least apparently, 5d Yang-Mills theories are non-renormalizable.
So one may wonder whether one can do any consistent quantum calculation at all
with these 5d theories. One attitude towards this issue is to restrict one's interest
to certain supersymmetric observables. There are examples in which
nonrenormalizable low energy effective descriptions of UV complete theories can be used
to calculate a selected set of observables in a consistent manner. The basic idea is that
extra symmetries can make nonrenormalizable theories more predictive than generic
expectations. Especially, quantum fluctuations for supersymmetric
observables often experience huge boson/fermion cancelations so that a priori wild
quantum fluctuations of non-renormalizable theories can be consistently controlled.
There are many
examples of this. Just to mention some, firstly Nekrasov's instanton partition function
on $\mathbb{R}^4\times S^1$ \cite{Nekrasov:2002qd,Nekrasov:2003rj,Nekrasov:2004vw} can be
calculated for non-renormalizable 5d gauge theories. See also \cite{Kim:2011mv} for
the usage of this partition function in 5d maximal SYM on $\mathbb{R}^4\times S^1$ to study
the 6d $(2,0)$ theory on $\mathbb{R}^4\times T^2$.  Also, studies of black hole partition
function from 4d $\mathcal{N}=2$ supergravity \cite{Dabholkar:2010uh} have been made.
Often, the underlying framework turns out to be the `localization' of the supersymmetric
path integrals, which states that a path integral enjoying certain fermionic symmetry
is secretly a Gaussian path integral around a set of supersymmetric saddle points.
UV quantum fluctuations come into a good control for the Gaussian path integrals.
Also, supersymmetric observables calculated this way can often be directly understood
from string theory.

Although this attitude is what we shall basically assume throughout this paper,
we also extensively consider the 5d maximal SYM as a special case, for which
our study might have implications on recent discussions on possible UV finiteness
of this theory \cite{Douglas:2010iu}. See also \cite{Bern:2012di}.

Our strategy above was recently applied to the study of the index of 6d $(2,0)$ theory \cite{Kinney:2005ej,Bhattacharya:2008zy} counting local BPS operators, which
can also be regarded as a supersymmetric partition function on $S^5\times S^1$ with various
twists along the Euclidean time circle $S^1$. The key idea is to study the supersymmetric
partition function of a suitable SYM theory on $S^5$, which can be regarded as a naive
dimensional reduction of the 6d theory along the circle \cite{Kim:2012av}. See also
\cite{Hosomichi:2012ek} for the on/off-shell formulations of general 5d $\mathcal{N}=1$ gauge theory on $S^5$, and \cite{Kallen:2012cs,Kallen:2012va,Kallen:2012zn,Imamura:2012xg} for
earlier and following works which study closely related subjects.\footnote{Similar $S^5$ partition function was also studied
for 5d theories with different gauge groups and matters \cite{Jafferis:2012iv},
with 5d UV superconformal fixed points. The $S^4\times S^1$
index for the same class of theories was studied in \cite{Kim:2012gu}. Also,
the $S^5\times S^1$ index was studied from an appropriate supersymmetric
gauge theory on $\mathbb{CP}^2\times S^1$ \cite{Kim:2012tr}.} In this paper, we continue
to study the 6d index from gauge theories on $S^5$, with emphasis on the full
microscopic derivation of the nonperturbative part of the $S^5$ partition function.

The setting and results of \cite{Kim:2012av} will be reviewed and extended in
section 2.1. Just to explain salient aspects which are relevant here, the QFT on $S^5$
in \cite{Kim:2012av} came with two dimensionless real parameters. One is the ratio of the
Yang-Mills gauge coupling $g_{YM}^2$ and the 5-sphere radius $r$, which we call $\beta$,
and another is the ratio $m$ of $r^{-1}$ and the mass for the adjoint hypermultiplet.
These are related to two chemical potentials of the 6d index \cite{Kim:2012av}. The path
integral of this theory localizes to a Gaussian one around a set of saddle points.
The saddle points are characterized as follows. Firstly, the saddle points are
characterized by self-dual Yang-Mills instantons on the $\mathbb{CP}^2$ base of $S^5$
in Hopf fibration. The Hopf fiber is picked by selecting a supercharge that one uses
to calculate the path integral, which is either chosen
arbitrarily (for the calculation of unrefined indices of \cite{Kim:2012av}) or uniquely fixed
(when one considers the most general index with all chemical potentials turned on).
As the saddle point configuration is homogeneous along the Hopf
fiber, this is an instanton string or instanton loop winding the Hopf fiber circle of
$S^5$. On top of this, there is also a Hermitian matrix which completes the parametrization
of the saddle point space, given by the value of an adjoint scalar in the 5d theory. So one
first has to evaluate the determinant of quadratic fluctuations with
fixed scalar VEV in various instanton sectors, and then integrate over the Hermitian
matrix to get the partition function. The perturbative measure of the matrix
integral at zero instanton number was calculated in
\cite{Kallen:2012cs,Kallen:2012va,Kim:2012av}. In some special cases, nonperturbative
contributions are suggested in \cite{Kim:2012av} with various consistency checks.

One main goal of this paper is to provide a derivation of the non-perturbative
partition function, with a highly nontrivial generalization by including
all allowed chemical potentials in the 6d index.

The instantons appearing in our localization are self-dual (meaning not
anti-self-dual) on $\mathbb{CP}^2$, in the convention that the K\"{a}hler
2-form $J_{\mu\nu}$ of $\mathbb{CP}^2$ is anti-self-dual \cite{Kim:2012av}.
There exist ADHM like constructions for the instanton moduli space on $\mathbb{CP}^2$
\cite{bryan} for certain gauge groups, even though we shall not use this formalism in our
studies. The study of this paper is rather based on the so-called equivariant localization
and index theorems, which we review in section 2. The equivariant deformation
utilizes the isometries on the instanton moduli space and refines the partition function.
The deformation in our case will use the $U(1)^2$ Cartans of the
$SU(3)$ isometry of $\mathbb{CP}^2$, and comes with two parameters which we call
$\epsilon_1,\epsilon_2$. The deformation lifts the position moduli of the instantons
on $\mathbb{CP}^2$. With this, one can use equivariant generalizations of Atiyah-Singer
index theorems to calculate the BPS modes on $S^5$ which contribute
to the $S^5$ partition function. $\epsilon_1,\epsilon_2$ play the role of
chemical potentials conjugate to $U(1)^2$ charges. In particular,
one can use the equivariant localization to calculate such indices, which states
that the index can be calculated by studying the indices around the fixed points of the
$U(1)^2$ isometries. See section 2.2 for the details. As there exist 3 fixed points on
$\mathbb{CP}^2$, the determinant factorizes into 3 factors inside a matrix integral.
Each part is obtained from an instanton calculus on
$\mathbb{R}^4$ with a $U(1)$ fiber, i.e. the Hopf fiber of $S^5$. We should note that,
for such a localization of instantons to happen (so that we can luckily use the
instanton calculus on $\mathbb{R}^4$), it is important for these
instantons to be in the self-dual sector, excluding the K\"{a}hler 2-form $J$.
This is because an anti-self-dual instanton configuration proportional
to $J$ will never be localized to fixed points, as $J$ is $SU(3)$ invariant.

On flat space, the equivariant parameters $\epsilon_1,\epsilon_2$ are introduced
as a sort of IR regulator for the noncompact $\mathbb{R}^4$ moduli space of instantons
coming from their positions, meaning that $\epsilon_1,\epsilon_2\rightarrow 0$
limit is divergent \cite{Nekrasov:2002qd}. As $\mathbb{CP}^2$ is compact, in our case
one can of course turn them off and obtain the undeformed partition functions and
6d indices, which reduce properly to the results obtaineed previously
\cite{Kallen:2012cs,Kallen:2012va,Kim:2012av}. However, it is also very useful to keep the
$\epsilon_1,\epsilon_2$ parameters as extra chemical potentials in the
6d index for the rotations on $S^5$. Together with the other two parameters
$\beta\sim\frac{g_{YM}^2}{r}$ and $m$ of the theory, they provide 4 chemical potentials
in total, which form the maximal set admitted by the 6d $(2,0)$ superconformal index.

After a general derivation of the perturbative/non-perturbative partition function
on $S^5$ with all four parameters kept, we make closer studies of this quantity as
the 6d superconformal index. To study the 6d index from our results, there are two
technical challenges that one has to overcome, which we partially achieve in this
paper in some special cases.

The first challenge is that our expression naturally takes the form of a weak-coupling
expansion at small coupling $\beta\ll 1$, because our result is organized as an infinite
series coming from various instanton sectors weighted
by a factor like $e^{-\frac{4\pi^2 k}{\beta}}\ll 1$ for $k$ instantons.
On the other hand, the index nature of the partition function is visible by an
expansion with fugacities, most importantly with $e^{-\beta}$ which is conjugate
to an energy-like conserved charge
on $S^5\times\mathbb{R}$. The expansion with $e^{-\beta}\ll 1$ is in the strong-coupling
regime from 5d viewpoint. To get to the strong coupling regime from our instanton series,
one has to sum over all instantons and re-expand the exact expression at large $\beta$.
So far, we managed to do so in two special cases, in which we have a good technical control
over the instanton series. Firstly, with 5d maximal SYM, corresponding in 6d to tuning some
chemical potentials, the instanton series is significantly simplified so that we recast it
using the Dedekind eta function. It is easy to re-expand this function using its modular
property and study the 6d index. Secondly, the Abelian instanton sum
is considerably simpler than the general non-Abelian one, in that it does not depend on
the Hermitian matrix for the scalar expectation value. In this
case, one can directly obtain the 6d index by cleverly mimicking the techniques
explored by Gopakumar-Vafa \cite{Gopakumar:1998ii}, rewriting the partition function
into an integral similar to Schwinger's proper time integral, and then re-expanding
it at strong coupling. The resulting strong-coupling expansion can be compared with the
Abelian 6d $(2,0)$ index directly computed using the free tensor supermultiplets
\cite{Bhattacharya:2008zy}. They completely agree with each other. Although this
calculation is trivial in 6d, it is a nontrivial example which
illustrates that our 5d approach is properly working.

Another technical challenge of getting the 6d index from the 5d partition function
is that our general expression for the partition function takes the form of a matrix integral
with a nontrivial measure. This is sometimes called the `Coulomb branch localization'
\cite{Doroud:2012xw,Benini:2012ui}, as the matrix coming from a nonzero scalar is formally
analogous to the Coulomb branch of QFT vacuum on flat space. In this paper, we mainly study
the Abelian case in which the integral is trivial, or the case with maximal SUSY in which
the integral is simplified.

At least for $U(N)$ gauge group, which admits a Fayet-Iliopoulos deformation which lifts
the Coulomb branch, there could be another type of localization for gauge theories, in which
the saddle points are given by a discrete sum over points on the `Higgs branch' of the QFT.
This was recently shown to be possible for supersymmetric partition functions on $S^2$
\cite{Doroud:2012xw,Benini:2012ui}. Also, at the level of supersymmetric quantum
mechanics, similar idea has been applied to provide alternative derivations of the instanton
partition function on $\mathbb{R}^4\times S^1$ or $\mathbb{R}^4$ \cite{Kim:2011mv} and
vortex partition function on $\mathbb{R}^2\times S^1$ or $\mathbb{R}^2$ \cite{Kim:2012uz}.
More `standard' derivations of these partition functions for instantons
\cite{Nekrasov:2002qd} and vortices \cite{Shadchin:2006yz} can be regarded as
`Coulomb branch localizations' in that a complex mechanics variable in the vector
supermultiplet remains unlocalized. The `Higgs branch localization' of \cite{Kim:2011mv,Kim:2012uz} sums
over discrete saddle points. Of course the details of Coulomb vs. Higgs branch localizations
depend on situations. However, if such an alternative localization is possible on $S^5$
without any matrix integral, perhaps the index structure of the $S^5$ partition function
could be made more manifest.\footnote{We thank Jaume Gomis and Davide Gaiotto
for suggesting and emphasizing this possibility.}

Overcoming the two technical challenges in full generality, if possible, would be
useful not just for studying the 6d $(2,0)$ theory but also for studying various $(1,0)$
superconformal theories on $S^5\times S^1$ from 5d gauge theories on $S^5$. For instance,
the 5d $Sp(N)$ gauge theory with $N_f=8$ fundamental hypermultiplets and one hypermultiplet
in anti-symmetric representation is a circle reduction of the 6d $(1,0)$ SCFT with
$E_8$ global symmetry. The 6d theory lives on a stack of $N$ M5-branes near the
Horava-Witten 9-plane, whose circle reduction yields $N$ D4-branes near an O8-plane
and $N_f=8$ D8-branes. It should be interesting to study this model and check, for instance,
the enhanced $E_8$ global symmetry which is not visible in the 5d theory. Similar
$E_n$ type enhancements of global symmetries of 5d SCFT were studied in \cite{Kim:2012gu}
for $n\leq 6$ using the index \cite{Bhattacharya:2008zy} on $S^4\times S^1$.

The remaining part of the paper is organized as follows. In section 2, we first explain
the 5d field theory setting for calculating the 6d $(2,0)$ superconformal index. Then
using equivariant indices, we compute the perturbative and instantonic contributions to
the $S^5$ partition function. In section 3, we provide three applications of our results
to study the 6d index. In section 3.1, we show that the Abelian index obtained from
our 5d instanton calculation perfectly agrees with the index for the 6d free tensor
supermultiplet. In section 3.2, we study a special unrefined index for the ADE gauge
theory keeping only one chemical potential, which is computed from the maximal SYM on $S^5$.
The $A_n$ and $D_n$ cases can be fully studied, while the $E_n$ cases are studied only up
to perturbative level. The $A_n$ and $D_n$ indices completely agree with the large
$N$ gravity dual indices on $AdS_7$. We briefly comment on the cases with
non-ADE gauge groups, having in mind the interpretation as twisted indices of 6d ADE theories.
Section 3.3 explains the expectation values of Wilson loops on $S^5$ and compare them
with the dual M2-brane Wilson surfaces in $AdS_7$. Section 4 explains possible
small ambiguities in our 5d approach and how we suggest to fix them. Two appendices
explain the technical details of instanton calculus and the structure of our off-shell
QFT from supergravity.

As we were finalizing the preparation of this paper, we encountered
\cite{Lockhart:2012vp,Imamura:2012bm} which partly overlap with our work. The formula in
\cite{Lockhart:2012vp} with three factors of topological string partition functions
is similar to our three factors of localized instanton contributions on $\mathbb{CP}^2$.
Also, our section 3.1 for the Abelian 6d index overlaps with \cite{Lockhart:2012vp}.
\cite{Imamura:2012bm} derived the perturbative partition function that we also study as
a part of our result.

\section{The partition function on $S^5$}

\subsection{QFT set-up and geometries}

In this subsection, we review and generalize the results on 5d SYM on (squashed)
$S^5$, from which we shall calculate the 5d partition function.

The 5d gauge theory on $S^5$ of our interest was presented in \cite{Kim:2012av}.
The suggestion was that the partition function of 6d $(2,0)$ on $S^5\times S^1$,
or the superconformal index \cite{Bhattacharya:2008zy} counting local operators of
this theory on $\mathbb{R}^6$, can be computed from the 5d SYM theory which one can
regard as its dimensional reduction to $S^5$. Let us start by explaining the 6d
$(2,0)$ superconformal symmetry. The $OSp(6,2|4)$ symmetry of this theory has
$6$ Cartans in the bosonic subgroup $SO(6,2)\times Sp(4)\sim SO(6,2)\times SO(5)$.
The Cartans of $SO(5)$ are called $R_1$, $R_2$, which rotate the two orthogonal
2-planes of internal $\mathbb{R}^5$ in a way that spinors carry $\pm\frac{1}{2}$
charges. The angular momenta of $SO(6)\subset SO(6,2)$ are called $j_1,j_2,j_3$,
which rotate $S^5$ in a way that they rotate the three 2-planes of $\mathbb{R}^6$
with half-integral  charges for spinors. The energy (or conformal dimension
for local operators) $E$ is the $SO(2)\subset SO(6,2)$ charge, made dimensionless
by multiplying $r$. The states counted by our index saturate the BPS bound $E\geq
2(R_1+R_2)+j_1+j_2+j_3$ \cite{Bhattacharya:2008zy}. See the beginning of our section 3
for the details of this 6d index.

In \cite{Kim:2012av}, the 6d index with two chemical potentials $\beta$, $\beta m$
conjugate to $E-\frac{R_1+R_2}{2}$ and $\beta m=R_1-R_2$ was considered. $\beta=\frac{g_{YM}^2}{2\pi r}$ is related to the 5d Yang-Mils coupling. $m$ is
$r$ times the mass of the adjoint hypermultiplet in 5d $\mathcal{N}=1$ SYM theory.
The derivation of the 5d theory in \cite{Kim:2012av} relied on the strategy of
first reducing the Abelian 6d theory on $S^5\times S^1$ to SYM on $S^5$, and then
trying a non-Abelian completion while securing the desired supergroup
$SU(4|1)$.\footnote{When $m=\pm\frac{1}{2}$, symmetry enhances to $SU(4|2)$ with
$16$ real SUSY, realizing maximal SYM on $S^5$.} See \cite{Kim:2012av} for the details.

The localization of the partition function of this theory
was explained in \cite{Kim:2012av}. By picking a particular Hermitian SUSY for the
calculation, one first has the saddle point given as follows. One has a constant scalar
among 5 adjoint scalars in the theory, which should be exactly integrated over after the
localized path integral for other modes is done. Also, the saddle point admits nontrivial
instanton configurations on the $\mathbb{CP}^2$ base of $S^5$, being homogeneous on the
$S^1$ Hopf fiber. The Hopf fiber on the round $S^5$ is picked by selecting a SUSY among
8 (in \cite{Hosomichi:2012ek,Kallen:2012va,Kim:2012av}) or among 16 (in \cite{Kim:2012av}).
One may either understand this choice as being arbitrary, or unique if one
has in mind the squashing of $S^5$ we explain below. See also \cite{Imamura:2012xg}.

One can make a step further to include more chemical potentials of the 6d index into
the parameters of the 5d action. As we explain in section 3, the remaining two chemical
potentials are for the two angular momenta $j_1-j_2$, $j_1-j_3$ in $SU(3)\in SO(6)$ which
acts on the $\mathbb{CP}^2$ base of $S^5$. The natural way of incorporating the two
parameters is by suitably squashing $S^5$. One can write down a gauge theory on this
space with 2 Hermitian supercharges. Gauge theories on a class of squashed $S^5$ were studied
in \cite{Imamura:2012xg}. In this paper, we view the squashing of $S^n$ realized as
dimensionally reducted chemical potentials for angular momenta on $S^n$, in the index
on $S^n\times S^1$. This has been emphasized, for instance, in \cite{Kim:2012gu} which
discusses the squashed $S^4$ partition function as a reduction of an index on
$S^4\times S^1$. This agrees with the computation on the squashed $S^4$
of `ellipsoid type' in \cite{Hama:2012bg}.

In this paper, rather then discussing the gauge theory on squashed $S^5$ and its
superalgebra there in full detail, we shall mostly work by deriving only some aspects of
this squashed theory that we need in our calculation. Most information that we need
shall be derivable by considering Abelian 5d theories on the squahsed sphere, as the
localization calculation often relies on the quadratic part of the action with a
straightforward non-Abelian generalization (at least when saddle point studies
are concerned). The quantum fluctuations are controlled solely by the structure of the
$\mathcal{Q}^2$ superalgebra which we use in the localization calculus, which is again
obtained from Abelian theory and then straightforwardly generalized.
The on-shell Abelian theory can easily be obtained by reducing the Abelian 6d
tensor theory and dualizing it in 5d. For the 5d non-Abelian theories, we mostly
assume that a non-Abelian generalization of the 5d Abelian theory
exists, with same superalgebra. For instance, this has been rigorously checked to be
true for gauge theories on $S^5$ without squashing. Especially, our detailed calculation
would be carried out around the fixed points of isometries on the squashed $S^5$, for
which we use the equivariant localization technique. For this, it suffices to know the
local QFT's on $\mathbb{R}^4$ with a $U(1)$ Hopf fibration. However, in appendix B we
set up the off-shell supergravity derivation of the QFT on squashed $S^5$, to
derived some facts we need. This formulation should admit one to derive the full QFT
straightforwardly.

The $U(1)^3$ rotation on $\mathbb{C}^3$ (which contains our $S^5$ given by $|Z_1|^2+|Z_2|^2+|Z_3|^2=1$) acts as
\begin{equation}
  (Z_1,Z_2,Z_3)\rightarrow (Z_1e^{ia_1},Z_2e^{ia_2},Z_3e^{ia_3})\ .
\end{equation}
The overall $U(1)$ rotation of all $Z_i$ shifts the Hopf fiber coordinate,
which is wrapped by the instanton loops and thus acts trivially on the saddle
points. As we shall introduce chemical potentials for the $SU(3)$ acting on
$\mathbb{CP}^2$ only, we impose the constraint $a_1+a_2+a_3=0$.
One may take the following two combinations
\begin{equation}
  \epsilon_1=a_1-a_3\ ,\ \ \epsilon_2=a_2-a_3\ ,
\end{equation}
which will play the role of Omega deformation parameters around a fixed point.
On $\mathbb{CP}^2$, there are 3 fixed points of $U(1)^2$ on $\mathbb{CP}^2$,
which are
\begin{equation}
  Z\equiv(Z_1,Z_2,Z_3)=(1,0,0)\ ,\ \ (0,1,0)\ ,\ \ (0,0,1)
\end{equation}
up to a phase rotation which is not in $\mathbb{CP}^2$.
Around each fixed point, $\mathbb{CP}^2$ is approximated by $\mathbb{R}^4$
with Omega deformation parameters given by
\begin{eqnarray}
  (1,0,0)&:&(a_2-a_1,a_3-a_1)=(\epsilon_2-\epsilon_1,-\epsilon_1)\nonumber\\
  (0,1,0)&:&(a_3-a_2,a_1-a_2)=(-\epsilon_2,\epsilon_1-\epsilon_2)\nonumber\\
  (0,0,1)&:&(a_1-a_3,a_2-a_3)=(\epsilon_1, \epsilon_2)\ .
\end{eqnarray}
The instanton partition function will take
the form of a product of three instanton partition functions on twisted
$\mathbb{R}^4\times S^1$, where $S^1$ is the Hopf fiber circle. This
is similar to the localization of instantons on $S^4$ around its north
and south poles \cite{Pestun:2007rz}.

Being the chemical potentials of $SU(3)$ rotations, $a_1,a_2,a_3$ or
$\epsilon_1,\epsilon_2$ can be understood as twisting the 6d geometry. Let us
consider this geometry and part of the bosonic action which should be
important to us. The $U(1)^2\subset SU(3)$ chemical potentials $a_i\equiv(a,b,c)$
subject to $a+b+c=0$ are encoded in the 6d geometry as
\begin{equation}\label{squash-S5}
  ds_6^2=r^2\left[dn_1^2+n_1^2\left(d\phi_1\!+\!\frac{ia_1}{r}d\tau\right)^2\!\!
  +dn_2^2+n_2^2\left(d\phi_2\!+\!\frac{ia_2}{r}d\tau\right)^2\!\!+dn_3^2+n_3^2
  \left(d\phi_3\!+\!\frac{ia_3}{r}d\tau\right)^2\right]+d\tau^2
\end{equation}
where $n_1^2+n_2^2+n_3^2=1$. We replaced $a_i$ above by $ia_i$ with real $a_i$,
as the 6d chemical potentials we introduce are real, which deform the action in a
complex way. The above complex metric is nothing but reflecting a complex
deformation of our QFT. To KK reduce on the circle $\tau\sim\tau+2\pi r_1$,
one rewrites the metric as
\begin{eqnarray}
  ds_6^2&=&r^2\left[dn_i^2+n_i^2d\phi_i^2+\frac{(a_in_i^2d\phi_i)^2}{1-n_i^2a_i^2}
  \right]+(1-n_i^2a_i^2)\left(d\tau+\frac{irn_i^2a_id\phi_i}{1-n_i^2a_i^2}
  \right)^2\ .
\end{eqnarray}
We are also interested in the metric near the three fixed points, at which one
of the $n_i$'s are $1$ and other two $0$. For example, let us consider the fixed point with
$n_1=1$. The results for the other fixed points are similar, with the roles of
$a_1,a_2,a_3$ changed. The metric locally looks like
\begin{equation}\label{fixed-metric}
  ds_6^2=\left[ds^2(\mathbb{R}^4)+\frac{r^2d\phi_1^2}{1-a_1^2}
  \right]+(1-a_1^2)\left(d\tau+\frac{ira_1d\phi_1}{1-a_1^2}\right)^2\ ,
\end{equation}
where the factor of $\mathbb{R}^4$ is provided by $n_2,n_3,\phi_2,\phi_3$.

Now we consider the deformation of the superalgebra by this squashing.
Before squashing, the Hermitian supercharge in $SU(4|1)$ that we choose
in $S^5\times S^1$ satisfies the following algebra
(up to varios gauge transformations when $\mathcal{Q}^2$ acts on gauge non-invariant
objects)
\begin{equation}
  \mathcal{Q}^2\sim E-\frac{2(R_1+R_2)+j_1+j_2+j_3}{r}\ .
\end{equation}
$E\sim-\frac{\partial}{\partial\tau}$ is the translation along $S^1$ with
dimension of mass.
The effect of twisting (\ref{squash-S5}) is to covariantize all time derivatives
in the 6d (Abelian) equations of motion and SUSY by
\begin{equation}
  \frac{\partial}{\partial \tau}\rightarrow \frac{\partial}{\partial \tau}
  -\frac{ia_i}{r}\frac{\partial}{\partial\phi_i}\ \ \ {\rm or}\ \ \
  E\rightarrow E-\frac{a_i}{r}j_i\ .
\end{equation}
As we further twist the time translation by chemical potentials
$\beta$, $\beta m$ conjugate to $E-\frac{R_1+R_2}{2r}$, $\frac{R_1-R_2}{r}$,
there is an extra shift of energy by $\frac{R_1+R_2}{2}-m(R_1-R_2)$.
After this shift, reducing the 6d theory to $S^5$ will force us to
restrict to $E\sim-\frac{\partial}{\partial\tau}=0$, leaving us with the
5d algebra
\begin{equation}\label{SUSY-localize}
  -\mathcal{Q}^2\sim\frac{3}{2r}(R_1+R_2)+m(R_1-R_2)+\frac{1}{r}
  \sum_{i=1}^3(1\!+\!a_i)j_i\ .
\end{equation}
We shall use this $\mathcal{Q}$ on $S^5$ to localize the path integral.

One can also consider the reduced 5d action obtained from the 6d $(2,0)$ theory
along the $\tau$ circle in the 6d geometry (\ref{squash-S5}) or (\ref{fixed-metric}).
One would first obtain constant warp factors on the 5d metric from the $(1-a^2)$ factor
of the $\tau$ circle, which will affect various terms in the
5d SYM (or in the Abelian case, supersymmetric Maxwell-like theory). Secondly, the line
element $d\tau+\frac{irad\phi_1}{1-a^2}$ at the last term of (\ref{fixed-metric}) induces
a new term in the 5d action when one reduces the 6d $(2,0)$ theory. This is because the
(imaginary) 1-form $C\sim\frac{irad\phi_1}{1-a^2}$ appearing in the line element for
the M-theory circle can be regarded as a Ramond-Ramond 1-form in type IIA
perspective. Namely, the Wess-Zumino coupling between `the RR bulk field'
$C_\mu$ and the `5d worldvolume field strength' $F$ of the form
$\int_{S^5}C\wedge F\wedge F$ is induced from the twisting of the M-theory metric.
Of course, our metric (\ref{fixed-metric})
is not the full M-theory metric and the above argument in the honest sense applies to
the 11d metric pull-backed to 6d worldvolume of M5-branes. However, taking Abelian 6d
$(2,0)$ theory on any curved space and reducing it on a circle factor, this
heuristic argument can be shown to hold rigorously in the 6d context
\cite{Aganagic:1997zk}. Based on these ideas, we capture the aspects of our gauge
theories on $S^5$ or twisted $\mathbb{R}^4\times S^1$ coming from squashing. We discuss
the basic setting of localization and the classical aspects of this calculus here,
leaving the quadratic quantum fluctuations to the next subsection.

With a symmetry $\mathcal{Q}$ which squares to a bosonic symmetry of the path integral,
the corresponding supersymmetric path integral admits a deformation of the action by
\begin{equation}
  S\rightarrow S+t\{\mathcal{Q},V\}
\end{equation}
for any $V$ which is invariant under $\mathcal{Q}^2$, without changing the result
of the integral. One simple choice of $V$ is $V\sim\mathcal{Q}$, as
\begin{equation}
  [\mathcal{Q}^2,V]\sim[\mathcal{Q}^2,\mathcal{Q}]=0\ .
\end{equation}
The last equation holds from the Jacobi identity. Thus, the deformation of the
action which one uses to localize the path integral is $t\mathcal{Q}^2$ with
$t\rightarrow\infty$. Therefore, in this setting, one has to calculate
the determinant of $\mathcal{Q}^2$ in the saddle point background.

In \cite{Hosomichi:2012ek,Kallen:2012cs,Kallen:2012va,Kim:2012av}, the saddle
points of the localization calculation were given by Yang-Mils self-dual instanons
on the $\mathbb{CP}^2$ base of Hopf fibration, with a covariantly constant
scalar $\phi$, before one turns on the squashing parameters.
After one turns on $\epsilon_1,\epsilon_2$, breaking $SU(3)$ into $U(1)^2$ in
the $\mathcal{Q}$-exact deformation $t\mathcal{Q}^2$, the
saddle points are further restricted by requiring that the deformed $\mathcal{Q}^2$
still vanishes at the saddle point. From the right hand side of the SUSY
algebra (\ref{SUSY-localize}), the instantons should be invariant under
\begin{equation}
  \sum_{i=1}^3(1+a_i)j_i
\end{equation}
among others. With all $a_i=0$, the component $j_1+j_2+j_3$ left the instanton
loop configuration invariant as the loop winds the Hopf fiber. With generic
$a_i$ satisfying $a+b+c=0$, the instanton's position moduli get further constrained.
On the triangle formed by $n_1^2,n_2^2,n_3^2$, one finds that
they have to be localized on one of the three fixed points of $U(1)^2$ described
in the previous subsection, with two of the $n_i$'s being zero. As the nonzero
field strength is localized to three points, the covariantly constant scalar $\phi$
at the saddle point is simply taken to be a constant scalar away from the fixed
point, similar to the results on $S^4$ \cite{Pestun:2007rz}.

We now decide the value of classical action at the saddle points.
We consider the sector with gauge fields and scalar mass terms, as these determine
the classical action which straightforwardly generalizes to non-Abelian case.
Firstly, in the notation of \cite{Aganagic:1997zk}, the
`type IIA dilaton' $\Phi$ and `RR gauge field' $C_\mu$
are defined from the 6d metric as
\begin{equation}
  ds_6^2=e^{-\frac{2}{3}\Phi}ds_5^2
  +e^{\frac{4}{3}\Phi}(d\tau+e^{-\Phi}C_\mu dx^\mu)^2\ .
\end{equation}
Rewriting (\ref{fixed-metric}) in this form, one obtains
\begin{equation}
 e^{\frac{4}{3}\Phi}=1-a^2\ ,\ \ C=\frac{irad\phi_1}{(1-a^2)^{\frac{1}{4}}}\ ,\ \
 ds_5^2=(1-a^2)^{\frac{1}{2}}\left[ds^2(\mathbb{R}^4)+\frac{r^2d\phi_1^2}{1-a^2}\right]\ .
\end{equation}
Or if one considers the full geometry, one would obtain
\begin{equation}
  e^{\frac{4}{3}\Phi}=1-n_i^2a_i^2\ ,\ \ ds_5^2=
  r^2(1-n_i^2a_i^2)^{\frac{1}{2}}\left[dn_i^2+n_i^2\phi_i^2
  +\frac{(a_in_i^2d\phi_i)^2}{1-n_i^2a_i^2}\right]\ .
\end{equation}
Our 5d action, which is the quadratic part of  their DBI action, is
\begin{equation}
  \frac{1}{4g_{YM}^2}\int d^5x\sqrt{g}\ e^{-\Phi}F_{\mu\nu}F^{\mu\nu}
  -\frac{(-i)}{2g_{YM}^2}\int e^{-\Phi}C\wedge F\wedge F\ .
\end{equation}
Compared to \cite{Aganagic:1997zk}, a factor $-i$ is multiplied to the WZ term
as we consider a Euclidean theory, with Lorentzian and Euclidean times are
related by $t=-i\tau$. Decomposing 5d indices to $i=1,2,3,4$ on $\mathbb{R}^4$
and $5$ for $\phi_1$, one finds the following $\mathbb{R}^4$ part of the gauge field action:
\begin{equation}
  \int d^5x\left(\frac{r}{4g_{YM}^2(1-a^2)}F_{ij}F_{ij}
  -\frac{ra}{8g_{YM}^2(1-a^2)}\epsilon^{ijkl}F_{ij}F_{kl}\right)\ .
\end{equation}
where $i,j$ indices on $\mathbb{R}^4$ are contracted with $\delta_{ij}$.
It is the saddle point value of this action under self-dual gauge fields on
local $\mathbb{R}^4$ which will be important for the calculation of our partition
function. Imposing the self duality condition, one obtains
\begin{equation}
  S=\frac{1}{4g_{YM}^2(1+a)}\int rd\phi_1\int d^4xF_{ij}F_{ij}=
  2\pi r\cdot\frac{4\pi^2}{g_{YM}^2(1+a)}=\frac{4\pi^2}{\beta(1+a)}\ .
\end{equation}
The factor of $\frac{1}{1+a}$ multiplying the standard instanton action
$\frac{4\pi^2}{\beta}$ at the first fixed point has to be replaced by
$\frac{1}{1+b}$ and $\frac{1}{1+c}$ for the second and third fixed point, respectively.
Therefore, the instanton expansions around the three fixed points have to be
made in
\begin{equation}
  e^{-\frac{4\pi^2}{\beta(1+a_i)}}\ \ (\ll 1)\ \ \ {\rm for}\ i=1,2,3
\end{equation}
at the three fixed points labeled by $i=1,2,3$.

In \cite{Kim:2012av,Kallen:2012va}, it was also important to obtain the value of
classical action for nonzero constant scalar $\phi$ in the vector multiplet (and
also an auxliliary field $D$ proportional to $\phi$) in the localization calculation.
On the round $S^5$, the classical action at the saddle point was
\begin{equation}\label{gaussian}
  e^{-S_0(\phi)}\ \ \ {\rm with}\ \ \ S_0=\frac{2\pi^2{\rm tr}\lambda^2}{\beta}\ .
\end{equation}
Here $\beta=\frac{g_{YM}^2}{2\pi r}$ and $\lambda=r\phi$. This is obtained by
plugging in the saddle point value $\phi$ and $D^3=\frac{i}{r}\phi$, in the
notation of \cite{Kim:2012av}, into their quadratic
classical action terms. Note that the off-shell action and SUSY were important for
obtaining the correct coefficient, as the above saddle point value for $D^3$ is
different from its on-shell value $D^3=-\frac{i}{r}\phi$. To get the correct
generalization of this Gaussian measure after squashing, one
should redo the off-shell analysis of SYM on the squashed $S^5$.

In appendix B, we study the off-shell supergravity method
\cite{Festuccia:2011ws,Hosomichi:2012ek} to find that the Gaussian
measure is generalized to
\begin{equation}\label{gaussian2}
  e^{-S_0}\ \ \ {\rm with}\ \ \
  S_0=\frac{2\pi^2{\rm tr}\lambda^2}{\beta(1+a)(1+b)(1+c)}
\end{equation}
with our squashing with $a+b+c=0$. We also provide a nontrivial support
that the coefficient of the Gaussian should be given as (\ref{gaussian2}).
This can be seen from a careful study of
the 6d Abelian index that we study in section 3.1. Note that the only place this
Gaussian coefficient appears in the Abelian case is the overall prefactor of the
partition function. After a careful regularization, we confirm that the above
factor is compatible with the known 6d index.

\subsection{Equivariant indices and the partition function}

Now with the classical result at the saddle points, we calculate the determinant
of the operator $\mathcal{Q}^2$ from the Gaussian flugcuations in $t\rightarrow\infty$
limit. Without squashing, this determinant in the zero instanton sector was
calculated either by brutal calculation with $S^5$ spherical harmonics
\cite{Kim:2012av} or more efficiently by using appropriate index
theorems \cite{Kallen:2012cs,Kallen:2012va}.

We use equivariant index theorems to calculate the determinant on
squashed $S^5$, in general instanton sector. As in \cite{Kallen:2012cs},
we view $S^5$ as a $U(1)$ Hopf fibration over $\mathbb{CP}^2$ and use appropriate
Atiyah-Singer index theorems on $\mathbb{CP}^2$. The index theorems used in
\cite{Kallen:2012cs,Kallen:2012va} can be generalized to include
more equivariant parameters $\epsilon_1,\epsilon_2$, as we explain below.
The relevant index for the 5d $\mathcal{N}=1$ vector supermultiplet is that for the
so-called self-dual complex. The hypermultiplet index is that for the Dirac complex.
See, for instance, \cite{Gomis:2011pf} for a summary. Both indices will be related
to the index for the Dolbeault operator $\bar\partial_V$ on $\mathbb{CP}^2$,
covariantized by appropriate vector bundles appearing on the right hand side of
(\ref{SUSY-localize}). In paticular, $V$ includes the $U(1)$ bundle from the
Hopf fiber. The index for the Dolbeault complex is given by
\begin{equation}\label{dolbeault}
  {\rm ind}(\bar\partial_V)=\int {\rm Td}(T{\mathbb{CP}^2})\wedge e^{c_1(V)}\ ,
\end{equation}
where ${\rm Td}(T\mathbb{CP}^2)$ is the Todd class for the holomorphic tangent
bundle of $\mathbb{CP}^2$.
We shall explain various factors above (in our equivariant version) later.
Here let us just leave a comment that, on $\mathbb{CP}^2$ with a Hopf fibration,
one includes $c_1\sim tJ$ with the K\"{a}hler 2-form on $\mathbb{CP}^2$ for the
$U(1)$ bundle, when one considers the index in a sector with $U(1)$ charge $t$
on $\mathbb{CP}^2$. The values of $t$ for various fields can be
read off from the right hand side of (\ref{SUSY-localize}), namely the value
of $\frac{3}{2}(R_1+R_2)+(j_1+j_2+j_3)$. $j_1+j_2+j_3$ takes integer
values. So for the vector multiplet, with $R_1=R_2=0$, $t$ is valued in integers.
For the hypermultiplet, there are two scalars with $(R_1,R_2)=(1,0), (0,1)$.
So $t$ should be shifted by a half-integer, say by $\frac{3}{2}$.

It will turn out that the full nonperturbative forms of these indices in instanton
backgrounds are most easily derived after turning on nonzero $\epsilon_1,\epsilon_2$.
In particular, the equivariant localization theorem that we shall employ admits us
to use the ADHM instanton calculus on flat space \cite{Nekrasov:2004vw}, with a simple
generalization. To address the equivariant indices of our interest, one first
replaces all the cohomologies used in the index by the equivariant ones with
nonzero $\epsilon_1,\epsilon_2$. Namely, the exterior derivative $d$ is replaced by
\begin{equation}
  D\equiv d+i_v
\end{equation}
with a Killing vector $v$ on $\mathbb{CP}^2$. In our case, $v$ is given by
\begin{equation}
  v\sim a_1\frac{\partial}{\partial\phi_1}+a_2\frac{\partial}{\partial\phi_2}+
  a_3\frac{\partial}{\partial\phi_3}\ .
\end{equation}
The closed forms for $d$ are no longer closed with $D$. So one has to find
deformed differential forms closed in $D$ to get the equivariant indices.
Among others, the K\"{a}hler 2-form $J$ of $\mathbb{CP}^2$ can be deformed to an
equivariantly closed form as follows. In the homogeneous coordinates $Z_i$ in the
previous subsection, one obtains
\begin{equation}
  J=\frac{i}{2}dZ_i\wedge d\bar{Z}_i\ .
\end{equation}
$d$ is generalized to
\begin{equation}
  D\equiv d+i_v\ ,\ \ v=-2a_1(iZ_1\partial_1-i\bar{Z}_1\bar\partial_1)
  -2a_2(iZ_2\partial_2-i\bar{Z}_2\bar\partial_2)-2a_3
  (iZ_3\partial_3-i\bar{Z}_3\bar\partial_3)\ .
\end{equation}
One obtains
\begin{equation}\label{hamiltonian}
  DJ=-a_1(Z_1d\bar{Z}_1+\bar{Z}_1dZ_1)+\cdots\equiv -d\mu\ \ \ \ \left({\rm with}
  \ \ \mu\equiv a_1|Z_1|^2+a_2|Z_2|^2+a_3|Z_3|^2\right)\ .
\end{equation}
So one finds the following equivariant generalization
$J_\epsilon$ of $J$:
\begin{equation}\label{equivariant-kahler}
  J_\epsilon=J+\mu\ ,\ \ DJ_\epsilon=0\ .
\end{equation}
$J_\epsilon$ replaces $J$ in $c_1(V)$ in the equivariant indices.
As explained in the previous subsection, our
convention in this paper is to use imaginary Omega deformations. So we
replace all $a_i$'s to $ia_i$'s in the above formulae.
The equivariant version of Todd class is given as follows.
The ordinary Todd class of a complex $n$ dimensional space is given using
the $n$ Chern roots $\lambda_1,\lambda_2,\cdots,\lambda_n$ by
\begin{equation}
  {\rm Td}=\prod_{i=1}^n\frac{\lambda_i}{e^{\lambda_i}-1}\ .
\end{equation}
On $\mathbb{R}^{2n}$, with deformations $\epsilon_1,\epsilon_2,\cdots$,
the $i$'th Chern root is shifted by an addition of a 0-form $i\epsilon_i$ in our
convention. Thus, the $0$-form component of the equivariant Todd class on $\mathbb{R}^{4}$ is
\begin{equation}
  {\rm Td}_\epsilon=\frac{(i\epsilon_1)(i\epsilon_2)}{(e^{i\epsilon_1}-1)(e^{i\epsilon_2}-1)}\ ,
\end{equation}
which will be all we need to calculate our indices.

The equivariant versions of various indices mentioned above have the same
forms, with ordinary characteristic classes used to define them replaced
by their equivariant versions. The extra $\epsilon$ parameters play the role of
chemical potentials for $j_1-j_3$, $j_2-j_3$ appearing on the right hand side
of $\mathcal{Q}^2$ in (\ref{SUSY-localize}). These indices will all
take the form of
\begin{equation}\label{EI-formal}
  {\rm ind}=\sum_in_ie^{-w_i}\ .
\end{equation}
Here $i$ labels various eigenvalues of $\mathcal{Q}^2$ for BPS modes captured
by the index, $w_i$ is the corresponding eigenvalue, the integers $n_i$ are
the index degeneracies of BPS modes with given eigenvalue. These data are
related to the 1-loop determinant of $\mathcal{Q}^2$ by
\begin{equation}\label{det-formula}
  {\rm det}=\prod_iw_i^{-n_i}\ .
\end{equation}
The eigenvalue $w_i$ will contain various continous parameters like $\epsilon_i$,
$m$, $\beta$, as well as the saddle point value $\lambda$ of the vector multiplet
scalar. In the instantonic sectors, one would have extra martix $\phi$ in the
adjoint representation of the instanton gauge symmetry. After obtaining this
1-loop determinant, one should suitably integrate $\phi$ and $\lambda$
to get the final (squashed) $S^5$ partition function.

The final ingredient that we need to calculate our indices is the
so-called equivariant localization. It states that the integral of an equivariantly
closed form $\alpha_\epsilon$ is given by \cite{Szabo:1996md}
\begin{equation}\label{equivariant-localization}
  \int \alpha_\epsilon=\sum_{p}\frac{\alpha_\epsilon(x_p)}{\chi(x_p)}\ ,
\end{equation}
where $p$ labels the fixed points $x_p$ of the isometry $v$ on $\mathbb{CP}^2$.
$\alpha_\epsilon(x_p)$ is the 0-form value of $\alpha_\epsilon$ at the $p$'th fixed point,
and $\chi(x_p)$ is the 0-form value of the equivariant Euler class there.
As we already explained, there are three fixed points of $v$. The local values of
$\alpha_\epsilon$ are those on $\mathbb{R}^4$, which we explained for important
characteristic classes above. The Euler class of $\mathbb{R}^4$ with
parameters $\epsilon_1,\epsilon_2$ is $\chi=(i\epsilon_1)(i\epsilon_2)$.
Now with equivariant localizaiton, calculation of $\alpha_\epsilon(x_p)$ almost
boils down to that of various indices in the instanton background on local
$\mathbb{R}^4\times S^1$ near the fixed points, except that the circle is nontrivially
fibered so that $e^{c_1}$ for the $U(1)$ bundle has to be multiplied. From
(\ref{hamiltonian}), (\ref{equivariant-kahler}), the 0-form value of $J_{\epsilon}$
at the $i$'th saddle point (with $|Z_i|=1$) is $ia_i$. This provides a
multiplicative factor
\begin{equation}\label{fibration-factor}
  e^{c_1(V)}\ \rightarrow\ \ e^{ia_i t}
\end{equation}
at the $i$'th saddle point, when the $U(1)$ charge of the BPS modes is $t$.
For hypermultiplets, $t$ is replaced by $t+\frac{3}{2}$
with an integer $t$.

Apart from the above factor from $U(1)$ fibration, the remaining
factor of $\alpha_\epsilon(x_p)$ is almost the instanton index on
$\mathbb{R}^4\times S^1$ \cite{Nekrasov:2004vw,Gomis:2011pf} with minor changes.
At an $\mathbb{R}^4$ around a fixed point with Omega background given by
$\epsilon_1,\epsilon_2$, the self-dual/Dirac indices that we need are related to
the Dolbeault index (\ref{dolbeault}) by \cite{Gomis:2011pf}
\begin{equation}\label{relation-indices}
  {\rm ind}_{\rm SD}=\frac{1+\epsilon^{i(\epsilon_1+\epsilon_2)}}{2}\
  {\rm ind}(\bar\partial_V)\ \ ,\ \ \ {\rm ind}_{\rm Dirac}=e^{i\frac{\epsilon_1+\epsilon_2}{2}}
  \frac{e^{im}+e^{-im}}{2}\ {\rm ind}(\bar\partial_V)\ .
\end{equation}
The factor $e^{im}+e^{-im}$
comes from $m(R_1-R_2)$ on the right hand side of (\ref{SUSY-localize}), for the two
hypermultiplet scalars with $(R_1,R_2)=(1,0), (0,1)$. The Dolbeault index for instantons
on $\mathbb{R}^4\times S^1$ was studied in detail in \cite{Nekrasov:2004vw}.
At a fixed point with $\epsilon_1,\epsilon_2$, one obtains \cite{Nekrasov:2004vw}
\begin{equation}\label{equivariant-dolbeault}
  {\rm Ch}(\mathcal{E}){\rm Td}(T\mathbb{C}^2)\ \stackrel{x_p}{\longrightarrow}\ \
  {\rm Ch}(\mathcal{E})\ \frac{1}{(e^{i\epsilon_1}-1)(e^{i\epsilon_2}-1)}\ ,
\end{equation}
where the second factor divides the 0-form value of the Todd class by the Euler
class, as required in (\ref{equivariant-localization}). ${\rm Ch}(\mathcal{E})$ is
the equivariant Chern character for the so-called
`universal bundle' \cite{Nekrasov:2004vw}, which depends on $\lambda$ for the gauge
symmetry of the theory, $\phi$ for the instanton gauge symmetry, among others.
${\rm Ch}(\mathcal{E})$ is given for the classical gauge groups with various
representations in the
second reference of \cite{Nekrasov:2004vw}, where ADHM formalisms of instantons
are known. Here we explain it mostly for $U(N)$, leaving the detailed formulae for
other classical gauge groups $SO(N)$, $Sp(N)$ in appendix A.1. The result for the
adjoint representation of $U(N)$ is given by
\begin{equation}\label{equivariant-chern}
  \hspace*{-1cm}\frac{{\rm Ch}(\mathcal{E})}{(e^{i\epsilon_1}-1)(e^{i\epsilon_2}-1)}=
  \frac{{\rm tr}_{{\rm adj}_N}\left(e^{\i\lambda}\right)}
  {(e^{i\epsilon_1}-1)(e^{i\epsilon_2}-1)}-e^{-i\frac{\epsilon_1+\epsilon_2}{2}}
  \left({\rm tr}_N(e^{i\lambda}){\rm tr}_{\bar{k}}(e^{i\phi})+c.c.\right)
  +(1-e^{-i\epsilon_1})(1-e^{-i\epsilon_2}){\rm tr}_{{\rm adj}_k}(e^{i\phi})
\end{equation}
in the $k$ instanton sector. $k$ and $\bar{k}$ denote the fundamental and
anti-fundamental representations of the instanton gauge symmetry $U(k)$, while
$N$, $\bar{N}$ denote those of $U(N)$. This form is to be used for calculating
the determinant for the vector multiplet, as well as an adjoint hypermultiplet
which is our main interest. For the hypermultiplet in general representation
$R$ of the gauge group, the systematic method for constructing ${\rm Ch}(\mathcal{E})$
was explained in the second reference of \cite{Nekrasov:2004vw}, with many examples.
The second and third terms on the right hand side of (\ref{equivariant-chern}) come
from the zero modes in the instanton background, which depend sensitively on the choice
of $R$ \cite{Nekrasov:2004vw}. However, the first term is simply the
perturbative contribution to the index, which for general $R$ is replaced by
\begin{equation}\label{pert-general}
  \frac{{\rm tr}_{R}\left(e^{\i\lambda}\right)}
  {(e^{i\epsilon_1}-1)(e^{i\epsilon_2}-1)}\ .
\end{equation}
Of course (\ref{pert-general}) also works for exceptional groups.
We explain the structure of this perturbative part in more detail in section 2.3.

Briefly commenting on the results for other classical gauge groups, various
modifications are made as follows. Firstly, for the $G=SO(N),Sp(N)$ gauge theory,
the $k$ instanton gauge symmetry is $\hat{G}_k\equiv Sp(k),O(k)$, respectively.
For the vector or hypermultiplet fields in adjoint representations, the characters
for ${\rm adj}_N$ and ${\rm adj}_k$ appearing in (\ref{equivariant-chern}) are replaced
by those of the adjoint representation of $G$ and appropriate representations of
$\hat{G}_k$, as explained in appendices A.1 and A.2.
The fundamental representations at the second term of (\ref{equivariant-chern})
are replaced by fundamental representations of $G$, $\hat{G}_k$. For hypermultiplets in
general representation $R$ of $G$, the instanton part of ${\rm Ch}(\mathcal{E})$
again changes a lot due to different instanton zero mode structures. We again refer to
\cite{Nekrasov:2004vw} for more details on general $R$.

Collecting (\ref{equivariant-localization}), (\ref{fibration-factor}),
(\ref{relation-indices}), (\ref{equivariant-dolbeault}), (\ref{equivariant-chern}),
one can write down the full $U(N)$ equivariant indices for various fields in
$U(N)$ adjoint representation, in the background of self-dual instantons on $S^5$.
Firstly, let us specialize to a contribution from the fixed point on $\mathbb{CP}^2$
with $|Z_3|=1$, with Omega background $\epsilon_1=a-c$, $\epsilon_2=b-c$.
Other two contributions can be easily obtained by permuting $a,b,c$.
The vector multiplet index is given by
\begin{eqnarray}\label{vector-equiv-U(N)}
  I_{\rm vector}^{U(N)}&=&-\frac{1+e^{i(\epsilon_1+\epsilon_2)}}{2}
  \sum_{t=-\infty}^\infty e^{it(1+c)/r}\left[\frac{}{}\right.
  \frac{{\rm ch}_{{\rm adj}_N}(e^{i\lambda})}{(1-e^{i\epsilon_1})(1-e^{i\epsilon_2})}\\
  &&\hspace{2.5cm}\left.\frac{}{}-e^{-i\frac{\epsilon_1+\epsilon_2}{2}}\left(
  {\rm tr}_N(e^{i\lambda}){\rm tr}_{\bar{k}}(e^{i\phi})+c.c.\right)
  +(1-e^{-i\epsilon_1})(1-e^{-i\epsilon_2}){\rm ch}_{{\rm adj}_{k}}(\phi)\right]\nonumber
\end{eqnarray}
where $t$ is an integer for $j_1+j_2+j_3$ charge along the $U(1)$ fiber.
$r^{-1}$ plays the role of its chemical potential.
Similarly, the hypermultiplet index in adjoint representation is given by
\begin{eqnarray}\label{hyper-equiv-U(N)}
  I^{U(N)}_{\rm hyper}&=&\frac{e^{i(m+\frac{3}{2}(1+c))}+e^{i(m+\frac{3}{2}(1+c))}}{2}
  e^{i\frac{\epsilon_1+\epsilon_2}{2}}\sum_{t=-\infty}^\infty e^{it(1+c)/r}
  \left[\frac{}{}\right.  \frac{{\rm ch}_{{\rm adj}_N}(e^{i\lambda})}
  {(1-e^{i\epsilon_1})(1-e^{i\epsilon_2})}\\
  &&\hspace{2.5cm}\left.\frac{}{}-e^{-i\frac{\epsilon_1+\epsilon_2}{2}}\left(
  {\rm tr}_N(e^{i\lambda}){\rm tr}_{\bar{k}}(e^{i\phi})+c.c.\right)
  +(1-e^{-i\epsilon_1})(1-e^{-i\epsilon_2}){\rm ch}_{{\rm adj}_k}(\phi)\right]\ .\nonumber
\end{eqnarray}
All $1+c$ factors in these expressions come from the chern class
for the $U(1)$ bundle given by (\ref{fibration-factor}). Also, for the hypermultiplet,
the factor $\frac{3}{2}(1+c)$ which shifts the mass $m$ comes from an appropriate
half-integral shifts of the $U(1)$ charge appearing in (\ref{SUSY-localize}),
as explained after (\ref{dolbeault}). The above form can be achieved by suitably
shifting the integer $t$ in the summation, which we find to be convenient.
See appendices A.1 and A.2 for more indices with other classical groups.

At this point, we should carefully explain various terms above more concretely.
The fundamental characters and $U(N)$ adjoint characters appearing in the expressions
are standard:
\begin{equation}
  {\rm ch}_{{\rm adj}_N}=\sum_{i,j=1}^Ne^{i(\lambda_i-\lambda_j)}\ ,\ \
  {\rm ch}_N=\sum_{i=1}^Ne^{i\lambda_i}=\left({\rm ch}_{\bar{N}}\right)^\ast\ ,\ \
  {\rm ch}_k=\sum_{I=1}^ke^{i\phi_I}=\left({\rm ch}_{\bar{k}}\right)^\ast\ ,
\end{equation}
apart from the subtle property of the perturbative part including
${\rm ch}_{{\rm adj}_N}$ that we explain in detail in section 3.3.\footnote{The
perturbative index with the above character includes the $N$ saddle
point values $\lambda_i$ themselves which we do not Gaussian integrate. We should
exclude them from the index.} As for the hypermultiplet, the adjoint character
of $U(k)$ in the last term is also given in the standard way as
${\rm ch}_{{\rm adj}_k}=\sum_{I,J=1}^ke^{i(\phi_I-\phi_J)}$.
For the vector multiplet part, the precise definition of the last term
is given after expanding $\epsilon_1,\epsilon_2$ dependent factors as
\begin{equation}
  -(1-e^{-i\epsilon_1})(1-e^{-i\epsilon_2}){\rm ch}_{{\rm adj}_{k}}(e^{i\phi})
  \rightarrow-\sum_{I\neq J}^ke^{i(\phi_I-\phi_J)}+
  (e^{-i\epsilon_1}+e^{-i\epsilon_2}-e^{-i(\epsilon_1+\epsilon_2)})
  \sum_{I,J=1}^ke^{i(\phi_I-\phi_J)}\ .
\end{equation}
The reason why the contribution with $I=J$ is excluded becomes clear when one
understands the meaning of all $4$ terms. The first term with negative coefficient
comes from the $U(k)$ gauge invariance constaint in the nonzero $\phi_I$ background:
as the latter manifesetly preserves only $U(1)^k$, the remainders
appear as constraints with $I\neq J$. The next two terms
come from the bosonic generators of the ADHM data in $U(k)$ adjoint representation.
(Bi-fundamental ADHM data are encoded in the term with fundamental characters above.)
The last term comes from the complex ADHM constaints imposed on the generators.

We now consider the determinant of the form (\ref{det-formula}). To motivate the
final expression, let us first comment on some schematic structures.
As the eigenvalue $t$ run over integers, the determinant contains an
infinite product over $t$, with eigenvalues taking the form of
\begin{equation}\label{product-schematic}
  \frac{t(1+c)}{r}+\left({\rm expressions\ containing}\ \epsilon_1,\epsilon_2,
  m,\phi,\lambda\right)\ .
\end{equation}
This infinite product after regularization becomes a sine function, meaning that going
from the equivariant index to the determinant is essentially like taking Plethystic
exponential, apart from the factor $(1+c)$ multiplying $t$. The last factor is an
effect coming from the circle fibration over $\mathbb{R}^4$.
The arguments of the sine functions in
the determinant are all divided by $1+c$ compared to the determinants for instantons
on $\mathbb{R}^4\times S^1$, absorbing all factors of $r$ to other parameters.
Another difference is that the hypermultiplet has its mass factor effectively shifted
to $\frac{3(1+c)}{2}$,
coming from the shift of $t$ eigenvalue by $t+\frac{3}{2}$ from its nonzero R-charge
in the algebra. Apart from these, the resulting measure is very similar to the instanton
determinant on $\mathbb{R}^4\times S^1$ so that we can essentially borrow the techniques
for the instanton calculus there \cite{Nekrasov:2002qd,Nekrasov:2003rj,Nekrasov:2004vw}.
Remembering the above difference and taking the product over $t$,
the determinant from a fixed point with $|Z_3|=1$ takes the form of
\begin{equation}
  Z_{\rm pert}^{(3)}Z_{\rm inst}^{(3)}\ .
\end{equation}
The perturbative part $Z_{\rm pert}$ coming from the first terms of
(\ref{vector-equiv-U(N)}) and (\ref{hyper-equiv-U(N)}) will be explained in
more detail in section 3.3, as one should carefully expand the denominators
into infinite series. The remaining terms in the indices are finite series apart
from the $t$ summation, so one obtains a finite number of sine factors
in the determinant. The result is
\begin{eqnarray}\label{U(N)-contour-1}
  Z_{\rm inst}^{(3)}&=&\frac{(1+c)^{-k}}{k!}\oint\left[\prod_{I=1}^k\frac{d\phi_I}{2\pi}\right]
  \prod_{I=1}^k\prod_{i=1}^N\frac{\sin\pi\frac{\phi_I-\lambda_i-m-\frac{3(1+c)}{2}}{1+c}
  \sin\pi\frac{\phi_I-\lambda_i+m+\frac{3(1+c)}{2}}{1+c}}{
  \sin\pi\frac{\phi_I-\lambda_i-\epsilon_+}{1+c}
  \sin\pi\frac{\phi_I-\lambda_i+\epsilon_+}{1+c}}\\
  &&\times\prod_{I\neq J}\sin\pi\frac{\phi_{IJ}}{1+c}\prod_{I,J}
  \frac{\sin\pi\frac{\phi_{IJ}-2\epsilon_+}{1+c}}{
  \sin\pi\frac{\phi_{IJ}-\epsilon_1}{1+c}\sin\pi\frac{\phi_{IJ}-\epsilon_2}{1+c}}\cdot
  \frac{\sin\pi\frac{\phi_{IJ}+m+\frac{3(1+c)}{2}+\epsilon_-}{1+c}
  \sin\pi\frac{\phi_{IJ}+m+\frac{3(1+c)}{2}-\epsilon_-}{1+c}}{
  \sin\pi\frac{\phi_{IJ}+m+\frac{3(1+c)}{2}+\epsilon_+}{1+c}
  \sin\pi\frac{\phi_{IJ}+m+\frac{3(1+c)}{2}-\epsilon_+}{1+c}}\nonumber
\end{eqnarray}
where $\phi_{IJ}\equiv\phi_I-\phi_J$, $\epsilon_\pm\equiv\frac{\epsilon\pm\epsilon_2}{2}$.
The overall factor of $(1+c)^{-k}$ comes from a mismatch of the factors of $(1+c)$
pulled out from (\ref{product-schematic}), coming from the lack of the $I=J$ terms
in the fisrt factor on the second line of (\ref{U(N)-contour-1}). At this stage,
we already integrated over the $\phi_I$ variable for the $U(k)$ gauge symmetry.
This expression takes a form very close to the instanton partition function on
$\mathbb{R}^4\times S^1$. The rules for the contour and the classification of poles
are explained in \cite{Nekrasov:2002qd}. Using $\epsilon_1=a-c$, $\epsilon_2=b-c$
and $a+b+c=0$, one can rewrite the above contour integral formula as
\begin{eqnarray}\label{U(N)-contour-2}
  Z_{\rm inst}^{(3)}&=&\frac{(1+c)^{-k}}{k!}\oint\left[\prod_{I=1}^k\frac{d\phi_I}{2\pi}\right]
  \prod_{I=1}^k\prod_{i=1}^N\frac{\sin\pi\frac{\phi_I-\lambda_i-m-\frac{3(1+c)}{2}}{1+c}
  \sin\pi\frac{\phi_I-\lambda_i+m+\frac{3(1+c)}{2}}{1+c}}{
  \sin\pi\frac{\phi_I-\lambda_i-\epsilon_+}{1+c}
  \sin\pi\frac{\phi_I-\lambda_i+\epsilon_+}{1+c}}\\
  &&\times\prod_{I\neq J}\sin\pi\frac{\phi_{IJ}}{1+c}\prod_{I,J}
  \frac{\sin\pi\frac{\phi_{IJ}+3c}{1+c}}{
  \sin\pi\frac{\phi_{IJ}-a+c}{1+c}\sin\pi\frac{\phi_{IJ}-b+c}{1+c}}\cdot
  \frac{\sin\pi\frac{\phi_{IJ}+m-\frac{1}{2}+a}{1+c}
  \sin\pi\frac{\phi_{IJ}+m-\frac{1}{2}+b}{1+c}}{
  \sin\pi\frac{\phi_{IJ}+m+\frac{3}{2}}{1+c}
  \sin\pi\frac{\phi_{IJ}+m+\frac{3}{2}+3c}{1+c}}\nonumber
\end{eqnarray}
Again, the results for other gauge groups are given in appendix A.1.

The integration over the $k$ eigenvalues $\phi_I$ can be explicitly done
for the $U(N)$ theory, yielding a simple formula for the residues.
The correct contour for this integral is known from \cite{Nekrasov:2002qd}, also
reviewed in detail in \cite{Kim:2011mv}. The poles contributing to this integral
are classified for the $U(N)$ theory by the $N$-colored Young diagrams with $k$ boxes.
The $N$-colored Young diagram is given by the set of $N$ Young diagrams
$Y=(Y_1,Y_2,\cdots, Y_N)$ with the total number of boxes being $k$.
The residue at this pole for $U(N)$ group takes the following form \cite{Bruzzo:2002xf}:
\begin{equation}\label{U(N)-residue}
  Z_Y^{(3)}=\prod_{i,j=1}^N\prod_{s\in Y_i}
  \frac{\sin\pi\frac{E_{ij}+m+\frac{3(1+c)}{2}-\epsilon_+}{1+c}
  \sin\pi\frac{E_{ij}-m-\frac{3(1+c)}{2}-\epsilon_+}{1+c}}
  {\sin\pi\frac{E_{ij}}{1+c}\sin\pi\frac{E_{ij}-2\epsilon_+}{1+c}}
\end{equation}
with
\begin{equation}
  E_{ij}=\lambda_i-\lambda_j-\epsilon_1h_i(s)+\epsilon_2(v_j+1)\ .
\end{equation}
Here, $s$ labels the boxes in the $i$'th Young diagram $Y_i$.
$h_i(s)$ is the distance from the box $s$ to the edge on the right side of
$Y_i$ that one reaches by moving horizontally. $v_j(s)$ is the distance from
$s$ to the edge on the bottom side of $Y_j$ that one reaches by moving vertically.
See \cite{Bruzzo:2002xf,Kim:2011mv} for more explanations with examples.
Collecting all residues, and also summing over different instanton numbers, the full
instanton partition function at this fixed point is given by
\begin{equation}
  Z_{\rm inst}^{(3)}=\sum_{k=0}^\infty e^{-\frac{4\pi^2k}{\beta(1+c)}}
  \sum_{Y;\ |Y|=k}Z_Y^{(3)}
\end{equation}
with the $k$ instanton weight $e^{-\frac{4\pi^2k}{\beta(1+c)}}$ explained in section 2.1.

Similar to the above contribution $Z_{\rm pert}^{(3)}Z_{\rm inst}^{(3)}$
from the third fixed point $|Z_3|=1$, there are contributions
$Z_{\rm pert}^{(1)}Z_{\rm inst}^{(1)}$ and $Z_{\rm pert}^{(2)}Z_{\rm inst}^{(2)}$
from the first and second fixed points on $\mathbb{CP}^2$. These two are obtained
from the third one by replacing $a,b,c$ by $b,c,a$ and $c,a,b$, respectively:
namely, $a$, $b$, $c$ are playing the special roles in the formulae for the first,
second, third fixed points, respectively. The integration over the $U(N)$
eigenvalues $\lambda_i$ has to be made with this determinant and the classical
contribution (\ref{gaussian2}) in the measure. The $S^5$ partition function
takes the following form:
\begin{equation}
  Z(\beta,m,\epsilon_1,\epsilon_2)=\frac{1}{N!}\int_{-\infty}^{\infty}
  \left[\prod_{i=1}^\infty d\lambda_i\right]e^{-\frac{2\pi^2{\rm tr}\lambda^2}
  {\beta(1\!+\!a)(1\!+\!b)(1\!+\!c)}}Z_{\rm pert}^{(1)}Z_{\rm inst}^{(1)}\cdot
  Z_{\rm pert}^{(2)}Z_{\rm inst}^{(2)}\cdot Z_{\rm pert}^{(3)}Z_{\rm inst}^{(3)}\ .
\end{equation}
All factors generically depend on $\beta,m,\epsilon_1,\epsilon_2$ nontrivially,
which we suppressed in the above expression for simplicity. The perturbative
determinants are carefully considered in section 3.3. There we shall be careful
about the overall normalization issue, which will eventually make the overall
$\frac{1}{N!}$ factor above meaningful. This factor can be understood as a division
by the order of $U(N)$ Weyl group. An important feature that one should remember
is that the round sphere limit $\epsilon_1,\epsilon_2\rightarrow 0$ is smooth. This is
true only after combining all factors from the three fixed points. In particular, there are
many ways of taking the limit of the two parameters. The existence of the smooth limit
implies that same result is obtained irrespective of the relative rate
at which the two parameters are sent to zero.

\subsection{More on perturbative determinant}

Let us consider the perturbative determinant in more detail. In this subsection,
we consider general gauge group $G$. Contrary to the instanton part of the index,
the perturbative part is
an infinite series in $e^{i\epsilon_1}$, $e^{i\epsilon_2}$
at each fixed point. At each value of $t$, the index summed over all three
fixed points again becomes a finite series, related to the fact that the
$\epsilon_1,\epsilon_2\rightarrow 0$ limit of the total index is smooth.
There are two possible ways of organizing the net perturbative contribution.
One is to identify the finite series
for given $t$ after summing over three indices, and then calculating the
determinant. Another way is to first expand each perturbative index
in a definite order of $e^{i\epsilon_1}$, $e^{i\epsilon_2}$, calculating
the determinant of each and then multiplying the three. Although the latter
apparently looks inefficient, it actually provides a useful expression
of $Z_{\rm pert}$ so that we explain both in turn.

Collecting the perturbative determinants from the previous
subsection, the net perturbative equivariant index for vector multiplet
at given value of $t$ is given by
\begin{eqnarray}\label{pert-index-vector}
  I_{{\rm vector},t}^{\rm pert}(\epsilon_1,\epsilon_2)&=&
  -e^{i\frac{t}{r}}\ \frac{I_{{\rm Dol},t}^{{\rm pert},+}+
  I_{{\rm Dol},t}^{{\rm pert},-}}{2}\ {\rm ch}_{\rm adj}(e^{i\lambda})\\
  I_{{\rm Dol},t}^{{\rm pert},+}(\epsilon_1,\epsilon_2)
  &=&\frac{e^{-it(\epsilon_1+\epsilon_2)/3}}
  {(1-e^{i\epsilon_1})(1-e^{i\epsilon_2})}+\frac{e^{it(2\epsilon_1-\epsilon_2)/3}}
  {(1-e^{i(\epsilon_2-\epsilon_1)})(1-e^{-i\epsilon_1})}+\frac{e^{it(2\epsilon_2-\epsilon_1)/3}}
  {(1-e^{-i\epsilon_2})(1-e^{i(\epsilon_1-\epsilon_2)})}\nonumber\\
  I_{{\rm Dol},t}^{{\rm pert},-}(\epsilon_1,\epsilon_2)
  &=&\frac{e^{-it(\epsilon_1+\epsilon_2)/3}e^{i(\epsilon_1+\epsilon_2)}}
  {(1-e^{i\epsilon_1})(1-e^{i\epsilon_2})}+\frac{e^{it(2\epsilon_1-\epsilon_2)/3}
  e^{i(\epsilon_2-2\epsilon_1)}}{(1-e^{i(\epsilon_2-\epsilon_1)})(1-e^{-i\epsilon_1})}
  +\frac{e^{it(2\epsilon_2-\epsilon_1)/3}e^{i(\epsilon_1-2\epsilon_2)}}
  {(1-e^{-i\epsilon_2})(1-e^{i(\epsilon_1-\epsilon_2)})}\nonumber\\
  &=&I_{{\rm Dol},-t}^{{\rm pert},+}(-\epsilon_1,-\epsilon_2)\ .\nonumber
\end{eqnarray}
$I^{{\rm pert},+}_{{\rm Dol},t}$ is essentially the perturbatve part of the
Dolbeault index. Similarly, for the hypermultiplet in representation $R$,
one obtains at given $t$
(after a little rearrangement)
\begin{equation}\label{pert-index-hyper}
  I_{{\rm hyper},t}^{\rm pert}=e^{i\frac{t}{r}}\
  \frac{e^{i(m+\frac{3}{2})}I_{{\rm Dol},t}^{{\rm pert},+}+e^{-i(m+\frac{3}{2})}
  I^{{\rm pert},-}_{{\rm Dol},t}}{2}\ {\rm ch}_R(e^{i\lambda})\ .
\end{equation}
We inserted $a=\frac{\epsilon_2-2\epsilon_1}{3}$, $b=\frac{\epsilon_1-2\epsilon_2}{3}$,
$c=-\frac{\epsilon_1+\epsilon_2}{3}$ in these expressions.

If we expand $I_{{\rm Dol},t}^{{\rm pert},\pm}$, which determines all the above
indices, the sum of three factors yields a simple finite series. In terms of
$a,b,c$, one finds
\begin{equation}\label{pert-expand-vector}
  \hspace*{-1.3cm}e^{it/r}I^{{\rm pert},+}_{{\rm Dol},t}=
  \left\{\begin{array}{lll}
  e^{it/r}\sum_{p,q,r\geq 0}^{p+q+r=t}e^{i(pa+qb+rc)}\!\!&\!\!=
  \sum_{p,q,r\geq 0}^{p+q+r=t}e^{i\left(p(\frac{1}{r}+a)+q(\frac{1}{r}+b)
  +r(\frac{1}{r}+c)\right)}&(t\geq 0)\\
  0\!\!&\!\!&(t=\!-1,\!-2)\\
  e^{it/r}\sum_{p,q,r\geq 0}^{p+q+r=-(t\!+\!3)}\!e^{-i(pa+qb+rc)}\!\!&\!\!=
  \sum_{p,q,r\geq 0}^{p+q+r=-(t\!+\!3)}\!e^{-i\left((p\!+\!1)(\frac{1}{r}\!+\!a)
  +(q\!+\!1)(\frac{1}{r}\!+\!b)+(r\!+\!1)(\frac{1}{r}\!+\!c)\right)}&
  (t\leq -3)
  \end{array}\right..
\end{equation}
As for $I^{{\rm pert},-}_{{\rm Dol},t}(a,b,c)=
I^{{\rm pert},+}_{{\rm Dol},-t}(-a,-b,-c)$, one finds
\begin{equation}\label{pert-expand-hyper}
  \hspace*{-1.3cm}e^{it/r}I^{{\rm pert},-}_{{\rm Dol},t}=
  \left\{\begin{array}{lll}
  e^{it/r}\sum_{p,q,r\geq 0}^{p+q+r=t-3}e^{i(pa+qb+rc)}\!\!&\!\!=
  \sum_{p,q,r\geq 0}^{p+q+r=t-3}e^{i\left((p\!+\!1)(\frac{1}{r}\!+\!a)
  +(q\!+\!1)(\frac{1}{r}\!+\!b)+(r\!+\!1)(\frac{1}{r}\!+\!c)\right)}&(t\geq 3)\\
  0\!\!&\!\!&(t=\!1,\!2)\\
  e^{it/r}\sum_{p,q,r\geq 0}^{p+q+r=-t}\!e^{-i(pa+qb+rc)}\!\!&\!\!=
  \sum_{p,q,r\geq 0}^{p+q+r=-t}\!e^{i\left(p(\frac{1}{r}\!+\!a)
  +q(\frac{1}{r}\!+\!b)+r(\frac{1}{r}\!+\!c)\right)}&
  (t\leq -3)
  \end{array}\right..
\end{equation}
Note that $I^{{\rm pert},+}_{{\rm Dol},t}$ with $t\geq 0$ is simply the degeneracy
for rank $t$ holomorphic polynomials in $\mathbb{C}^3$. The
$\epsilon_1,\epsilon_2\rightarrow 0$ limit agrees with the perturbative
index studied in \cite{Kallen:2012cs},
\begin{equation}\label{index-reduced}
  \lim_{\epsilon_1,\epsilon_2\rightarrow 0}I^{{\rm pert},+}_{{\rm Dol},t}=
  \frac{(t+1)(t+2)}{2}\ .
\end{equation}
The spectra of $I_{{\rm Dol},t}^{{\rm pert},\pm}$ are completely
mapping to each other with a sign flip on the eigenvalues. Since the remaining
factor ${\rm ch}_{\rm adj}$ is real, the contributions of two factors in the
numerator of the vector multiplet index (\ref{pert-index-vector}) yield
same factors. The same is true for the hypermultiplet if the representation
$R$ is real, but not in general.

With this information,
one can immediately write down the determinant as the following formal product
(redefining $ra_i$ to be new $a_i$)\footnote{By abusing the notion of
roots, in this subsection we definite $\alpha$ to run over all weights  of generators,
including Cartans. Also, the scalar expectataion value $\lambda$ ins 5d QFT is real,
so that one will have to replace $\lambda\rightarrow i\lambda$ to use these determinants
in field theory.}:
\begin{equation}
  \hspace*{-1cm}{\rm det}_{\rm V}=\prod_{\alpha\in{\rm root}}\prod_{p,q,r=0}^\infty\!\!
  \left(\frac{}{}\!p(1\!+\!a)\!+\!q(1\!+\!b)\!+\!r(1\!+\!c)\!+\!\alpha(\lambda)\right)
  \left(\!\frac{}{}\!(p\!+\!1)(1\!+\!a)\!+\!(q\!+\!1)(1\!+\!b)\!+\!
  (r\!+\!1)(1\!+\!c)\!+\!\alpha(\lambda)\right)\ .
\end{equation}
One has remember that the index theorem is simply counting the BPS modes,
so whether we are Gaussian integrating them or not is what we should decide. In
particular, the mode in the first factor with $p=q=r=0$ is not to be kept when
$\alpha=0$, as they are constant scalar modes $\lambda$ which we should integrate
over exactly at the final stage. For the hypermultiplet, one obtains
\begin{eqnarray}
  \hspace*{-1.7cm}{\rm det}_{\rm H}\!&\!=\!&\!\!\prod_{\mu\in{\rm weight}}\prod_{p,q,r=0}^\infty\!\!
  \left(\!p(1\!+\!a)\!+\!q(1\!+\!b)\!+\!r(1\!+\!c)\!+\!m\!+\!\frac{3}{2}
  +\!\mu(\lambda)\right)^{-\frac{1}{2}}\!\!\left(\frac{}{}\!p(1\!+\!a)\!+\!
  q(1\!+\!b)\!+\!r(1\!+\!c)\!+\!m\!+\!\frac{3}{2}-\!\mu(\lambda)\right)^{-\frac{1}{2}}
  \nonumber\\
  \hspace*{-1.7cm}&&\times\left(\!\frac{}{}\!(p\!+\!1)(1\!+\!a)\!+\!(q\!+\!1)(1\!+\!b)\!+\!
  (r\!+\!1)(1\!+\!c)\!-\!m\!-\!\frac{3}{2}\!+\!\mu(\lambda)\right)^{-\frac{1}{2}}\nonumber\\
  \hspace*{-1.7cm}&&\times\left(\!\frac{}{}\!(p\!+\!1)(1\!+\!a)\!+\!(q\!+\!1)(1\!+\!b)\!+\!
  (r\!+\!1)(1\!+\!c)\!-\!m\!-\!\frac{3}{2}\!-\!\mu(\lambda)\right)^{-\frac{1}{2}}\ .
\end{eqnarray}
When the representation $R$ is real, we can replace $-\mu(\lambda)$ in
the product by $\mu(\lambda)$, after which one obtains
\begin{eqnarray}
  \hspace*{-1.7cm}{\rm det}_{\rm H}\!&\!=\!&\!\!\prod_{\mu\in{\rm weight}}\prod_{p,q,r=0}^\infty\!\!
  \left(\!p(1\!+\!a)\!+\!q(1\!+\!b)\!+\!r(1\!+\!c)\!+\!m\!+\!\frac{3}{2}
  +\!\mu(\lambda)\right)^{-1}\nonumber\\
  \hspace*{-1.7cm}&&\times\left(\!\frac{}{}\!(p\!+\!1)(1\!+\!a)\!+\!(q\!+\!1)(1\!+\!b)\!+\!
  (r\!+\!1)(1\!+\!c)\!-\!m\!-\!\frac{3}{2}\!+\!\mu(\lambda)\right)^{-1}\ .
\end{eqnarray}
For the theory with an adjoint hypermultiplet, which is our main interest,
one can use the last formula. All infinite products have to be regularized,
after which one obtains a generalization of the triple sine function. To naturally do so,
it should be desirable to rewrite the above $p,q,r$ prodcut (which is democratic in
the $a,b,c$) as an infinite product over $t$ and further to a finite product in
two integers, like the restricted sums which appear in (\ref{pert-expand-vector}),
(\ref{pert-expand-hyper}). The $t$ product could be regularized using zeta functions.
One should obtain same pre-factors as those studied in the
$\epsilon_1,\epsilon_2\rightarrow 0$ limit \cite{Kallen:2012cs,Kallen:2012va,Kim:2012av}.

When $a\!=\!b\!=\!c\!=\!0$, only $p+q+r\equiv t$ appears in the product. The perturbative
determinant cam be written as (we changed
$\lambda\rightarrow i\lambda$ to go to QFT convention)
\begin{eqnarray}
  {\det}_V&=&\prod_\alpha\prod_{t=0}^\infty(t+i\alpha(\lambda))^{\frac{(t+1)(t+2)}{2}}
  (t+3+i\alpha(\lambda))^{\frac{(t+1)(t+2)}{2}}=
  \prod_{\alpha}\alpha(\lambda)\prod_{t=1}^\infty
  \left(t+i\alpha(\lambda)\right)^{t^2+2}\nonumber\\
  {\det}_H&=&\prod_\mu\prod_{t=0}^\infty\left(t+\frac{3}{2}+m+i\mu(\lambda)
  \right)^{-\frac{1}{2}\cdot\frac{(t+1)(t+2)}{2}}
  \left(t+\frac{3}{2}+m-i\mu(\lambda)\right)^{-\frac{1}{2}\cdot\frac{(t+1)(t+2)}{2}}\\
  &&\times\left(t+\frac{3}{2}-m+i\mu(\lambda)\right)^{-\frac{1}{2}\cdot\frac{(t+1)(t+2)}{2}}
  \left(t+\frac{3}{2}-m-i\mu(\lambda)\right)^{-\frac{1}{2}\cdot\frac{(t+1)(t+2)}{2}}\ .\nonumber
\end{eqnarray}
$\det_V$ was calculated in \cite{Kallen:2012cs}.
When the hypermultiplet representation is real, one obtains
\begin{equation}
  {\det}_H=\prod_\mu\prod_{t=0}^\infty\left(t+\frac{3}{2}+m+i\mu(\lambda)
  \right)^{-\frac{(t+1)(t+2)}{2}}\left(t+\frac{3}{2}-m+i\mu(\lambda)
  \right)^{-\cdot\frac{(t+1)(t+2)}{2}}\ .
\end{equation}
This agrees with the result of \cite{Kim:2012av}, with the parameter $\Delta$ in
that paper related to $m$ here by $\Delta=m+\frac{1}{2}$. It was also noted that,
at $m=\pm\frac{1}{2}$ when we expect to have SUSY enhancement on $S^5$ ,
$\det_V\det_H$ experiences further cancelation so that the net measure becomes
the product of $\sinh$ functions \cite{Kim:2012av}. This form will be used
in section 3.2 to study a particular class of 6d indices.

It would also be useful later to have a different form of the above product,
obtained by directly expanding each fixed point index and recasting them into
a Plethystic-like exponential. For instance, we expand all denominators of
the indices in power series of $e^{i\epsilon_1}$, $e^{i\epsilon_2}$ assuming
${\rm Im}\ \epsilon_1>{\rm Im}\ \epsilon_2>0$. By expanding in different orders,
one would be able to get similar expressions with the roles of $a,b,c$ exchanged.
After some calculation, one obtains the following product form for $\det_V$:
\begin{eqnarray}\label{pert-expand-vector-2}
  {\rm det}_V&=&\!\!\prod_\alpha\prod_{t=-\infty}^{\infty}\prod_{j,k=0}^\infty
  \!\!\Big((1\!+\!a)t\!+\!j\epsilon_1\!+\!k\epsilon_2\!+\!\alpha(\lambda)\!\Big)^{\frac{1}{2}}
  \!\Big((1\!+\!a)t\!+\!(j\!+\!1)\epsilon_1\!+\!(k\!+\!1)\epsilon_2\!+\!
  \alpha(\lambda)\!\Big)^{\frac{1}{2}}\\
  &&\quad \left.\times \ \Big((1\!+\!b)t\!+\!j\epsilon_1\!+\!k(\epsilon_1\!-\!\epsilon_2)\!+\!
  \alpha(\lambda)\Big)^{\frac{1}{2}}
  \Big((1\!+\!b)t\!+\!(j\!+\!1)\epsilon_1\!+\!(k\!+\!1)(\epsilon_1\!-\!\epsilon_2)\!+\!
  \alpha(\lambda) \Big)^{\frac{1}{2}}\right.\nn \\
  &&\quad\times \ \Big((1\!+\!c)t\!+\!(j\!+\!1)\epsilon_2\!+\!k(\epsilon_1\!-\!\epsilon_2)\!+\!
  \alpha(\lambda)\Big)^{-\frac{1}{2}}
  \Big((1\!+\!c)t\!+\!j\epsilon_2\!+\!(k\!+\!1)(\epsilon_1\!-\!\epsilon_2)\!+\!
  \alpha(\lambda)  \Big)^{-\frac{1}{2}}\ .\nn
\end{eqnarray}
The variable $t$ in the product is the same $t$ we used for the index. $j,k$ are for
the infinite series expansions of various denominators in the indices. We emphasize
again here that, in the above product over $t,j,k$, the first factor on the first line
and the first factor on the second line with $t=j=k=0$ has to be excluded in the product
when $\alpha\!=\!0$, corresponding to the Cartans.

One can factor out
all the $(1+a)t$, $(1+b)t$, $(1+c)t$ terms in the product, and then take a product
over $t$ to obtain an expression which becomes and infinite product over $j,k$ with
sine functions. After rearrangements, and pulling out the formally divergent prefactors,
one obtains (with given $\alpha$)
\be\label{vector-pert}
    \hspace*{-1.5cm}{\rm det}_V&=& \prod_{t=1}^\infty\prod_{j,k=0}^\infty\left(\frac{(1\!+\!a)(1\!+\!b)}{1\!+\!c}t\right)^2
    \times\prod_{j,k=0}^\infty\frac{(1\!+\!a)(1\!+\!b)}{\pi(1+c)}\nn \\
    \hspace{-.5cm}&&\times \left[\prod_{j,k}'\frac{\sin\pi\frac{j\epsilon_1\!+\!k\epsilon_2\!+\!\lambda}{1+a}\sin\pi\frac{(j\!+\!1)\epsilon_1\!+\!(k\!+\!1)\epsilon_2\!+\!\lambda}{1+a}
    \sin\pi\frac{j\epsilon_1\!+\!k(\epsilon_1\!-\!\epsilon_2)\!+\!\lambda}{1+b}\sin\pi\frac{(j\!+\!1)\epsilon_1\!+\!(k\!+\!1)(\epsilon_1\!-\!\epsilon_2)\!+\!\lambda}{1+b}}
    {\sin\pi\frac{(j\!+\!1)\epsilon_2\!+\!k(\epsilon_1\!-\!\epsilon_2)\!+\!\lambda}{1+c}\sin\pi\frac{j\epsilon_2\!+\!(k\!+\!1)(\epsilon_1\!-\!\epsilon_2)\!+\!\lambda}{1+c}}\right]^{1/2} \nn \\
   \hspace{-.5cm} &=&\mathcal{N} \left[\prod_{j,k}'\frac{\left(1\!-\!e^{2\pi i\frac{j\epsilon_1\!+\!k\epsilon_2\!+\!\lambda}{1+a}}\right)
    \left(1\!-\!e^{2\pi i\frac{(j\!+\!1)\epsilon_1\!+\!(k\!+\!1)\epsilon_2\!+\!\lambda}{1+a}}\right)
    \left(1\!-\!e^{2\pi i\frac{j\epsilon_1\!+\!k(\epsilon_1\!-\!\epsilon_2)\!+\!\lambda}{1+b}}\right)
    \left(1\!-\!e^{2\pi i\frac{(j\!+\!1)\epsilon_1\!+\!(k\!+\!1)(\epsilon_1\!-\!\epsilon_2)\!+\!\lambda}{1+b}}\right)}
    {\left(1\!-\!e^{2\pi i\frac{(j\!+\!1)\epsilon_2\!+\!k(\epsilon_1\!-\!\epsilon_2)\!+\!\lambda}{1+c}}\right)
    \left(1\!-\!e^{2\pi i\frac{j\epsilon_2\!+\!(k\!+\!1)(\epsilon_1\!-\!\epsilon_2)\!+\!\lambda}{1+c}}\right)}\right]^{1/2} \nn \\
   \hspace{-.5cm} &=&\mathcal{N}\ {\rm PE}'\left[-\frac{\left(1+e^{2\pi i\frac{\epsilon_1\!+\!\epsilon_2}{1+a}}\right)e^{2\pi i \frac{\lambda}{1+a}}}{2\!\left(1\!-\!e^{2\pi i\frac{\epsilon_1}{1+a}}\right)\!\!\left(1\!-\!e^{2\pi i\frac{\epsilon_2}{1+a}}\right)}
    -\frac{\left(1+e^{2\pi i\frac{2\epsilon_1\!-\!\epsilon_2}{1+b}}\right)e^{2\pi i \frac{\lambda}{1+b}}}{2\!\left(1\!-\!e^{2\pi i\frac{\epsilon_1}{1+b}}\right)\!\!\left(1\!-\!e^{2\pi i\frac{\epsilon_1-\epsilon_2}{1+b}}\right)}
    +\frac{\left(e^{2\pi i\frac{\epsilon_2}{1+c}}+e^{2\pi i\frac{\epsilon_1\!-\!\epsilon_2}{1+c}}\right)e^{2\pi i \frac{\lambda}{1+c}}}{2\!\left(1\!-\!e^{2\pi i\frac{\epsilon_2}{1+c}}\right)\!\!\left(1\!-\!e^{2\pi i\frac{\epsilon_1-\epsilon_2}{1+c}}\right)}\right] \nn \\
   \hspace{-.5cm} &=&\mathcal{N}\ {\rm PE}'\left[\frac{e^{\pi i\frac{b+c-2a}{1+a}}\!+\!e^{-\pi i\frac{b+c-2a}{1+a}}}{8\sin\pi\frac{b-a}{1+a}\sin\pi\frac{c-a}{1+a}}e^{2\pi i\frac{\lambda}{1+a}}
    +(a,b,c\rightarrow b,c,a)+(a,b,c\rightarrow c,a,b)\right]
\ee
where
\be
    \mathcal{N}_V\!\!&\!\!=\!\!&\!\!\mathcal{M}_\alpha
    \prod_{j,k=0}^\infty\left[\prod_{t=1}^\infty\left(\frac{(1\!+\!a)(1\!+\!b)}{1\!+\!c}t\right)^2\cdot
    \frac{(1\!+\!a)(1\!+\!b)}{2\pi i(1+c)} \right. \\
    &&\!\!\left.\cdot\ {\rm exp}\left[-\pi i\frac{(2j\!+\!1)\epsilon_1\!+\!(2k\!+\!1)\epsilon_2\!}{1+a}-\pi i\frac{(2j\!+\!1)\epsilon_1\!+\!(2k\!+\!1)(\epsilon_1\!-\!\epsilon_2)\!}{1+b} +\pi i\frac{(2j\!+\!1)\epsilon_2\!+\!(2k\!+\!1)(\epsilon_1\!-\!\epsilon_2)\!}{1+c}\right]\right] \nn
\ee
with
\begin{equation}
  \mathcal{M}_\alpha=\left\{
  \begin{array}{ll}
  1&{\rm if}\
  \alpha\ {\rm is\ a\ root}\\
  \frac{1}{(1+a)^{\frac{1}{2}}(1+b)^{\frac{1}{2}}}&{\rm if}\
  \alpha=0,\ {\rm for\ a\ Cartan}
  \end{array}\right.\ .
\end{equation}
Here, PE denotes the Plethystic-like exponential, regarding all expression like
$\frac{\epsilon_i}{1+a_i}$, $\frac{a_i}{1+a_j}$, $\frac{\lambda}{1+a_i}$
as `chemical potentials.' The primes in the product only applies when $\alpha=0$,
corresponding to states in the Cartan. The prime in this case excludes the contributions from
`zero eigenvalues' for $t=j=k=0$, and the primes in PE also exclude the corresponding $-1$ 's
in the exponential, which will yield $0$. Also, the structure of the prefactor $\mathcal{N}$
needs explanation. the factor $\frac{(1+a)(1+b)}{\pi(1+c)}$ on the fist
line inside the $j,k$ product appears by extracting the $\frac{1+a_i}{\pi}$ factors from the
eigenvalues in (\ref{pert-expand-vector-2}) with $t=0$, as they are providing the leading
linear factor $x$ in $\sin(\pi x)=\pi x\prod_{t\neq 0} \left(1+\frac{x}{t}\right)$.
However, note that one does not have such sine factors for $j=k=0$: for this, there are
no $1+a$ and $1+b$ factors to be pulled out into $\mathcal{N}$. This explains
the division by $(1+a)^{\frac{1}{2}}(1+b)^{\frac{1}{2}}$ in $\mathcal{M}_\alpha$.
Later, when one considers a gauge theory with
one adjoint hypermultiplet, all the prefactors $\mathcal{N}$ from vector and hypermultiplets
cancel out except for this $\mathcal{M}_\alpha$ part, as there are no zero eigenvalues in
the hypermultiplet to be removed from the product.
The exclusion of $-1$ in the exponent of PE for the Cartan components
can be expressed more manifestly by writing
\begin{equation}
  {\rm det}_V=\mathcal{N}_V\ {\rm PE}\left[\frac{e^{\pi i\frac{b+c-2a}{1+a}}\!+\!e^{-\pi i\frac{b+c-2a}{1+a}}}{8\sin\pi\frac{b-a}{1+a}\sin\pi\frac{c-a}{1+a}}
  +(a,b,c\rightarrow b,c,a)+(a,b,c\rightarrow c,a,b)+1\right]\ .
\end{equation}
In the expansion with ${\rm Im}\ \epsilon_1>{\rm Im}\ \epsilon_2>0$ that
we have assumed, the $+1$ at the end should be divided into $+\frac{1}{2}+\frac{1}{2}$
and each $+\frac{1}{2}$ has to combine with the first and second term in the exponent
to make the expansions free of the constant term $-\frac{1}{2}$'s (i.e. zero eigenvalues).
For the hypermultiplet, very similar expansion can be made without worrying about
the zero eigenvalues, even for $t=j=k=0$. After a similar algebra, one obtains
\begin{equation}
  {\det}_H=\mathcal{N}_H^{-1}\ {\rm P.E}\left[-\frac{e^{\pi i\frac{2m+3(1+a)}{(1+a)}}+e^{-\pi i\frac{2m+3(1+a)}{(1+a)}}}{8\sin\pi\frac{b-a}{1+a}\sin\pi\frac{c-a}{1+a}}e^{2\pi i\frac{\lambda}{1+a}} +(a,b,c\rightarrow b,c,a)+(a,b,c\rightarrow c,a,b)\right]\qquad
\end{equation}
where $\mathcal{N}_H$ takes exactly the same form as $\mathcal{N}_V$ above,
except for the absence of $\mathcal{M}_\alpha$ factor.

We also explain the form of the net perturbative determinant for the theory with
one adjoint hypermultiplet, combining $\det_V$ and $\det_H$, which will be our main
interest in the next section.
Firstly, the prefactors $\mathcal{N}_V$, $\mathcal{N}_H$ almost cancel but not exactly,
due to the absence of $t=j=k=0$ zero modes in the vector multiplet. The net pre-factor is
\begin{equation}\label{prefactor}
  \frac{\mathcal{N}_V}{\mathcal{N}_H}=\left[
  \frac{1}{(1+a)^{\frac{1}{2}}(1+b)^{\frac{1}{2}}}\right]^{r}
\end{equation}
where $r$ is the rank of the gauge group. The prefactor in this expression is asymmetic
in the exchange of $a,b,c$, simply because we are expanding $Z_{\rm pert}$ in an
asymmetric way. Such an exchange symmetry should be there, although implicit in the
expressions. The remaining part can also be
organized as
\be\label{pert-index-rewritten}
    \hspace*{-1.3cm}&&\frac{\mathcal{N}_V}{\mathcal{N}_H}{\rm PE}^\prime\left[\frac{-e^{\pi i\frac{2m+3(1+a)}{(1+a)}}\!-\!e^{-\pi i\frac{2m+3(1+a)}{(1+a)}}\!+\!e^{\pi i\frac{b+c-2a}{1+a}}\!+\!e^{-\pi i\frac{b+c-2a}{1+a}}}{8\sin\pi\frac{b-a}{1+a}\sin\pi\frac{c-a}{1+a}}\sum_{\alpha}
    e^{2\pi i\frac{\alpha(\lambda)}{1+a}} +(a,b,c\ \ {\rm permutation})\right] \nn \\
    \hspace*{-1.3cm}&&=\frac{\mathcal{N}_V}{\mathcal{N}_H}
    {\rm exp}\left[\sum_{p=1}^\infty\frac{1}{2p}\left(f(p,a,b,c)\sum_\alpha
    e^{\frac{2\pi ip\alpha(\lambda)}{1+a}}+(a,b,c\rightarrow b,c,a)+(a,b,c\rightarrow c,a,b)
    {\rm ``-1"}\right)\right]
\ee
with
\begin{equation}\label{f-pert}
  f(p,a,b,c)\equiv\frac{\sin\frac{p\pi(m-\frac{1}{2}+b)}{1+a}
  \sin\frac{p\pi(m-\frac{1}{2}+c)}{1+a}}{\sin\frac{p\pi(a-b)}{1+a}
  \sin\frac{p\pi(a-c)}{1+a}}\ .
\end{equation}
The ``$-1$'' subtraction from the absent zero modes is to be made only when $\alpha=0$.
In the expansion prescription of denominators that we are advocating, this $-1$ has
to combine with the third term in the exponent to make it finite. The last structure,
together with the prefactor (\ref{prefactor}) will be important for comparing our
Abelian partition function with the 6d Abelian index in section 3.1.

\section{Applications}

In this section, we shall study our $S^5$ partition function as the
6d $(2,0)$ superconformal index on $S^5\times  S^1$. Before studying specific
cases, let us start by explaining the general structure of this index.

The 6d $(2,0)$ theory, with $OSp(6,2|4)$ symmetry, has
$SO(6,2)\times SO(5)$ bosonic symmetry with $6$ Cartans $R_1$, $R_2$,
 $j_1$, $j_2$, $j_3$, $E$ as explained in section 2.1.
Among the $32$ supercharges, $16$ Poincare supercharges $Q$'s
have $E=\frac{1}{2}$, $(R_1,R_2)=(\pm\frac{1}{2},\pm\frac{1}{2})$ as a 4-component
$SO(5)$ spinor, and finally $(j_1,j_2,j_3)=(\pm\frac{1}{2},\pm\frac{1}{2},\pm\frac{1}{2})$
with the constraint that the product of three eigenvalues is negative.
Conformal supercharges $S$'s are Hermitian conjugate to $Q$'s, so they have
conjugate charge contents to $Q$'s. We choose
$Q^{R_1,R_2}_{j_1,j_2,j_3}=Q^{+,+}_{-,-,-}\equiv Q$ and its conjugate
$S^{R_1,R_2}_{j_1,j_2,j_3}=S^{-,-}_{+,+,+}\equiv S$, and consider an index which
counts states (or operators) annihilated by them \cite{Bhattacharya:2008zy}.
The superconformal algbera yields
\begin{equation}
  \{Q,S\}\sim E-2(R_1+R_2)-(j_1+j_2+j_3)\ .
\end{equation}
The states counted by the index that we shall define in a moment
saturate the following BPS energy bound, given by their five Cartan charges:
\begin{equation}\label{BPS-bound}
  E\geq 2(R_1+R_2)+(j_1+j_2+j_3)\ .
\end{equation}
So for these BPS states, we only have $5$ independent charges labeling them. Among
these, only four of them can be used to define a Witten index
\cite{Bhattacharya:2008zy}. This is because the Witten index demands that
only the combinations of these $5$ independent charges which commute with $Q,S$
be used to weight the states. We may take the following $4$ linear
combinations:
\begin{equation}
  E-\frac{R_1+R_2}{2}\ ,\ \ R_1-R_2\ ,\ \ j_1-j_3\ ,\ \ j_2-j_3\ .
\end{equation}
The superconformal index is defined by
\begin{equation}
  I(\beta,m,\epsilon_1,\epsilon_2)={\rm Tr}\left[(-1)^Fe^{-\beta^\prime\{Q,S\}}
  e^{-\beta(E-\frac{R_1+R_2}{2})}e^{\beta m(R_1-R_2)}
  e^{-\gamma_1(j_1-j_3)}e^{-\gamma_2(j_2-j_3)}\right]\ .
\end{equation}
where the trace is taken over the Hilbert space. $\beta^\prime$ is a regulator
which does not appear in the index. As for the chemical potentials
$\gamma_1,\gamma_2$, one may introduce $a,b,c$ (subject to the
constraint $a+b+c=0$) and rewrite
\begin{equation}
  e^{-\gamma_1(j_1-j_3)}e^{-\gamma_2(j_2-j_3)}=e^{-\beta(aj_1+bj_2+cj_3)}\
\end{equation}
inside the trace, relating $\gamma_1=\beta a$, $\gamma_2=\beta b$. The
parameters $a,b,c$ are the squashing parameters in the 5d theory that we
have been discussing in section 2. Also, the $\beta,m$
are interpreted on $S^5$ as the square of gauge coupling $g_{YM}^2=2\pi\beta$
and the hypermultiplet mass \cite{Kim:2012av}.

\subsection{Index for the Abelian 6d $(2,0)$ theory}

The index for Abelian 6d $(2,0)$ theory can be calculated directly in 6d,
as the theory is free. In this sense, this theory is somewhat trivial. However,
it provides a highly nontrivial testing ground for tour 5d approach,
and especially to concretely see how it works.

Apparently, note that the dimensionally reduced Abelian 5d theory is also free.
In particular, this means that the perturbative part of the partition function
is given by a simple Gaussian matrix integral as there are no perturbative
interactions at all. However, the instanton part is often subtle even for the
Abelian theories, as one is often required to consider singular small instantons
in the quantum theory to do the correct physics \cite{Witten:1995gx}. Strictly
speaking, these small instantons are beyond the rigorous scope of our QFT in
a narrow sense. However, there have been various proposals to regularize the
singular instantons (including the Abelian instantons that we discuss here) by
introducing certain UV regulators. For instance, in the $U(N)$ theory, noncommutative
deformation was shown to admit such a regularization at short distance
\cite{Nekrasov:1998ss}, making even Abelian instantons to be regular QFT solitons.
We think these small instantons could be providing (maybe a small amount of) UV
completion data beyond the 5d QFT to make the latter useful to study 6d SCFT.
See section 4 for the summary of all such ambiguities in our approach.
Anyway, these small instantons make even the Abelian 5d theory to be nontrivial
enough to reproduce the 6d physics.

We also note that studies of the Abelian theories in 5d/6d are recently made
on $T^5$ and $T^6$ \cite{Dolan:2012wq,Bak:2012ct}. In particular, the approach of
\cite{Bak:2012ct} for studying the Abelian instantons on spatial $T^4$ may be useful
even for studying our non-Abelian small instantons on $\mathbb{CP}^2$ with a $U(1)$ fiber.

In our convention for the chemical potential, the 6d Abelian index is given in terms
of the following `letter index' \cite{Bhattacharya:2008zy}:
\begin{equation}\label{letter}
  f(\beta,m,a,b,c)=\frac{e^{-\frac{3\beta}{2}}(e^{\beta m}+e^{-\beta m})-
  e^{-2\beta}(e^{\beta a}+e^{\beta b}+e^{\beta c})+e^{-3\beta}}
  {(1-e^{-\beta(1+a)})(1-e^{-\beta(1+b)})(1-e^{-\beta(1+c)})}\ .
\end{equation}
One can quickly understand various terms and factors as follows. The free 6d
tensor supermultiplet contains $5$ real scalars in $SO(5)$ vector,
$16$ fermions $\psi^{R_1,R_2}_{j_1,j_2,j_3}$ in $SO(5)$ spinor and $SO(6)$ chiral
spinor (same chirality as the Poincare SUSY), and a self-dual 3-form tensor.
Among them, the BPS fields with respect to $Q,S$ defined above are given
as follows: two complex scalars which have $(R_1,R_2)$ charges
$(1,0)$ and $(0,1)$, three fermions $\psi^{+,+}_{-,+,+}$, $\psi^{+,+}_{+,-,+}$,
$\psi^{+,+}_{+,+,-}$, and none from the 3-form field. The first two terms in the
numerator of (\ref{letter}) come from two BPS scalars, and  the next three terms
from the three BPS fermions. The three factors in the denominator come from
the multiplications of three BPS derivatives, which carry
scale dimension $E=1$ and $(j_1,j_2,j_3)$ charges $(1,0,0)$, $(0,1,0)$, $(0,0,1)$,
respectively. So one is only left to understand the last term of the numerator.
This comes from a BPS constraint from fermion equation of motion, as one component of
their Dirac equation turns out to be constructible from BPS fermions and derivatives
only. Note that $\slash\hspace{-0.24cm}\partial\psi$ is anti-chiral, and one of its
$(j_1,j_2,j_3)$ charges is $(\frac{1}{2},\frac{1}{2},\frac{1}{2})$. This, together
with its scale dimension $E=\frac{7}{2}$ and $R_1=R_2=\frac{1}{2}$, saturates the
BPS bound (\ref{BPS-bound}). The full Abelian index is the Plethystic
exponential of $f$:
\begin{eqnarray}\label{Abelian-full-index}
  \hspace*{-1cm}I&=&PE[f]\equiv\exp\left[\sum_{p=1}^\infty\frac{1}{p}f(p\beta,m,a,b,c)\right]\\
  \hspace*{-1cm}&=&\prod_{n_1=0}^\infty\prod_{n_1=0}^\infty\prod_{n_3=0}^\infty
  \frac{(1-q^{2+n_1+n_2+n_3}\zeta_1^{n_1-1}\zeta_2^{n_2}\zeta_3^{n_3})
  (1-q^{2+n_1+n_2+n_3}\zeta_1^{n_1}\zeta_2^{n_2-1}\zeta_3^{n_3})
  (1-q^{2+n_1+n_2+n_3}\zeta_1^{n_1}\zeta_2^{n_2}\zeta_3^{n_3-1})}
  {(1-yq^{\frac{3}{2}+n_1+n_2+n_3}\zeta_1^{n_1}\zeta_2^{n_2}\zeta_3^{n_3})
  (1-y^{-1}q^{\frac{3}{2}+n_1+n_2+n_3}\zeta_1^{n_1}\zeta_2^{n_2}\zeta_3^{n_3})
  (1-q^{3+n_1+n_2+n_3}\zeta_1^{n_1}\zeta_2^{n_2}\zeta_3^{n_3})}\nonumber
\end{eqnarray}
where $q\equiv e^{-\beta}$, $y\equiv e^{-\beta m}$,
$\zeta_i\equiv (e^{-\beta a}, e^{-\beta b}, e^{-\beta c})$ for $i=1,2,3$ (satisfying
$\zeta_1\zeta_2\zeta_3=1$).

To the above $I$, which just counts BPS excitations, one can multiply the `index
Casimir energy' factor \cite{Kim:2012av}, which is the summations of all zero point
charges multiplied by the chemical potentials. With the chemical potentials
$\beta$, $\beta m$, $\beta a_i$ ($i=1,2,3$), this factor is an
exponential of $\beta$ times some quantity which does not depend on $\beta$.
For instance, in the free theory, the calculation can be done as follows, by introducing
a UV regulator $\beta^\prime$ as usual for zero point quantities:
\begin{equation}
  \lim_{\beta^\prime\rightarrow 0}
  {\rm tr}_{\rm modes}\left[(-1)^F\left(E-\frac{R_1+R_2}{2}-m(R_1-R_2)+a_ij_i\right)
  e^{-\beta^\prime\left(E-\frac{R_1+R_2}{2}-m(R_1-R_2)+a_i j_i\right)}\right]\ ,
\end{equation}
where the regulating exponential is introduced in the most general form,
subject to the constraint that it has to commute with $Q,S$ which are symmetries of
our index. The linear appearances of $m,a_i$ labels different kinds of
zero point charges. Once we do this calculation for the free theory,
eliminating or renormalizing the divergences proportional to the inverse powers of
$\beta^\prime$, one obtains The `Casimir energy' factor
\begin{equation}\label{casimir-ambiguous}
  e^{-\beta\epsilon_0}\equiv
  \exp\left[\frac{\beta}{24}\left(1+\frac{2abc+
  (1-ab-bc-ca)\delta+\delta^2}{(1+a)(1+b)(1+c)}\right)\right]
\end{equation}
where $\delta\equiv\frac{1}{4}-m^2$. The fact that this quantity depends on
the chemical potential ratios $m$, $a_i$ in a complicated way, meaning not linear,
is quite puzzling. This is because one naturally expects the exponent to be linear
in all chemical potentials $\beta$, $\beta m$, $\beta a_i$, in which case their
coefficients can be interpreted as zero point charges. For instance, in the 3d
indices on $S^2\times S^1$, such an expected behavior perfectly showed up after
a localization calculation \cite{Kim:2009wb}. As the expressions are not linear,
it is not clear how one should interpret this quantity. It might have to do with
some wrong regularization/renormalization that we have done. However, a curious
fact is that (\ref{casimir-ambiguous}) will be precisely reproduced from the
5d calculus, as we shall explain. Note that, in the case with maximal SYM on $S^5$,
demanding the regulator to commute with all $16$ SUSY uniquely fixes the possible
regulator to be $e^{-\beta(E-R_1)}$ for the index studied in \cite{Kim:2012av}.
As there are no more chemical potentials ratios one can introduce, the result
does not depend on the continuous parameters of the theory.
We shall discuss it more in section 4.

The indices (\ref{letter}) or (\ref{Abelian-full-index}) naturally admits a
strong coupling expansion in 5d perspective if we interpret $\beta$ as the
5d gauge coupling, as we expand the index in $q=e^{-\beta}\ll 1$ for $\beta\gg 1$.
On the other hand, to see the 5d gauge theory structures that we derived at
weak coupling, one has to expand the index at $\beta\ll 1$. In general, given an
expression which
manifestly shows an index structure in $q=e^{-\beta}$, performing a high temperature
$\beta$ expansion to all orders is difficult. However, when the expression manifestly
takes the form of a Plethystic exponential, one can borrow the techniques used by
Gopakumar-Vafa \cite{Gopakumar:1998ii} which provides a conversion between strong
and weak coupling expansions. \cite{Gopakumar:1998ii}
first proposed this when the index is given by the MacMahon function, but the same
technique applies to our Abelian index.

To perform the dual expansion easily, we start by noting that the letter index
(\ref{letter}) can be rearranged to following form:
\begin{eqnarray}\label{letter-massage}
  f&=&\frac{e^{-\beta(1+c)}}{1-e^{-\beta(1+c)}}-
  \frac{e^{-\beta(1+c)}(1-e^{-\beta(\frac{1}{2}+m-c)})(1-e^{-\beta(\frac{1}{2}-m-c)})}
  {(1-e^{-\beta(1+a)})(1-e^{-\beta(1+b)})(1-e^{-\beta(1+c)})}\nonumber\\
  &=&\frac{e^{-\beta(1+c)}}{1-e^{-\beta(1+c)}}-
  \frac{\sinh\frac{\beta(1+2m-2c)}{4}\sinh\frac{\beta(1-2m-2c)}{4}}
  {2\sinh\frac{\beta(1+a)}{2}\sinh\frac{\beta(1+b)}{2}\sinh\frac{\beta(1+c)}{2}}\ .
\end{eqnarray}
Of course this decomposition obscures the manifest symmetry under $a,b,c$ permutation,
which shall be again made manifest after we finalize the weak-coupling expansion.
The Plethystic exponential of the first term is proportional to the Dedekind eta
function, whose weak coupling expansion is easily done by using its modular property.
Calling this piece of index $I_{(1)}$, one obtains
\begin{equation}\label{I1}
  I_{(1)}\equiv e^{-\frac{\beta}{24}}\eta(e^{-\beta(1+c)})
  =\left(\frac{\beta}{2\pi}\right)^{\frac{1}{2}}(1+c)^{\frac{1}{2}}
  e^{-\frac{\beta}{24}+\frac{\pi^2}{6\beta(1+c)}}\exp\left[\sum_{p=1}^\infty\frac{1}{p}\
  \frac{e^{-\frac{4\pi^2 p}{\beta(1+c)}}}{1-e^{-\frac{4\pi^2 p}{\beta(1+c)}}}\right]
\end{equation}
where $\eta(q)\equiv q^{-\frac{1}{24}}\prod_{n=1}^\infty(1-q^n)^{-1}=q^{-\frac{1}{24}}
PE[\frac{q}{1-q}]$.

Now we explain how to dualize the Plethystic exponential of the second factor of
(\ref{letter-massage}), which we call $I_{(2)}$. We write $\log I_{(2)}$ as
\begin{equation}\label{GV-rewriting}
  \hspace*{-0.7cm}-\sum_{p=1}^{\infty}\frac{1}{p}\frac{\sinh\frac{p\beta(\frac{1}{2}+m-c)}{2}
  \sinh\frac{p\beta(\frac{1}{2}-m-c)}{2}}{2\sinh\frac{p\beta(1+a)}{2}\sinh\frac{p\beta(1+c)}{2}
  \sinh\frac{p\beta(1+c)}{2}}=-\sum_{n=-\infty}^{\infty}\int_\epsilon^{x_{\rm max}}\frac{ds}{s}
  \frac{\sinh\frac{s(\frac{1}{2}+m-c)}{2}\sinh\frac{s(\frac{1}{2}-m-c)}{2}}{2\sinh\frac{s(1+a)}{2}
  \sinh\frac{s(1+c)}{2}\sinh\frac{s(1+c)}{2}}e^{\frac{2\pi ins}{\beta}}\ ,
\end{equation}
where $\epsilon$ is a small positive number to regulate the $s$ integral divergent
around $s=0$. To show this identity, one first uses sum over $n$ on the right hand side
before performing the integral, and uses the delta function identity
\begin{equation}
  \sum_{n=-\infty}^\infty e^{\frac{2\pi ins}{\beta}}=
  \beta\sum_{p=-\infty}^\infty\delta(s-p\beta)
\end{equation}
to obtain the left hand side. The sum over $p$ is restricted to $p\geq 1$
as the integral variable $s$ is positive, leading to the left hand side.
Although each integral with given $n$ is divergent as $\epsilon\rightarrow 0$, the
divergences summed over $n$ should cancel out as the left hand side is finite.
This is the same manipulation as done in
\cite{Gopakumar:1998ii}. On the mathematics side, treatments similar to this have
been developed after a new proof of Dedekind eta function's modular property in
\cite{siegel}.

On the right hand side of (\ref{GV-rewriting}), we first separate out the piece
in the integral which diverges at $s=0$. We rearrange the right hand side as
follows:
\begin{eqnarray}\label{finite-integral}
  &&\sum_{n=-\infty}^\infty\int_0^\infty\frac{ds}{s}\left[\frac{\left(\frac{1}{2}-c\right)^2-m^2}
  {s(1+a)(1+b)(1+c)}-\frac{\sinh\frac{s(\frac{1}{2}+m-c)}{2}
  \sinh\frac{s(\frac{1}{2}-m-c)}{2}}{2\sinh\frac{s(1+a)}{2}\sinh\frac{s(1+b)}{2}
  \sinh\frac{s(1+c)}{2}}\right]e^{\frac{2\pi ins}{\beta}}\nonumber\\
  &&-\frac{\left(\frac{1}{2}-c\right)^2-m^2}{s(1+a)(1+b)(1+c)}\sum_{n=-\infty}^\infty
  \int_\epsilon^\infty\frac{ds}{s^2}e^{\frac{2\pi ins}{\beta}}\ .
\end{eqnarray}
The second line is easily calculated by redoing the $n$ summation first,
which yields
\begin{equation}\label{O(1/beta)}
  -\frac{\pi^2}{6\beta}\frac{\left(\frac{1}{2}-c\right)^2-m^2}{(1+a)(1+b)(1+c)}\ .
\end{equation}
As for the first line,
we have sent $\epsilon\rightarrow 0^+$ to ease the analysis below. Even though
each integrand at given $n$ is finite so that one might think that the meaning
of the integral is unambiguous at $s=0$, this is not true after the infinite sum
over $n$. Namely, had one been taken the summation first, it would have yielded
a delta function $\sum_{p=-\infty}^\infty\beta\delta(s-p\beta)$. With $\epsilon>0$,
we have considered the delta functions with $p\geq 1$ only, meaning that we
understand the above expression as excluding the delta function contribution at $s=0$.
However, to use the contour integral technique later, it will be convenient to include
this contribution and subtract it out. Understanding first line of
(\ref{finite-integral}) this way, we have to subtract the contribution coming from
half of the $\beta\delta(s)$ contribution at $p=0$ as follows:
\begin{eqnarray}\label{O(beta)}
  &&-\frac{\beta}{2}\left[\frac{\left(\frac{1}{2}-c\right)^2-m^2}{s^2(1+a)(1+b)(1+c)}
  -\frac{\sinh\frac{s(\frac{1}{2}+m-c)}{2}\sinh\frac{s(\frac{1}{2}-m-c)}{2}}
  {2s\sinh\frac{s(1+a)}{2}\sinh\frac{s(1+b)}{2}
  \sinh\frac{s(1+c)}{2}}\right]_{s=0}\nonumber\\
  &&=-\frac{\beta}{24}\ \frac{2abc+(1-ab-bc-ca)\delta+\delta^2}
  {(1+a)(1+b)(1+c)}\ .
\end{eqnarray}

Now with all the integrals on the first line of (\ref{finite-integral})
including $s=0$ in the normal sense, can evaluate the integrals as follows.
We consider the term with $n=0$ later, which is independent of $\beta$.
The terms with $n\neq 0$ are organized as
\begin{eqnarray}
  &&\sum_{n\neq 0}\int_0^{\infty}\frac{ds}{s}\left[\frac{\left(\frac{1}{2}-c\right)^2-m^2}
  {s(1+a)(1+b)(1+c)}-\frac{\sinh\frac{s(\frac{1}{2}+m-c)}{2}
  \sinh\frac{s(\frac{1}{2}-m-c)}{2}}{2\sinh\frac{s(1+a)}{2}\sinh\frac{s(1+b)}{2}
  \sinh\frac{s(1+c)}{2}}\right]e^{\frac{2\pi ins}{\beta}}\\
  &&=\sum_{n=1}^\infty\int_{-\infty}^{\infty}\frac{ds}{s}
  \left[\frac{\left(\frac{1}{2}-c\right)^2-m^2}
  {s(1+a)(1+b)(1+c)}-\frac{\sinh\frac{s(\frac{1}{2}+m-c)}{2}
  \sinh\frac{s(\frac{1}{2}-m-c)}{2}}{2\sinh\frac{s(1+a)}{2}\sinh\frac{s(1+b)}{2}
  \sinh\frac{s(1+c)}{2}}\right]e^{\frac{2\pi ins}{\beta}}\ .\nonumber
\end{eqnarray}
We evaluate it by using contour integrals. We enclose the
real axis of the $s$ plane with a large semicircle on the upper half plane,
as this semi-circle yields zero contribution. One obtains
\begin{equation}\label{abelian-instanton}
  \sum_{p=1}^{\infty}\frac{1}{p}\left[\frac{e^{-\frac{4\pi^2 p}{\beta(1+a)}}}
  {1-e^{-\frac{4\pi^2 p}{\beta(1+a)}}}f(p,a,b,c)+\frac{e^{-\frac{4\pi^2 p}{\beta(1+b)}}}
  {1-e^{-\frac{4\pi^2 p}{\beta(1+b)}}}f(p,b,c,a)+\frac{e^{-\frac{4\pi^2 p}{\beta(1+c)}}}
  {1-e^{-\frac{4\pi^2 p}{\beta(1+c)}}}\left(f(p,c,a,b)-1\right)\right]
\end{equation}
with
\begin{equation}
  f(p,a,b,c)\equiv\frac{\sin\frac{p\pi(m-\frac{1}{2}+b)}{1+a}\sin\frac{p\pi(m-\frac{1}{2}+c)}
  {1+a}}{\sin\frac{p\pi(a-b)}{1+a}\sin\frac{p\pi(a-c)}{1+a}}
\end{equation}
after some rearrangement. The three terms in the summation come from three
classes of poles at $s=\frac{2\pi pi}{1+a},\frac{2\pi pi}{1+b},\frac{2\pi pi}{1+c}$
with $p\geq 1$, respectively. The function $f(p,a,b,c)$ is symmetric
under the exchange of its last two arguments $b,c$. Note that
this function is the same as $f(p,a,b,c)$ appearing in (\ref{f-pert}) while we
were rewriting the perturbative partition function. This part now became manifestly
invariant under $a,b,c$ exchanges. This part also takes the form of instanton
expansions at three fixed points.

The integral with $n=0$ can also
be evaluated in the same way. As the integrand with $n=0$ is an even function in $s$,
we can integrate over $-\infty<s<\infty$ and multiply $\frac{1}{2}$. Using
the same contour, the result is
\begin{eqnarray}\label{abelian-pert-1}
  &&\frac{1}{2}\int_{-\infty}^{\infty}\frac{ds}{s}\left[
  \frac{\frac{9}{4}-m^2}{s(1+a)(1+b)(1+c)}-\frac{\sinh\frac{s(\frac{3}{2}+m)}{2}
  \sinh\frac{s(\frac{3}{2}-m)}{2}}{2\sinh\frac{s(1+a)}{2}\sinh\frac{s(1+b)}{2}
  \sinh\frac{s(1+c)}{2}}\right]\\
  &&\stackrel{\rm residues}{\longrightarrow}
  \sum_{p=1}^{\infty}\frac{f(p,a,b,c)}{2p}+\frac{f(p,b,c,a)}{2p}
  +\frac{f(p,c,a,b)-1}{2p}\ .\nonumber
\end{eqnarray}
This expression also formally looks invariant under $a,b,c$ exchanges,
but one has to be careful. As the
last sum of $-\frac{1}{2p}$ is divergent, one cannot separate it out from
the rest and claim that the first three summations are $a,b,c$ symmetric.
Actually, this expression is not $a,b,c$ symmetric in a subtle way.
To see why, it is again useful to recall that the above expression has the
same form as the perturbative partition function that we rewrote in
(\ref{pert-index-rewritten}). Considering the Abelian perturbative index,
one finds that the exponential on the second line of (\ref{pert-index-rewritten})
exactly takes the same form as our sum of residues (\ref{abelian-pert-1}).
If one multiplies $\frac{\mathcal{N}_V}{\mathcal{N}_H}
=(1+a)^{-\frac{1}{2}}(1+b)^{-\frac{1}{2}}$ for the Abelian theory to this
exponential, one obtains the full perturbative determinant of the Abelian theory
which should now be $a,b,c$ symmetric. We thus identify
\begin{equation}\label{abelian-pert-2}
  \exp\left[\sum_{p=1}^{\infty}\frac{f(p,a,b,c)}{2p}+\frac{f(p,b,c,a)}{2p}
  +\frac{f(p,c,a,b)-1}{2p}\right]=(1+a)^{\frac{1}{2}}(1+b)^{\frac{1}{2}}
  \ Z_{\rm pert}^{U(1)}\ .
\end{equation}
This finishes the evaluations of all ingredients in the weak-coupling expansion.

Collecting all, (\ref{I1}), (\ref{O(1/beta)}), (\ref{O(beta)}),
(\ref{abelian-instanton}), (\ref{abelian-pert-2}), one obtains
\begin{eqnarray}\label{final-abelian}
  I\!\!&\!=\!&\!\!\left[\frac{\beta}{2\pi}(1\!+\!a)(1\!+\!b)(1\!+\!c)\right]^{\!\!\frac{1}{2}}
  \!\!e^{-\beta\epsilon_0-\frac{\beta}{24}\left[1+\frac{2abc+(1-ab-bc-ca)(\frac{1}{4}-m^2)
  +(\frac{1}{4}-m^2)^2}{(1+a)(1+b)(1+c)}\right]+\frac{\pi^2}{6\beta}\left(\frac{1}{1+c}-
  \frac{(\frac{1}{2}-c)^2-m^2}{(1+a)(1+b)(1+c)}\right)}\\
  &&\times Z_{\rm pert}\exp\left[\sum_{p=1}^\infty\frac{1}{p}\left(\frac{e^{-\frac{4\pi^2 p}{\beta(1+a)}}}
  {1-e^{-\frac{4\pi^2 p}{\beta(1+a)}}}f(p,m,a,b,c)+(a,b,c\rightarrow b,c,a)
  +(a,b,c\rightarrow c,a,b)\right)\right]\ .\nonumber
\end{eqnarray}
We have included the Casimir energy factor $e^{-\beta\epsilon_0}$ given by
(\ref{casimir-ambiguous}). Although we already mentioned after (\ref{casimir-ambiguous})
that its interpretation is confusing to us, $-\beta\epsilon_0$ in the exponent
exactly cancels another $\mathcal{O}(\beta)$ term on the fisrt line of (\ref{final-abelian}).
Saying it differently, the 5d calculation is completely reproducing $\epsilon_0$,
as we know that perturbative contribution does not have nontrivial $\beta$ dependence
in 5d. All terms now look manifestly $a,b,c$ symmetric, perhaps except for the
$\mathcal{O}(\beta^{-1})$ term. in the exponent of the first line. However,
this is actually $a,b,c$ symmetric. It may be rewritten, for instance, as
\begin{equation}
  \frac{\pi^2}{6\beta}\left(\frac{1}{1+c}-
  \frac{(\frac{1}{2}-c)^2-m^2}{(1+a)(1+b)(1+c)}\right)=\frac{\pi^2}{6\beta}\
  \frac{\frac{3}{4}-\frac{a^2+b^2+c^2}{2}+m^2}{(1+a)(1+b)(1+c)}
\end{equation}
Actually, this is nothing but the high temperature leading form of the index
(\ref{final-abelian}) that one can easily derive by expanding the letter
index (\ref{letter}) at $\beta\ll 1$.

One can identify each factor as an appropriate contribution in the 5d calculation
that we derived in section 2. The factor
\begin{equation}
  \left[\frac{\beta}{2\pi}(1+a)(1+b)(1+c)\right]^{\frac{1}{2}}
\end{equation}
on the first line is the result of the integral over the single eigenvalue of
$\lambda$ with the Gaussian measure (\ref{gaussian2}), as this variable does not
have any other measure in the Abelian case. $Z_{\rm pert}$ on the second line is
what we already identified as the perturbative determinant. The exponential on
the second line looks like a product of instanton contributions from the
three fixed points.

Indeed, one can see that each factor is $Z_{\rm inst}^{(i)}$
with $i=1,2,3$. For instance, let us just consider the last one:
\begin{equation}\label{abelian-inst-fixed}
  \exp\left[\sum_{p=1}^\infty\frac{1}{p}\frac{e^{-\frac{4\pi^2 p}{\beta(1+c)}}}
  {1-e^{-\frac{4\pi^2 p}{\beta(1+c)}}}f(p,m,c,a,b)\right]=PE\left[
  \frac{e^{-\frac{4\pi^2}{\beta(1+c)}}}{1-e^{-\frac{4\pi^2}{\beta(1+c)}}}\
  \frac{\sin\frac{\pi\left(\frac{b-c}{2}+m+\frac{3(1+c)}{2}\right)}{1+c}
  \sin\frac{\pi\left(\frac{b-c}{2}-m-\frac{3(1+c)}{2}\right)}{1+c}}
  {\sin\frac{\pi(c-a)}{1+c}\sin\frac{\pi(b-c)}{1+c}}
  \right]\ .
\end{equation}
Again the last PE is just a formal Plethystic, regarding
$e^{-\frac{4\pi^2 p}{\beta(1+c)}}$, $\frac{\pi a_i}{1+c}$, $\frac{\pi m}{1+c}$
as `chemical potentials.' We rearranged the arguments in $f$ to make the mass shift
$m+\frac{3(1+c)}{2}$ structure clear.

One can see that this factor agrees with $Z_{\rm pert}^{(3)}$ as follows. Firstly,
one has to recall that the instanton partition function on $\mathbb{R}^4\times S^1$
has been calculated in an alternative method, from a related
topological string partition function. On flat space, the object analogous to our
$Z_{\rm inst}^{(3)}$ is the instanton partition function of the $\mathcal{N}=2^\ast$
theory with hypermultiplet mass $m$. This can be realized as M-theory compactified
on a noncompact toric Calabi-Yau space to $\mathbb{R}^{4,1}$. The topological string
partition function on CY$_3$ calculates an index for the BPS states of the 5d theory
\cite{Gopakumar:1998ii}. This is exactly what the instanton partition functions are doing
on $\mathbb{R}^4\times S^1$. So one finds two different expressions for a given
partition function. In fact, for the Abelian 5d $\mathcal{N}=2^\ast$
theory, \cite{Iqbal:2008ra} uses some identities of topological vertices to recast the
topological string partition function into a Plethystic form. The instanton part of
the partition funciton is
\begin{equation}\label{abelian-inst-plethystic}
  PE\left[\frac{q}{1-q}\frac{\sin\left(\pi(\epsilon_-+m)\right)
  \sin\left(\pi(\epsilon_--m)\right)}{\sin\left(-\pi\epsilon_1\right)
  \sin\left(\pi\epsilon_2\right)}\right]\ ,
\end{equation}
with $\epsilon_\pm=\frac{\epsilon_1\pm\epsilon_2}{2}$. $q$ is the fugacity
for the instanton number and $\epsilon_1,\epsilon_2$ are the Omega deformation
parameters, or chemical potentials for rotations on $\mathbb{R}^4$.
The last ratio of 4 sine functions is the index for the center-of-mass
supermultiplet for one superparticle moving on $\mathbb{R}^4$, preserving the
SUSY of instantons. From this, and the expansion $\frac{q}{1-q}=q+q^2+q^3+\cdots$,
this form of index was used in \cite{Kim:2011mv} to prove that the Abelian instantons
on $\mathbb{R}^{4,1}$ form unique thoreshold bound states at all instanton number,
which is one of the key conjectures of M-theory \cite{Hull:1994ys}.

(\ref{abelian-inst-plethystic}) should also be the result one obtains from a direct
instanton calculus, after summing over all the residues for poles given by Young diagrams
(this time uncolored, as $N=1$), weighted by $q^k$ for $k$ instantons, similar to (\ref{U(N)-residue}) but on the flat space. Recall that the differences between (\ref{U(N)-residue}) and
the residue formula on the flat space (written for instance in \cite{Kim:2011mv})
are firstly that the factors $\frac{1}{1+c}$ are absent
in the arguments of sine's, and also that the mass shift $m\rightarrow m+\frac{3(1+c)}{2}$
is absent. The statement that the $U(1)$ instanton sum of the form
\begin{equation}\label{inst-schematic}
  Z_{\rm inst}=\sum_kq^k\sum_{Y;\ |Y|=k}Z_Y
\end{equation}
equals (\ref{abelian-inst-plethystic}) has a good physical reason to believe,
but has not been proved in full generality, as far as we are aware of.
It has been checked in \cite{Kim:2011mv} explicitly up to $k\leq 3$. Now,
accepting this identity, it is a simple corollary that the Abelian instanton sum
(\ref{U(N)-residue}) with $\frac{1}{1+c}$ factor and mass shift enjoys a
Plethystic-like form similar to (\ref{abelian-inst-plethystic}). The identification
of parameters are
\begin{equation}
  q\rightarrow\ e^{-\frac{4\pi^2}{\beta(1+c)}}\ ,\ \
  \epsilon_1\rightarrow a-c\ ,\ \ \epsilon_2\rightarrow b-c\ ,\ \
  m\rightarrow m+\frac{3(1+c)}{2}\ ,
\end{equation}
which precisely changes (\ref{abelian-inst-plethystic}) into (\ref{abelian-inst-fixed}).
This proves that the exponential on the second line of (\ref{Abelian-full-index}) is
indeed $Z_{\rm inst}^{(1)}Z_{\rm inst}^{(2)}Z_{\rm inst}^{(3)}$.

The only remaining factor to be understood is the term of order
$\mathcal{O}(\beta^{-1})$ in the exponent on the first line.
The term of order $\beta^{-1}$ has no way to come out from the study of
quantum fluctuation of dynamical fields on $S^5$. However, to correctly probe the
high temperature regime of a 6d index, such a factor is very natural as the index
degeneracy often grows fast (although not as much as that of the ordinary
partition function). In \cite{Kim:2012av}, it was argued that such terms with
negative power in $\beta$ should come from the constant shift of the $S^5$ action,
which couples the parameters like $g_{YM}^2\sim\beta$ in the theory to the background
curvature of $S^5$. Such constant couplings are never breaking the symmetries of SYM
on $S^5$. So as far as we can see at the moment, this seems to be a genuine ambiguity
in our 5d approach. However, one is fitting just one ambiguous coefficient to a favorable
value, after which the 6d physics of infinitely many BPS states is reproduced.
In particular, this single fitting
makes the whole partition function to take the form of an index at strong coupling.
See section 4 for more details on the structure of such curvature couplings.
Similar curvature couplings play important roles in \cite{Vafa:1994tf}, in which
S-duality of 4d $\mathcal{N}=4$ SYM on curved spaces is discussed. To have S-duality,
constant shifts to the action which takes exactly the same form as ours is
assumed.

Thus, up to the $\mathcal{O}(\beta^{-1})$ term which is ambiguous in 5d,
our 5d calculation completely reproduces the known Abelian 6d index.

\subsection{Indices from 5d maximal SYM and $AdS_7$ duals}

We consider the partition function at a special subset of the parameter space
$\beta,m,\epsilon_1,\epsilon_2$, yielding an unrefined index in 6d.
We set $m=\pm\frac{1}{2}$ (in the dimensionless unit after
multiplying $r$) and $\epsilon_1=\epsilon_2=0$. At these two points, one has
maximal SYM on $S^5$ \cite{Kim:2012av}. As the partition functions
for the two cases with $m=\pm\frac{1}{2}$ turn out to be the same,
we simply consider the case with $m=\frac{1}{2}$. At this point, a simple conjecture on the
instanton part of the partition function was made in \cite{Kim:2012av} for the $U(N)$
gauge group. With this conjecture, the matrix integral including the perturbative part
could be calculated exactly, yielding an index which survives nontrivial tests
and also predicts an $N^3$ scaling of the (index version of) vacuum Casimir energy
\cite{Kim:2012av}. In this subsection, we derive the conjectured closed form
expression for $Z_{\rm inst}$ for the $U(N)$ theory, and then derive a
similar closed form expression for $SO(2N)$. We also propose a simple form for
the instantonic correction for $E_n$ gauge group with maximal SUSY. We also elaborate
on the 6d index structure of the ADE partition function, including the perturbative
part, and discuss the structure of the partition function for the BCFG gauge groups.
The $A_n$ and $D_n$ partition functions completely agree with their gravity dual
indices on $AdS_7\times S^4$ and $AdS_7\times S^4/\mathbb{Z}_2$, respectively,
at large $N$.

We start from the perturbative part \cite{Kim:2012av}. Although the partition
function is generally a matrix integral with the factorized integrand consisting of
$Z_{\rm pert}$ and $Z_{\rm inst}$, in foresight we use the fact that
$Z_{\rm inst}$ is independent of $\lambda$ with maximal SUSY. So in this case,
the matrix integral can be done with the perturbative measure only. For gauge group
$G$, one obtains \cite{Kim:2012av}\footnote{
To deal with arbitrary gauge group in a unified way, we use the normalized trace
which is independent of the representation. We define
${\rm tr}\equiv\frac{1}{|\theta|^2x_R}{\rm tr}_R$. $|\theta|$ is the length of the long
root, the Dynkin index $x_R$ of representation $R$ is defined as
${\rm tr}_R(T^aT^b)=|\theta|^2x_R\delta^{ab}$ so that ${\rm tr}(T^aT^b)=\delta^{ab}$.
Here and below, we use the convention $|\theta|^2=2$. We use the 5d action with the
Yang-Mills term $\frac{1}{4g_{\rm YM}^2}{\rm tr}(F_{\mu\nu}F^{\mu\nu})$,
using this trace. The unit instanton's action is $4\pi^2/\beta$
for all gauge groups with $2\pi\beta\equiv g_{YM}^2$.}
\begin{equation}\label{pert-integral}
  Z_{\rm pert}=\frac{1}{|W|}\int d\lambda\prod_{\alpha\in\Delta_+}
  \left(2\sinh\frac{\alpha\cdot\lambda}{2}\right)^2e^{-\frac{2\pi^2}{\beta}{\rm tr}\lambda^2}
  =\left(\frac{\beta}{2\pi}\right)^{r/2}e^{\frac{\beta}{12}c_2|G|}\prod_{\alpha\in\Delta_+}
  2\sinh\left(\beta\frac{\alpha\cdot\rho}{2}\right)\ .
\end{equation}
$\Delta_+$ is the set of positive roots, $r$ is the rank, $c_2$ is the dual
Coxeter number, $|G|$ is the dimension of the semi-simple part of $G$, and $\rho$
is the Weyl vector. The necessary group theory data are summarized in
Table \ref{group}.

\begin{table}[t!]
$$
\begin{array}{c|ccccccccc}
  \hline &SU(N)&SO(2N\!+\!1)&Sp(N)&SO(2N)&E_6&E_7&E_8&F_4&G_2\\
  \hline r&N\!-\!1&N&N&N&6&7&8&4&2\\
  |G|&N^2\!-\!1&N(2N\!+\!1)&N(2N\!+\!1)&N(2N\!-\!1)&78&133&248&52&14\\
  c_2&N&2N\!-\!1&N\!+\!1&2N\!-\!2&12&18&30&9&4\\
  \hline
\end{array}
$$

\caption{Group theoretic data}\label{group}
\end{table}
One can evaluate this perturbative determinant for various groups.
We list the results for the classical groups. Firstly, for
simply-laced group $SU(N)$ (including the overall $U(1)$ factor to make it $U(N)$)
and $SO(2N)$, one finds
\begin{eqnarray}\label{pert-simply-laced}
  Z^{U(N)}_{\rm pert}\!&\!=\!&\!
  e^{\frac{\beta}{6}N(N^2-1)}\left(\frac{\beta}{2\pi}
  \right)^{\frac{N}{2}}\prod_{m=1}^{N-1}(1-e^{-m\beta})^{N-m}\\
  Z^{SO(2N)}_{\rm pert}\!&\!=\!&\!
  e^{\frac{\beta}{6}c_2|G|}\left(\frac{\beta}{2\pi}
  \right)^{\frac{N}{2}}\prod_{m=1}^{N-1}(1-e^{-m\beta})^{N-m}\\
  &&\times(1-e^{-(N-1)\beta})^{\left[\frac{N}{2}\right]}\prod_{m=1}^{N-2}\left[(1\!-\!e^{-m\beta})
  (1\!-\!e^{-(2N-m-2)\beta})\right]^{\left[\frac{m+1}{2}\right]}\ .\nonumber
\end{eqnarray}
For the non-simply-laced classical gauge groups, one obtains
\begin{eqnarray}
  Z^{SO(2N\!+\!1)}_{\rm pert}\!&\!=\!&\!
  e^{\frac{\beta}{6}c_2|G|}\left(\frac{\beta}{2\pi}
  \right)^{\frac{N}{2}}\prod_{m=1}^{N-1}(1-e^{-m\beta})^{N-m}\\
  &&\times(1-e^{-N\beta})^{\left[\frac{N}{2}\right]}\prod_{m=1}^{N-2}
  \left[(1\!-\!e^{-(m\!+\!1)\beta})(1\!-\!e^{-(2N-m-1)\beta})\right]^{\left[\frac{m+1}{2}\right]}
  \prod_{m=1}^N\left(1-e^{-\beta(m-\frac{1}{2})}\right)\nonumber\\
  Z^{Sp(N)}_{\rm pert}\!&\!=\!&\!
  e^{\frac{\beta}{6}c_2|G|}\left(\frac{\beta}{2\pi}
  \right)^{\frac{N}{2}}\prod_{m=1}^{N-1}(1-e^{-m\frac{\beta}{2}})^{N-m}\\
  &&\times(1-e^{-(N\!+\!1)\frac{\beta}{2}})^{\left[\frac{N}{2}\right]}
  \prod_{m=1}^{N-2}\left[(1\!-\!e^{-(m\!+\!2)\frac{\beta}{2}})
  (1\!-\!e^{-(2N-m)\frac{\beta}{2}})\right]^{\left[\frac{m+1}{2}\right]}
  \prod_{m=1}^N\left(1-e^{-m\beta}\right)\ ,\nonumber
\end{eqnarray}
where $[a]$ is the biggest integer no larger than $a$. Generally,
the right hand side of (\ref{pert-integral}) can be written as
\begin{equation}
  Z_{\rm pert}=\left(\frac{\beta}{2\pi}\right)^{\frac{r}{2}}e^{\frac{\beta}{12}c_2|G|
  +\beta(\rho,\rho)}\prod_{\alpha\in\Delta_+}(1-e^{-\beta(\alpha\cdot\rho)})=
  \left(\frac{\beta}{2\pi}\right)^{\frac{r}{2}}e^{\frac{\beta}{6}c_2|G|}
  \prod_{\alpha\in\Delta_+}(1-e^{-\beta(\alpha\cdot\rho)})\ ,
\end{equation}
where we used the Freudenthal-de Vries formula at the last step.

Apart from the overall $\left(\frac{\beta}{2\pi}\right)^{r/2}$ factor,
the perturbative expression already takes the form of an index.
For $U(N)$ and $SO(2N)$ cases in which we shall derive closed forms of instanton
corrections, these factors obstructing index interpretation will be canceled against
similar obstructing factors from the instanton parts, making the whole $S^5$
partition function an index. It should be exciting to study whether similar
cancelation happens for other gauge groups, especially having in mind the possible
interpretations of the non-simply-laced partition functions as twisted indices of
6d ADE theories. See the end of this subsection for comments.

Note also that in all cases, the `zero point energy' contribution all turns out to
be $-\frac{c_2|G|}{6}$. It was noted in \cite{Kim:2012av} from the 5d perturbation
theory viewpoint that this combination of numbers can be understood as
$f^{abc}f^{abc}=c_2|G|$ where $f^{abc}$ is the structure constant, which is the universal
group theory factor for all 2-loop vacuum bubble diagrams. Note that the Casimir energy
starts to appear from $\mathcal{O}(\beta)\sim\mathcal{O}(g_{YM}^2)$, which is
at 2-loop order. It is curious to find
that the same combination $c_2|G|$ appears as an anomaly coefficient of the 6d ADE
$(2,0)$ theory \cite{Intriligator:2000eq}.

As for the instanton part, we first consider the $U(N)$ case. We can use
the residue formula (\ref{U(N)-residue}) for the third fixed point, and similar
formulae for the other two fixed points by permuting $a,b,c$. Inserting the maximal
SUSY values $m=\frac{1}{2}$, $a,b,c=0$,  one finds a very simple structure of
these residues. Let us first stage our claim and then prove it.

Inserting $m=\frac{1}{2}$ in (\ref{U(N)-residue}) is simple. On the other hand,
the limit $\epsilon_1,\epsilon_2\rightarrow 0$ is somewhat tricky in that each
saddle point contribution is not smooth at the point $\epsilon_1=\epsilon_2=0$.
(It may even diverge.) However, the total instanton partition function
$Z_{\rm inst}^{(1)}Z_{\rm inst}^{(2)}Z_{\rm inst}^{(3)}$ is smooth at this point,
or at least believed to be so from general consideration as the partition function
is finite on round $S^5$. So we shall take this limit in a specific order which
simplifies the analysis. We shall first take $c\rightarrow 0$, keeping $a,b$
to be finite. We then take $a=-b$ to zero. If we take the limit in this order,
we claim that
\begin{equation}
  Z_{\rm inst}^{(1)},\ Z_{\rm inst}^{(2)}\rightarrow 1\ \ \ {\rm and}\ \ \
  Z_{\rm inst}^{(3)}\rightarrow\eta(e^{-\frac{4\pi^2}{\beta}})^{-N}=
  e^{\frac{N\pi^2}{6\beta}}\prod_{n=1}^\infty\frac{1}{(1-e^{-\frac{4\pi^2 n}{\beta}})^N}\ ,
\end{equation}
so that the net instanton contribution is
$Z_{\rm inst}=\eta(e^{-\frac{4\pi^2}{\beta}})^{-N}$ as conjectured in \cite{Kim:2012av}.\footnote{As explained in \cite{Kim:2012av} and section 3.1,
the overall $e^{\frac{N\pi^2}{6\beta}}$ is an ambiguous factor from the 5d perspective.
See section 4 for more discussions.} Just to make sure, we also checked this
claim in arbitrary possible order of the limit $\epsilon_1,\epsilon_2\rightarrow 0$,
for $N=1,2,3$ up to $k\leq 3$.
To prove the claim, it suffices to show that $Z_{\{Y_1,\cdots,Y_N\}}$ for all
Young diagrams become $1$ for the third fixed point, while $Z_Y$'s at the other two
fixed points become all $0$ (apart from the void case) in the limit. The last $0$'s
obviously yield the total instanton partition function to be $1$ at those fixed points.
As for $Z_{\rm inst}^{(3)}$, the instanton partition function with all $Z_{\{Y_1,\cdots,Y_N\}}=1$ is the generating function which counts the colored Young diagram with
$e^{-\frac{4\pi^2}{\beta}}$
being the fugacity of the box number $k$. This generating function is simply $\eta^{-N}$.
Note also that, as we shall be showing that all possible divergent terms in $Z_Y$ cancel out
to yield finite values within a fixed point (if the limit is taken in the above order),
one can simply set $a_i=0$ in $e^{-\frac{4\pi^2k}{\beta(1+a_i)}}\rightarrow e^{-\frac{4\pi^2k}{\beta}}$.

We first prove that $Z_Y$'s are identically $0$ at the first and second fixed points.
We start by noting that all $Z_Y$'s at the first fixed point $|Z_1|=1$ contain at
least one factor of
\begin{equation}\label{1st-com}
  \frac{\sin\pi\left(\frac{\frac{b-c}{2}+m+\frac{1+a}{2}}{1+a}\right)
  \sin\pi\left(\frac{\frac{b-c}{2}-m+\frac{1+a}{2}}{1+a}\right)}
  {\sin\pi\frac{-\epsilon_1}{1+a}\sin\pi\frac{\epsilon_2}{1+\alpha}}
  \ \ \stackrel{m=\frac{1}{2}}{\longrightarrow}\ \
  \frac{\sin\frac{\pi b}{1+a}\sin\frac{\pi c}{1+a}}
  {\sin\frac{\pi(a-b)}{1+a}\sin\frac{\pi(a-c)}{1+a}}\ ,
\end{equation}
where we used $a+b+c=0$ and periodicity of sine function at various places.
This basically comes from the fact that all instanton indices on $\mathbb{R}^4\times S^1$
contain a factor of index from the centor-of-mass supermultiplet \cite{Kim:2012av}, which
is (\ref{1st-com}) for the first fixed point. One can similarly show that $Z$'s at the second
fixed point has at least one factor like (\ref{1st-com}), with the role of $a$ and $b$
changed. Now take the limit $c\rightarrow 0$ first, keeping $a$ and $b$ finite satisfying
$a+b=0$. Then obviously the above factor is zero from the $\sin\frac{\pi c}{1+a}$ factor
in the numerator of (\ref{1st-com}). It is also easy to show that there are no other
divergent factors in (\ref{U(N)-residue}) as one takes this limit. So all $Z_Y$'s from
the first fixed point are zero, proving $Z_{\rm inst}^{(1)}=1$.
One can also show $Z_{\rm inst}^{(2)}=1$ in exactly the same way,
proving half of our claim.

Now let us consider the third fixed point in this limit. For $Z_Y$'s at this
fixed point, one obtains the following at $m=\frac{1}{2}$ after a manipulation
similar to (\ref{1st-com}):
\begin{equation}
  Z_{\{Y_1,Y_2,\cdots, Y_N\}}=
  \prod_{i,j=1}^N\prod_{s\in Y_i}
  \frac{\sin\pi\frac{\lambda_{ij}-(a-c)h_i(s)+(b-c)v_j(s)+b}{1+c}
  \sin\pi\frac{\lambda_{ij}-(a-c)h_i(s)+(b-c)v_j(s)-a}{1+c}}
  {\sin\pi\frac{\lambda_{ij}-(a-c)h_i(s)+(b-c)v_j(s)+(b-c)}{1+c}
  \sin\pi\frac{\lambda_{ij}-(a-c)h_i(s)+(b-c)v_j(s)-(a-c)}{1+c}}\ .
\end{equation}
Taking the limit $c\rightarrow 0$, keeping $a=-b$, one finds that
the factor inside the product becomes
\begin{equation}
  \frac{\sin\pi(\lambda_{ij}-ah_i(s)+bv_j(s)+b)
  \cdot\sin\pi(\lambda_{ij}-ah_i(s)+bv_j(s)-a)}
  {\sin\pi(\lambda_{ij}-ah_i(s)+bv_j(s)+b)
  \cdot\sin\pi(\lambda_{ij}-ah_i(s)+bv_j(s)-a)}=1\ .
\end{equation}
So all $Z_Y$'s at the third fixed point are $1$, completing
the proof of $Z_{\rm inst}=\eta(e^{-\frac{4\pi^2}{\beta}})^{-N}$.

We can extend the study of maximal SYM partition function for other classical
gauge groups $SO(N)$ and $Sp(N)$. The instanton part of the $SO(2N)$ gauge group
at the maximal SUSY point can be analyzed in a similar manner. As a simple residue
formula like (\ref{U(N)-residue}) is not available, one has to study directly the
limit from the $SO(2N)$ contour integral formula for the $Sp(k)$ eigenvalues
$\phi_I$. This analysis is explained in appendix A.2. The result is
\begin{equation}
  Z^{SO(2N)}_{\rm inst}=\eta(e^{-\frac{4\pi^2}{\beta}})^{-N}
\end{equation}
at the maiximal SUSY point $m=\frac{1}{2}$, $\epsilon_1,\epsilon_2\rightarrow 0$.

The instanton partition functions for other classical groups, $SO(2N\!+\!1)$
and $Sp(N)$ could also be studied in this limit. These could be useful to develop and justify
a twisted index interpretation of non-ADE partition functions, which we briefly
explain at the end of this subsection. The analysis is more involved for $Sp(N)$,
as the instanton gauge symmetry is $O(k)$ rather than $SO(k)$, demanding us to
consider various $\mathbb{Z}_2$ even and odd sectors. Also, the instanton indices for
exceptional gauge groups are not available from our ADHM approach, due to the
absence of such a formalism. However, from the simple result we obtained at the maximal
SUSY point for $U(N)$ and $SO(2N)$ series, it is tempting to conjecture that
\begin{equation}\label{E-conjecture}
  Z_{\rm inst}^{E_n}\stackrel{?}{=}\eta(e^{-\frac{4\pi^2}{\beta}})^{-n}=
  \left(\frac{2\pi}{\beta}\right)^{n/2}\eta(e^{-\beta})^n
\end{equation}
for $n=6,7,8$. Perhaps it may be possible to address this issue systematically,
using recent developments in the exceptional instanton countings \cite{Keller:2012da}.
More generally, with general gauge groups, one might wonder whether
\begin{equation}\label{instanton-conjecture}
  Z_{\rm inst}\stackrel{???}{=}\eta(e^{-\frac{4\pi^2}{\beta}})^{-r}=
  \left(\frac{2\pi}{\beta}\right)^{r/2}\eta(e^{-\beta})^r
\end{equation}
is true. However, the analysis of $Sp(N)$ instantons at $k=1$ show that
this formula cannot be generally true. See appendix A.2 for the details.

To conclude with solid parts of our discussions only, we derived
the following exact partition function for the $U(N)$ and $SO(2N)$ theories
at the maximal SUSY point $m=\frac{1}{2}$, $\epsilon_1\!=\!\epsilon_2\!=\!0$:
\begin{eqnarray}
  Z^{U(N)}&=&e^{\beta\left(\frac{N(N^2-1)}{6}+\frac{N}{24}\right)}
  \prod_{m=1}^N\frac{1}{(1-e^{-m\beta})^m}
  \prod_{m=N+1}^\infty\frac{1}{(1-e^{m\beta})^N}\nonumber\\
  Z^{SO(2N)}&=&e^{\beta\left(\frac{c_2|G|}{6}+\frac{N}{24}\right)}
  \prod_{m=1}^N\frac{1}{(1-e^{-m\beta})^m}
  \prod_{m=N+1}^\infty\frac{1}{(1-e^{m\beta})^N}\\
  &&\times(1-e^{-(N-1)\beta})^{\left[\frac{N}{2}\right]}\prod_{m=1}^{N-2}\left[(1\!-\!e^{-m\beta})
  (1\!-\!e^{-(2N-m-2)\beta})\right]^{\left[\frac{m+1}{2}\right]}\ ,\nonumber
\end{eqnarray}
which we interpret as the 6d index ${\rm Tr}\left[(-1)^Fe^{-\beta(E-R_1)}\right]$.
The following large $N$ indices can be obtained from the above
exact indices:
\begin{eqnarray}\label{large-N}
  Z^{U(\infty)}&=&\prod_{n=1}^\infty\frac{1}{(1-q^n)^n}=PE\left[\frac{q}{(1-q)^2}\right]\\
  Z^{SO(2\infty)}&=&\prod_{n=1}^\infty\frac{(1-q^{2n-1})^n(1-q^{2n})^n}{(1-q^n)^n}
  =PE\left[\frac{q^2}{(1-q)(1-q^2)}\right]\ ,\nonumber
\end{eqnarray}
where $PE$ again denotes Plethystic exponential with the fugacity $q=e^{-\beta}$.

\begin{table}[t!]
$$
\begin{array}{c|ccc|c}
  \hline &\epsilon&SO(6)&SO(5)&{\rm boson/fermion}\\
  \hline p\geq 1&2p&(0,0,0)&(p,0)&{\rm b}\\
  p\geq 1&2p+\frac{1}{2}&(\frac{1}{2},\frac{1}{2},\frac{1}{2})
  &(p-\frac{1}{2},\frac{1}{2})&{\rm f}\\
  p\geq 2&2p+1&(1,0,0)&(p-1,1)&{\rm b}\\
  p\geq 3&2p+\frac{3}{2}&(\frac{1}{2},\frac{1}{2},-\frac{1}{2})
  &(p-\frac{3}{2},\frac{3}{2})&{\rm f}\\
  \hline p=1&\frac{7}{2}&(\frac{1}{2},\frac{1}{2},-\frac{1}{2})
  &(\frac{1}{2},\frac{1}{2})&{\rm b\ (fermionic\ constraint)}\\
  \hline
\end{array}
$$
\caption{BPS Kaluza-Klein fields of $AdS_7\times S^4$ supergravity, upon $S^4$
KK reduction. At given $p$, the entries grouped by horizontal lines stay in
the same representation of $OSp(6,2|4)$.}\label{sugra}
\end{table}
One can also study these large $N$ indices from their gravity duals
on $AdS_7\times S^4$ and $AdS_7\times S^4/\mathbb{Z}_2$. The large $N$
limit of the $U(N)$ result was already shown to completely agree with the supergravity
index on $AdS_7\times S^4$ \cite{Kim:2012av}. Let us start by reviewing this proof.
The BPS gravity fields on $AdS_7$ after a KK reduction on $S^4$ are given by
Table \ref{sugra}. The $SO(6)$ charges denote
the values of $(j_1,j_2,j_3)$ for its highest weight state, and $SO(5)$
the values of $(R_1,R_2)$ for its highest weight state. One has to sum over all their
contributions to the index, weighting them by $\pm$ signs for bosons/fermions
and also by $q^{E-R_1}$. Multiplying to this the $AdS_7$ BPS wavefunction
factor $\frac{1}{(1-q)^3}$, one obtains the single particle gravity index in
$AdS_7\times S^4$. By reading off the information of Table \ref{sugra}, this was
calculated in \cite{Kim:2012av} to be
\begin{equation}
  \hspace*{-1cm}I_{\rm sp}(q)=\frac{1}{(1-q)^3}\left[\sum_{p=1}^\infty\sum_{n=0}^pq^{2p-n}
  -3\sum_{p=1}^\infty\sum_{n=1}^pq^{2p+1-n}+3\sum_{p=2}^\infty\sum_{n=1}^{p-1}q^{2p+1-n}
  -\sum_{p=3}^\infty\sum_{n=1}^{p-2} q^{2p+1-n}+q^3\right]=\frac{q}{(1-q)^2}\ .
\end{equation}
The multiparticle index obtained from $I_{\rm sp}$ yields the MacMahon function
\begin{equation}
  I_{\rm mp}(q)=PE\left[I_{\rm sp}(q)\right]
  =\prod_{n=1}^\infty\frac{1}{(1-q^n)^n}
\end{equation}
as the multiparticle gravity index on $AdS_7\times S^4$. This completely agrees
with the large $N$ limit of $U(N)$ index in (\ref{large-N}).

Now we consider the gravity dual of the $SO(2N)$ index. From Table \ref{sugra},
one has to appropriately discard the $\mathbb{Z}_2$ odd modes. As this is a parity
or inversion action for the `internal $\mathbb{R}^5$,' it suffices for us to consider
the representation of fields under $SO(5)$. Firstly, all BPS bosonic fields on the first
line of the table are chiral primaries. The states with $SO(5)$ charges $(R_1,R_2)=(p,0)$
can be viewed as being formed by rank $p$ polynomials of two holomorphic coordinates
of $\mathbb{R}^5$. As the $\mathbb{Z}_2$ parity acts as inversion on these coordinates,
the $\mathbb{Z}_2$ even fields are those with even $p$. Since the $\mathbb{Z}_2$ orbifold
preserves the whole $OSp(8|4)$ symmetry, $\mathbb{Z}_2$ even/odd natures
of fields should be grouped into those of representations. Since the first four lines
of Table \ref{sugra} belong to the same representation at given $p$, one should only keep
the fields with even $p$ on $AdS_7\times S^4/\mathbb{Z}_2$. The last line of the table
comes from a component of a constraint multiplet at $p=1$ which saturates our BPS bound.
As one can see from, say, table 3 in p.31 of \cite{Bhattacharya:2008zy}, this multiplet
contains a state with $E=4$, $(j_1,j_2,j_3)=(0,0,0)$ and $(R_1,R_2)=(1,0)$, which is
$\mathbb{Z}_2$ odd, meaning that this multiplet itself should be odd.
Thus the last line of our Table \ref{sugra} should be discarded.
Therefore, the single particle gravity index on $AdS_7\times S^4/\mathbb{Z}_2$ is given by
the above $AdS_7\times S^4$ summations with all $p$'s restricted to even integers:
\begin{eqnarray}
  I_{\rm sp}(q)&=&\frac{1}{(1-q)^3}\left[\sum_{m=1}^\infty\sum_{n=0}^{2m}q^{4m-n}
  -3\sum_{m=1}^\infty\sum_{n=1}^{2m}q^{4m+1-n}+3\sum_{m=1}^\infty\sum_{n=1}^{2m-1}q^{4m+1-n}
  -\sum_{m=2}^\infty\sum_{n=1}^{2m-2} q^{4m+1-n}\right]\nonumber\\
  &=&\frac{q^2}{(1-q)(1-q^2)}\ .
\end{eqnarray}
In the above summation, we have set $p=2m$. This is again in precise
agreement with the large $N$ limit (\ref{large-N}) of the $SO(2N)$ index.

Finally, let us briefly comment on the non-ADE partition functions that we obtained.
Of course, the possible 6d index interpretation of the $S^5$ partition function
critically depends on the choice of the theory: most importantly, the gauge group
and matter content. Thus while ADE theories with one adjoint hypermultiplet are
naturally expected to have index interpretations of 6d ADE theories, other theories
may not. However, for non-ADE simple gauge groups, there could possibly be
interpretations as `twisted' indices of 6d ADE theories, where the twisting is
the action of outer automorphism of the 6d ADE gauge group.\footnote{We thank
Yuji Tachikawa, and also Neil Lambert, Sung-Soo Kim for helpful comments on this.}
%
Following \cite{Tachikawa:2011ch}, one can consider a twisted circle compactification
of the 6d $(2,0)$ theory on $S^5\times S^1$. The twisting could be presumably inserting
a discrete `charge conjugation' operator to the 6d index, similar to the index discussed
in \cite{Zwiebel:2011wa}. Upon dimensional reduction to 5d, the zero modes of the
twisting can be in adjoint representation of non-ADE 5d gauge groups. The $\mathbb{Z}_2$
twisting of $A_{2N\!+\!1}$ $(2,0)$ theory is supposed to yield $SO(2N\!+\!1)$ 5d SYM,
$\mathbb{Z}_2$ of $A_{2N}$ and $D_{N\!+\!1}$ yielding $Sp(N)$ 5d SYM
(with different values for the discrete theta angle), $\mathbb{Z}_2$ of $E_6$
yielding $F_4$ 5d SYM, and finally $\mathbb{Z}_3$ of $D_4$ yielding $G_2$ 5d SYM.
These are S-dual to the low dimensional gauge groups that one obtains by twisted
reductions of higher dimensional Yang-Mills theories. See, for instance,
\cite{Kim:2004xx} for a summary.
So it would be interesting to see whether our non-ADE partition functions on $S^5$
can be quantitatively interpreted as such indices.

\subsection{Wilson loops and their large $N$ gravity duals}

One can also consider more nontrivial BPS observables
by inserting various operators on $S^5$. For instance, one can
insert BPS Wilson loops on $S^5$. Wilson loops of 5d SYM have been studied
in \cite{Young:2011aa}, related to the Wilson surface operators of the 6d $(2,0)$ theory.

For simplicity, let us stick to the supercharge $\mathcal{Q}$ that was chosen in
\cite{Hosomichi:2012ek,Kallen:2012va,Kim:2012av} to localize the path integral.
We also consider round $S^5$ only. Then the corresponding Killing spinor
satisfies, among others, the properties $\gamma^5\epsilon=\epsilon$ and
$\hat\gamma^3\epsilon=-\epsilon$ in the notation of \cite{Kim:2012av}, where $\gamma^5$
is the gamma matrix on $S^5$ with index on the Hopf fiber direction, and $\hat\gamma^3$
is the internal gamma matrix along the direction of vector multiplet scalar $\phi$.
Consider the following Wilson loop operator:
\begin{equation}
  W_R=\frac{1}{\dim R}{\rm tr}_R\left[P\exp\left(\oint ds(iA_\mu\dot{x}^\mu+
  \phi|\dot{x}|)\right)\right]\ .
\end{equation}
From the supersymmetry variation \cite{Kim:2012av}
\begin{equation}
  i\delta A_5+\delta\phi=\frac{i}{2}\left[\epsilon^\dag(\gamma^5+\hat\gamma^3)\lambda
  -\lambda^\dag(\gamma^5+\hat\gamma^3)\epsilon\right]\ ,
\end{equation}
the Wilson loop preserves $\mathcal{Q}$ when the curve $x^\mu(s)$ wraps the Hopf fiber,
located at a point on $\mathbb{CP}^2$ base. One can consider the localization of
the path integral with $W_R$ inserted. The analysis goes exactly in the same manner
as the case without the insertion of $W_R$, except that we have to include the saddle
point value of $W_R$ inside the matrix integral. The scalar $\phi$
appears in this saddle point value:
\begin{equation}
  W_R\rightarrow \frac{1}{\dim R}{\rm tr}_R\left(e^{2\pi\lambda}\right)\ \ \ \
  {\rm with}\ \lambda=r\phi\ .
\end{equation}
One can of course consider the general case, but it is illustrating to consider
the expectation value in the maximal SYM with $m=\frac{1}{2}$ and $\epsilon_1=\epsilon_2=0$.
To study the Wilson loop expectation value, one divides the path integral with $W_R$ insertion
by the partition function itself. Now, recall that the instanton contribution
again is independent of the matrix $\lambda$ with maximal SUSY, so that they can be taken
out of the integral. With the above normalization, the nonperturbative contributions
exactly cancel out and it suffices to consider the perturbative contribution only.
So one obtains the exact Wilson loop expectation value as follows:
\begin{equation}
  \langle W_R\rangle_{\rm exact}=\frac{1}{Z_{\rm pert}\dim R}\cdot
  \frac{1}{|W|}\int d\lambda\prod_{\alpha\in\Delta_+}
  \left(2\sinh\frac{\alpha\cdot\lambda}{2}\right)^2
  {\rm tr}_R\left(e^{2\pi\lambda}\right)e^{-\frac{2\pi^2}{\beta}{\rm tr}\lambda^2}\ .
\end{equation}
This takes precisely the same form as the matrix integral for the Wilson loops
in pure Chern-Simons theory on $S^3$ \cite{Kapustin:2009kz,Tierz:2002jj}, upon
relating our coupling with the Chern-Simons level $k$ by analytic continuation
$\beta=-\frac{2\pi i}{k}$. This is a continuation of the strange coincidence
between the perturbative maximal SYM partition function on $S^5$ and pure
Chern-Simons partition function on $S^3$ observed in \cite{Kim:2012av}, to the
expectation values of BPS Wilson loops.

No matter what the physical implication for this coincidence is, if there is anything
at all, this relation is useful as the Chern-Simons partition function and Wilson loops
are extensively studied in the literature. For instance, the expectation value of
Wilson loops in fundamental representation of $U(N)$ is given by
\cite{Witten:1988hf,Kapustin:2009kz}
\begin{equation}\label{fundamental-wilson}
  \langle W_N\rangle_{CS}=\frac{e^{-\frac{N\pi i}{k}}}{N}
  \frac{\sin\frac{\pi N}{k}}{\sin\frac{\pi}{k}}\ \ \longrightarrow\ \
  \langle W_N\rangle_{S^5}=\frac{e^{\frac{N\beta}{2}}}{N}
  \frac{\sinh\frac{N\beta}{2}}{\sinh\frac{\beta}{2}}\ ,
\end{equation}
where we plugged in $-\frac{\pi}{k}\rightarrow\frac{\beta}{2}$ to obtain
the $S^5$ expectation value. The large $N$ limit of this quantity
is also interesting. One finds that
\begin{equation}\label{wilson-large-N}
  \langle W_N\rangle_{S^5}=\frac{e^{\left(N-\frac{1}{2}\right)\beta}}{N}
  \ \frac{1-e^{-N\beta}}{1-e^{-\beta}}
  \ \stackrel{N\rightarrow\infty}{\longrightarrow}\ \ e^{N\beta}\ .
\end{equation}
The large $N$ limit we take is not the 't Hooft limit of 5d SYM, since we keep
$\beta$ fixed as we take $N\rightarrow\infty$. This is the same as the large $N$
limits we took for the partition functions in the previous subsection, as
$\beta$ is the chemical potential even in the large $N$ $AdS_7$ duals.

It would also be interesting to compare the above result with the gravity
dual observable in $AdS_7\times S^4$. The Wilson loop on $S^5$ will naturally
uplift to the Wilson surface operator on $S^5\times S^1$, which winds the Hopf fiber
of $S^5$ as well as wrapping the extra $S^1$. As the fundamental Wilson loop in
5d gauge theory is roughly the locus of the fundamental string worldsheet ending
on D4-branes, the Wilson surface is identified as the M2-branes ending on M5-branes.
So one is led to consider the Euclidean M2-brane minimal worldvolume in $AdS_7$
\cite{Rey:1998ik,Maldacena:1998im} with suitably compactified Euclidean time
(preserving SUSY), which asymptotes to Hopf fiber $\times\ S^1$
on boundary $S^5\times S^1$.

To calculate the M2-brane Euclidean action, one would first have to find the
classical solution and then further carefully regularize the divergent action.
However, it is easy to analyze the large $N$ scaling of this quantity from
simple considerations. The M2-brane action is evaluated by studying
\begin{equation}\label{M2-action}
  S_{M2}\sim\tau_{M2}\int_{\Sigma_3}\sqrt{\det g}
\end{equation}
where $g$ is the induced metric on the classical worldvolume solution
$\Sigma_3$, and $\tau_{M2}\sim\frac{1}{\ell_P^3}$ is the M2-brane tension related
to the 11d Planck length $\ell_P$. As the classical configuration of $\Sigma_3$
is determined by its dynamics in $AdS_7$, the only length scale which can affect
the classical solution and the action is the $AdS_7$ radius $\ell$, at least when
$\mathcal{O}(\beta)\sim 1$ so that the compactification of Euclidean time does not
provide another scale. So the only combination of dimensionful parameters which can appear is
\begin{equation}
  S_{M2}\sim\frac{\ell^3}{\ell_P^3}\sim N\ .
\end{equation}
This is in agreement with the large $N$ scaling of the Wilson loop expectation value
(\ref{wilson-large-N}) on $S^5$. Comparison of the order $1$ coefficients
will be an interesting question. The naive compactification
of the Euclidean time direction of global $AdS_7$ will presumably give the wrong
answer, as this is not a supersymmetric compactification which is required from
our index. If the M2-brane action calculated this way shows a mismatch in the
$\mathcal{O}(1)$ factor with (\ref{wilson-large-N}), this probably will be a similar
phonomenon as the discrepancy between the coeffcients of $N^3$ for the Casimir
energy of $AdS_7$ and the index version of it calculated in \cite{Kim:2012av,Kallen:2012zn}.

\section{Comments on ambiguities and maximal SUSY}

In the examples that we considered in sections 3.1 and 3.2, we encountered
(small) ambiguities that we had to fix by hand, to have the 5d partition function to
correctly reproduce the 6d physics. They were factors of the form $e^{\frac{A}{\beta}}$
with positive constants $A$, which are supposed to provide
the leading high temperature behaviors of the index free energies. From the structure
of the 5d partition function, it seems that such terms cannot arise from the physics
of dynamical fields on $S^5$, as this is neither perturbative correction nor
instanton correction, the latter taking the above form with negative $A$.
Supposing that the coefficient above is suitably fixed, we were able to capture
all degeneracy information for infinite towers of 6d BPS excitations.

Physically, one may regard such a term as coming from the coupling of the
parameters of the theory, such as $g_{YM}$, $m$ in our case, with background
geometry. This viewpoint was adopted in \cite{Kim:2012av}, motivated by a similar
observation that 4d twisted $\mathcal{N}=4$ SYM on various curved manifolds
exhibits S-duality only when the partition functions acquire nonzero contributions
from similar curvature couplings \cite{Vafa:1994tf}. For simplicity, let us
consider the round $S^5$ for which all curvature invariants reduces to
a polynomial of the scalar curvature.
Also putting the maximal SYM there for simplicity, the only parameter $g_{YM}^2$
has the dimension of length. So all possible curvature couplings yield
a constant shift to the action of the following form:
\begin{equation}\label{curvature-coupling}
  S_{\rm bkgd}=\sum_{n=0}^\infty\frac{A_n}{(g_{YM}^2)^{5-2n}}\int_{S^5}R^n\sim
  \frac{\alpha_{-5}}{\beta^5}+\frac{\alpha_{-3}}{\beta^3}+\frac{\alpha_{-1}}{\beta}
  +\alpha_1\beta+\alpha_3\beta^3+\cdots\ .
\end{equation}
Clearly, addition of such terms does not break any symmetry of the field theory.
It seems that the first three terms cannot appear from the dynamics of QFT fields on
$S^5$. The remaining infinite series in positive powers of $\beta$ takes the
form of perturbative contribution to the partition function. Addition of
such a series with arbitrary coefficients of course will spoil all the perturbative
considerations from the field theory. In fact, such additions will also spoil all
perturbative studies of QFT on curved manifolds like $S^3$ or $S^4$ in the literature.
So we only consider the first three terms when they seem to be inevitably necessary.

The above terms in negative powers of $\beta$ are naturally required
from the 6d physics because we have larger degeneracies of states with
growing energy. There are only finite number of coefficients that one can tune,
and our strategy is fitting these small ambiguities to get the information of
infinitely many BPS states. In the two examples (Abelian index and the index computed
from 5d maximal SYM) that we concretely studied, the only nonzero coefficient was
$\alpha_{-1}$. On the other hand, $\frac{\alpha_{-3}}{\beta^3}$ seems to be the
right asymptotic scalings for the 6d BPS partition function (counting BPS bosons
plus fermions) at high temperature because $\beta^{-3}\sim T^3$
would come from the three holomorphic
BPS derivatives of the 6d theory. For this reason, we expect that
$\alpha_{-5}$ should always be zero as the index cannot grow faster
than the BPS partition function. It would be interesting to see
in the general non-Abelian $S^5$ partition function whether nonzero $\alpha_{-3}$
is required, to make the full expression into an index. If that is the case, the
implication is that the index would not be experiencing too much cancelation
between boson/fermion so that there is a chance for the index degeneracy to be
close to the true BPS degeneracy of the theory. With enough technical control
over our general matrix integral and their strong-coupling expansion, whether
$\alpha_{-3}\neq 0$ or not will be predictable from 5d considerations.

When we generalize the above considerations to squashed $S^5$, there are more
quantities that one can construct from the background, other than the scalar
curvature $R\sim r^{-2}$, so that the analysis could be much more complicated.
Apart from various curvature invariants, we have already seen in section 2.1 that
some background scalar and vector field had to be nonzero. It will be interesting
to try to understand the structure of such terms from the off-shell supergravity
methods \cite{Festuccia:2011ws}, which were recently developed to conveniently
construct QFT with rigid SUSY on curved spaces. One should first try to understand
the $\mathcal{O}(\beta^{-1})$ term of the Abelian index in section 3.1.

Another source of possible ambiguity in our 5d calculation is that the index
version of Casimir energy is somewhat difficult to interpret in general.
If one can fully evaluate our general partition function and expand it at strong
coupling, the result will give some values of this Casimir energy factor
$e^{-\beta\epsilon_0}$, supposing that the strong coupling expansion indeed
takes the form of index. Indeed, our Abelian partition function did provide the expression
obtained directly in 6d in section 3.1. Also, \cite{Kallen:2012zn} computed the
leading large $N$ behavior of the perturbative free energy which takes the form
$\beta\epsilon_0$ with definite values of $\epsilon_0$. \cite{Kallen:2012zn}
considered the case with nonzero $\beta$, $m$. The Casimir energy here also
exhibited the mass dependence $\epsilon_0\propto\left(\frac{9}{4}-m^2\right)^2$
which is hard for us to interpret. In any case, the fact that both 5d and 6d
calculations yield the same nontrivial expression (\ref{casimir-ambiguous})
is interesting. Perhaps, one might have to carefully reconsider the regulator
dependence of the calculation of the $S^5$ partition function.

Finally, as explained at the beginning of section 3.1, the fact that
instanton moduli space has small instanton singularities could also be regarded
as a possible ambiguity in our study. One may regard the `natural' UV prescriptions
such as \cite{Nekrasov:1998ss} may be providing some UV completion beyond
the potentially dangerous 5d QFT, at least in the BPS sector.

\vskip 0.5cm

\hspace*{-0.8cm} {\bf\large Acknowledgements}

\vskip 0.2cm

\hspace*{-0.75cm} We thank Keshav Dasgupta, Davide Gaiotto, Jaume Gomis, Daniel Jafferis,
Sung-Soo Kim, Igor Klebanov, Zohar Komargodski, Neil Lambert, Kimyeong Lee, Sungjay Lee,
Alexander Maloney, Rob Myers, Jaemo Park, Leonardo Rastelli, Jorge Russo, Yuji Tachikawa
and Johannes Walcher for helpful discussions.
This work is supported by the BK21 program of the Ministry of Education, Science and Technology
of Korea (JK,SK), the National Research Foundation of Korea (NRF) Grants No. 2010-0007512
(JK, SK), 2012R1A1A2042474 (HK,JK,SK), 2012R1A2A2A02046739 (SK) and 2005-0049409 through
the Center for Quantum Spacetime (CQUeST) of Sogang University (JK,SK). The work of SK at
the Perimeter
Institute is supported by the Government of Canada through Industry Canada, and by the
Province of Ontario through the Ministry of Economic Development \& Innovation. SK
thanks the organizers and staffs of the program ``Exact results in gauge theory and their
applications'' at the Aspen Center for Physics, for the hospitality and the support by
the National Science Foundation under Grant No. PHYS-1066293.

\appendix

\section{Instanton countings for classical gauge groups}

\subsection{Indices and determinants for $SO(N)$}

$SO(2N\!+\!1)$ case is a slight generalization of the $SO(2N)$, at least
at some formal level in index considerations. So we only discuss the index and
partition fucntion for the $SO(2N)$ gauge theory here. Using the results of
\cite{Nekrasov:2004vw}, one can write down the $SO(2N\!+\!1)$ results
straightforwardly.

The fundamental characters for $SO(2N)$ and $Sp(k)$ are given as
\be
	\text{ch}_N = \sum_{i=1}^N (e^{i \lambda_i} + e^{-i \lambda_i}) \,, \hspace{1cm} \text{ch}_k = \sum_{i=1}^k (e^{i \phi_I} + e^{-i \phi_I}).
\ee
Since the adjoint characters for $SO(2N)$ ($Sp(k)$) are obtained by (anti)symmetrization, we have
\be
	\text{ch}_{\text{adj}_N} &=& \sum_{i<j}^N (e^{i \lambda_i + i \lambda_j} + e^{i \lambda_i - i \lambda_j} + e^{-i \lambda_i + i \lambda_j} + e^{-i \lambda_i - i \lambda_j}) + N \\
	\text{ch}_{\text{adj}_k} &=& \sum_{I<J}^k (e^{i \phi_I + i \phi_J} + e^{i \phi_I - i \phi_J} + e^{-i \phi_I + i \phi_J} + e^{-i \phi_I - i \phi_J}) + \sum_{i=1}^k (e^{2i \phi_I} + e^{-2i \phi_I}) + k.
\ee
Inserting them into (\ref{equivariant-chern}) with slight modification \cite{Nekrasov:2004vw},
we can compute the equivariant indices for a vector multiplet and an adjoint hypermultiplet.
Firstly, for the perturbative part, the indices are
\be
	I^\text{pert}_\text{vector} &=& -\frac{1 + e^{i(\epsilon_1 + \epsilon_2)}}{2} \sum_{t=-\infty}^{\infty} e^{it(1+c)/r} \left[\frac{\sum_{i<j}^N (e^{i \lambda_i + i \lambda_j} + e^{i \lambda_i - i \lambda_j} + e^{-i \lambda_i + i \lambda_j} + e^{-i \lambda_i - i \lambda_j}) + N}{(1-e^{i \epsilon_1}) (1-e^{i \epsilon_2})}\right]\nonumber\\
	I^\text{pert}_\text{hyper} &=&    \frac{e^{i\frac{\epsilon_1 + \epsilon_2}{2}}}{(1-e^{i \epsilon_1}) (1-e^{i \epsilon_2})} \left(\frac{e^{i(m + \frac{3}{2} (1+c))} +e^{-i(m + \frac{3}{2} (1+c))}}{2}\right)  \nn \\ && \times \sum_{t=-\infty}^{\infty} e^{it(1+c)/r} \left[\sum_{i<j}^N (e^{i \lambda_i + i \lambda_j} + e^{i \lambda_i - i \lambda_j} + e^{-i \lambda_i + i \lambda_j} + e^{-i \lambda_i - i \lambda_j}) + N\right]
\ee
for $\epsilon_1 = a-c$ and $\epsilon_2 = b-c$. The indices from instantonic contributions can be obtained in a similar way:
\be
	I^\text{inst}_\text{vector} &=& -\frac{1 + e^{i(\epsilon_1 + \epsilon_2)}}{2}  \left[ -e^{-i \frac{\epsilon_1 + \epsilon_2}{2}} \left(\sum_{i=1}^n \sum_{I=1}^k e^{i \lambda_i + i \phi_I} + e^{-i \lambda_i + i \phi_I} + e^{i \lambda_i - i \phi_I} + e^{-i \lambda_i - i \phi_I} \right)\right.\nn \\
	&&+ (1-e^{- \epsilon_1})(1-e^{- \epsilon_2}) \left\{\sum_{I<J}^k (e^{i \phi_I + i \phi_J} + e^{i \phi_I - i \phi_J} + e^{-i \phi_I + i \phi_J} + e^{-i \phi_I - i \phi_J}) + k\right\} \nn \\
	&& \left. + \sum_{i=1}^k (e^{2i \phi_I} + e^{-2i \phi_I} + e^{2i \phi_I - i\epsilon_1 - i\epsilon_2} + e^{-2i \phi_I- i\epsilon_1 - i\epsilon_2}) \right] \, \sum_{t=-\infty}^{\infty} e^{it(1+c)/r}
\ee
\be
	I^\text{inst}_\text{hyper} &=& \frac{e^{i\frac{\epsilon_1 + \epsilon_2}{2}}}{(1-e^{i \epsilon_1}) (1-e^{i \epsilon_2})} \left(\frac{e^{i(m + \frac{3}{2} (1+c))} +e^{-i(m + \frac{3}{2} (1+c))}}{2}\right) \sum_{t=-\infty}^{\infty} e^{it(1+c)/r} \nn \\
	&& \times \left[ -e^{-i \frac{\epsilon_1 + \epsilon_2}{2}} \left(\sum_{i=1}^n \sum_{I=1}^k e^{i \lambda_i + i \phi_I} + e^{-i \lambda_i + i \phi_I} + e^{i \lambda_i - i \phi_I} + e^{-i \lambda_i - i \phi_I} \right)\right.\nn \\
	&&+ (1-e^{- \epsilon_1})(1-e^{- \epsilon_2}) \left\{\sum_{I<J}^k (e^{i \phi_I + i \phi_J} + e^{i \phi_I - i \phi_J} + e^{-i \phi_I + i \phi_J} + e^{-i \phi_I - i \phi_J}) + k\right\} \nn \\
	&& \left. + \sum_{i=1}^k (e^{2i \phi_I} + e^{-2i \phi_I} + e^{2i \phi_I - i\epsilon_1 - i\epsilon_2} + e^{-2i \phi_I- i\epsilon_1 - i\epsilon_2}) \right].
\ee
From these indices, we can read off the 1-loop determinant and the instanton partition function takes the form of a contour integration as follows.
\be
\label{dncontour}
\begin{gathered}
  Z_k^{(3)} = \frac{(1+c)^{-k}}{k!}\oint \left[ \prod_{I=1}^k \frac{d \phi_I}{2 \pi} \right]
  \prod_{I,J} \frac{\sin{\pi\frac{\phi_{IJ} - 2\epsilon_+}{1+c}}}{\sin{\pi\frac{\phi_{IJ} -\epsilon_1}{1+c}} \sin{\pi\frac{\phi_{IJ}-\epsilon_2}{1+c}}}\frac{\sin{\pi\frac{\phi_{IJ} +m + \frac{3(1+c)}{2} + \epsilon_-}{1+c}}
  \sin{\pi\frac{\phi_{IJ} + m + \frac{3(1+c)}{2} - \epsilon_-}{1+c}}}
  {\sin{\pi\frac{\phi_{IJ} + m + \frac{3(1+c)}{2} + \epsilon_+}{1+c}} \sin{\pi\frac{\phi_{IJ} +m + \frac{3(1+c)}{2} - \epsilon_+}{1+c}}} \\
  \prod_{I\neq J} \sin{\pi\tfrac{\phi_{IJ}}{1+c}} \prod_{I<J} \frac{\sin{\pi\frac{\phi_I + \phi_J}{1+c}} \sin{\pi\frac{\phi_I+\phi_J - 2\epsilon_+}{1+c}}}{\sin{\pi\frac{\phi_I+\phi_J -\epsilon_1}{1+c}} \sin{\pi\frac{\phi_I+\phi_J-\epsilon_2}{1+c}}} \frac{\sin{\pi\frac{\phi_I+\phi_J +m + \frac{3(1+c)}{2} + \epsilon_-}{1+c}} \sin{\pi\frac{\phi_I+\phi_J + m + \frac{3(1+c)}{2} - \epsilon_-}{1+c}}}{\sin{\pi\frac{\phi_I+\phi_J + m + \frac{3(1+c)}{2} + \epsilon_+}{1+c} \sin{\pi\frac{\phi_I+\phi_J +m + \frac{3(1+c)}{2} - \epsilon_+}{1+c}}}} \\
  \prod_{I<J} \frac{\sin{\pi\frac{\phi_I + \phi_J}{1+c}} \sin{\pi\frac{\phi_I+\phi_J + 2\epsilon_+}{1+c}}}{\sin{\pi\frac{\phi_I+\phi_J +\epsilon_1}{1+c}} \sin{\pi\frac{\phi_I+\phi_J+\epsilon_2}{1+c}}}\frac{\sin{\pi\frac{\phi_I+\phi_J -m - \frac{3(1+c)}{2} + \epsilon_-}{1+c}} \sin{\pi\frac{\phi_I+\phi_J - m - \frac{3(1+c)}{2} - \epsilon_-}{1+c}}}{\sin{\pi\frac{\phi_I+\phi_J - m - \frac{3(1+c)}{2} + \epsilon_+}{1+c} \sin{\pi\frac{\phi_I+\phi_J -m - \frac{3(1+c)}{2} - \epsilon_+}{1+c}}}} \\
  \prod_{I}\prod_i \frac{\sin{\pi\frac{\phi_I + \lambda_i +m+ \frac{3(1+c)}{2}}{1+c}}  \sin{\pi\frac{\phi_I + \lambda_i -m - \frac{3(1+c)}{2}}{1+c}} \sin{\pi\frac{\phi_I - \lambda_i + m + \frac{3(1+c)}{2}}{1+c}}  \sin{\pi\frac{\phi_I - \lambda_i -m - \frac{3(1+c)}{2}}{1+c}}}{\sin{\pi\frac{\phi_I + \lambda_i - \epsilon_+}{1+c}} \sin{\pi\frac{\phi_I + \lambda_i + \epsilon_+}{1+c}} \sin{\pi\frac{\phi_I - \lambda_i - \epsilon_+}{1+c}} \sin{\pi\frac{\phi_I - \lambda_i + \epsilon_+}{1+c}}}\\
  \prod_{I} \frac{\sin{\pi\frac{2\phi_I}{1+c}} \sin{\pi\frac{2\phi_I}{1+c}} \sin{\pi\frac{2\phi_I - 2\epsilon_+}{1+c}}\sin{\pi\frac{2\phi_I + 2\epsilon_+}{1+c}}}{\sin{\pi\frac{2\phi_I + m + \frac{3(1+c)}{2} + \epsilon_+}{1+c}} \sin{\pi\frac{2\phi_I - m - \frac{3(1+c)}{2} + \epsilon_+}{1+c}} \sin{\pi\frac{2\phi_I +m + \frac{3(1+c)}{2} - \epsilon_+}{1+c}} \sin{\pi\frac{2\phi_I - m - \epsilon_+}{1+c}}}.
\end{gathered}
\ee
Note that we have not divided it by the full dimension of Weyl group yet, thus part of gauge redundancy still remains.

By choosing the proper contour, which is the same as $U(N)$ case, the integration of (\ref{dncontour}) is nothing but extracting residues from poles that can be classified in terms of extended $N$-colored Young diagrams. For $Z_k$ in $SO(2N)$ theory, we consider $k$ boxes arranged into $2N$ sectors; i.e., $Y = (Y_{1+},\, Y_{2+}, \,\dots | \, Y_{1-},\, Y_{2-}, \,\dots )$, where $Y_{i\pm}$ denotes the Young diagram with boxes belonging to the $i_\pm$'th sector. For example, if $\phi_I = \pm \lambda_i - \epsilon_+ - n \epsilon_1 - m \epsilon_2$ for nonnegative $n, m$, the $I$'th box is placed in the $Y_{i\pm}$ Young diagram. Due to the factor $\sin{\pi\frac{\phi_I+\phi_J + 2\epsilon_+}{1+c}}$, which comes from ADHM constraints, here applies the following restriction: Both $Y_{i+}$ and $Y_{i-}$ cannot be occupied at the same time. Collecting all residues with this rule, we can compute $Z_k$ correctly.

Now we encounter the remaining part of Weyl redundancy that connects $Y_{i+}$ and $Y_{i-}$ by mirroring. Hence we should mod out the set of mirrored configurations. This can be illustrated well by an example. Consider $Z_{k=2}$ for $SO(4)$. All possible residues of extended 2-colored Young diagrams are collected, with division by number of redundancies, as follows.
\be
   Z_2 &=& \tfrac{1}{2} \left\{ (\,\tiny\yng(2)\,,\,\cdot\, | \,\cdot\, , \,\cdot\,) + (\,\cdot\,,\,\cdot\, | \,\tiny\yng(2)\,, \,\cdot\,) \right\} + \tfrac{1}{2} \left\{ (\,\cdot\,,\,\tiny\yng(2)\, | \,\cdot\, , \,\cdot\,) + (\,\cdot\,,\,\cdot\, | \,\cdot\,, \,\tiny\yng(2)\,) \right\} \nn \\
   && +\tfrac{1}{2} \left\{ (\,\tiny\yng(1,1)\,,\,\cdot\, | \,\cdot\, , \,\cdot\,) + (\,\cdot\,,\,\cdot\, | \,\tiny\yng(1,1)\,, \,\cdot\,) \right\} + \tfrac{1}{2} \left\{ (\,\cdot\,,\,\tiny\yng(1,1)\, | \,\cdot\, , \,\cdot\,) + (\,\cdot\,,\,\cdot\, | \,\cdot\,, \,\tiny\yng(1,1)\,) \right\} \nn \\
   && + \tfrac{1}{4} \left\{(\,\tiny\yng(1)\,,\,\tiny\yng(1)\, | \,\cdot\, , \,\cdot\,) + (\,\tiny\yng(1)\,,\,\cdot\, | \,\cdot\, , \,\tiny\yng(1)\,) + (\,\cdot\,,\,\tiny\yng(1)\, | \,\tiny\yng(1)\, , \,\cdot\,) + (\,\cdot\,,\,\tiny\yng(1)\, | \,\cdot\, , \,\tiny\yng(1)\,) \right\}.
\ee
For the case of higher $N$ and $k$, $Z_k$ can be computed in a similar manner.

\subsection{Indices and determinants for $Sp(N)$}

The equivariant index for the perturbative part is obtained from that of $U(N)$ theory
by simply exchanging the $U(N)$ character to $Sp(N)$ character.
The vector multiplet contribution is
\be
    I^{{\rm pert}}_{{\rm vector}} &=& -\frac{1+e^{i(b+c-2a)}}{2(1-e^{i(b-a)})(1-e^{i(c-a)})}\sum_{t\in Z}e^{i\frac{t}{r}(1+a)} \nn \\
    && \quad \times
    \left[\sum_{i<j}^N\left(e^{i\lambda_i+i\lambda_j}+e^{-i\lambda_i-i\lambda_j}+e^{i\lambda_i-i\lambda_j}+e^{-i\lambda_i+i\lambda_j}\right)
    +\sum_{i=1}^N\left(e^{2i\lambda_i}+e^{-2i\lambda_i}\right)+N\right] \quad
\ee
and the index of the hypermultiplet  in the adjoint representation is
\be
    I^{{\rm pert}}_{{\rm hyper}} &=& \frac{e^{\frac{i}{2}(b+c-2a)}}{(1-e^{i(b-a)})(1-e^{i(c-a)})}\left(\frac{e^{i(m+\frac{3(1+a)}{2r})}+e^{-i(m+\frac{3(1+a)}{2r})}}{2}\right)\sum_{t\in Z}e^{i\frac{t}{r}(1+a)} \nn \\
    && \quad \times
    \left[\sum_{i<j}^N\left(e^{i\lambda_i+i\lambda_j}+e^{-i\lambda_i-i\lambda_j}+e^{i\lambda_i-i\lambda_j}+e^{-i\lambda_i+i\lambda_j}\right)
    +\sum_{i=1}^N\left(e^{2i\lambda_i}+e^{-2i\lambda_i}\right)+N\right] \quad
\ee

For the instanton part, one has to carefully consider the dual gauge group of the instanton moduli space which is $O(k)$ gauge group at $k$ instanton sector for $Sp(N)$ gauge theory.
The $O(k)$ action is generated by two disjoint components : $O(k)_+$ consists of the elements with determinant $+1$ and $O(k)_-$ consists of the elements with determinant $-1$.
The group action is generated by the following elements
\be
    e^{i\phi_+} = \left\{\begin{array}{l} {\rm diag}\left(e^{i\sigma_2\phi_1},\cdots,e^{i\sigma_2\phi_n}\right)\ {\rm for \ even}\ k \\
                                            {\rm diag}\left(e^{i\sigma_2\phi_1},\cdots,e^{i\sigma_2\phi_n},1\right)\ {\rm for \ odd}\ k
                    \end{array}\right.
\ee
for $O(k)_+$, and
\be\label{Okm-action}
    e^{i\phi_+} = \left\{\begin{array}{l} {\rm diag}\left(e^{i\sigma_2\phi_1},\cdots,e^{i\sigma_2\phi_{n-1}},1,e^{i\frac{\rho}{2}}\right)\ {\rm for \ even}\ k \\
                                            {\rm diag}\left(e^{i\sigma_2\phi_1},\cdots,e^{i\sigma_2\phi_n},e^{i\frac{\rho}{2}}\right)\ {\rm for \ odd}\ k
                    \end{array}\right.
\ee
for $O(k)_-$, where $\sigma_2$ is Pauli matrix and $\rho$ is the periodicity of $\phi_I$.
(When we use the usual periodicity $\phi_I\sim\phi_I+2\pi$, the $U(1)$ action $e^{i\rho/2}$ becomes $e^{i\pi}=-1$. This corresponds to $\mathbb{Z}_2$ quotient which is a member of $O(k)$ group action. However, for our case, I used the periodicity $\rho$ for $\phi_I$ as the final formula has the terms like $\sin\pi\frac{\phi_I+\cdots}{1+a}$
and $\phi_I$ is no longer periodic under $2\pi$ shift.)
The index $n$ is defined as $k=2n+\chi \ (\chi=0\ {\rm or}\ 1)$.

Firsly, the equivariant indices of the instanton part for $O(k)_+$ are
\be
    \hspace{-1cm}{\rm ind}^{k_+}_{{\rm vec}}\!\!&\!\!=\!\!&\!\!-\frac{1+e^{2i\epsilon_+}}{2}
        \Bigg[-e^{-i\epsilon_+}\sum_{I=1}^n\sum_{i=1}^N\left(e^{i\phi_I+i\lambda_i}+e^{-i\phi_I+i\lambda_i}+e^{i\phi_I-i\lambda_i}+e^{-i\phi_I-i\lambda_i}\right)
        -\chi e^{-i\epsilon_+}\sum_{i=1}^N\left(e^{i\lambda_i}+e^{-i\lambda_i}\right) \nn \\
    \hspace{-1cm}&&  +(1-e^{-i\epsilon_1})(1-e^{-i\epsilon_2})\!\!\left( \sum_{I<J}^n\left(e^{i\phi_I+i\phi_J}+e^{-i\phi_I+i\phi_J}+e^{i\phi_I-i\phi_J}+e^{-i\phi_I-i\phi_J}\right)
    +\chi\sum_{I=1}^n\left(e^{i\phi_I}+e^{-i\phi_I}\right)+n\right) \nn \\
    \hspace{-1cm}&& -\left(e^{-i\epsilon_1}+e^{-i\epsilon_2}\right)\left(\sum_{I=1}^n\left(e^{2i\phi_I}+e^{-2i\phi_I}\right)+\chi\right)\Bigg] \sum_{t\in Z} e^{it(1+a)}
\ee
for the vector multiplet and
\be
    \hspace{-1cm}{\rm ind}^{k_+}_{{\rm hyper}}\!\!&\!\!=\!\!&\!\!e^{i\epsilon_+}
        \Bigg[-e^{-i\epsilon_+}\sum_{I=1}^n\sum_{i=1}^N\left(e^{i\phi_I+i\lambda_i}+e^{-i\phi_I+i\lambda_i}+e^{i\phi_I-i\lambda_i}+e^{-i\phi_I-i\lambda_i}\right)
        -\chi e^{-i\epsilon_+}\sum_{i=1}^N\left(e^{i\lambda_i}+e^{-i\lambda_i}\right) \nn \\
    \hspace{-1cm}&&  +(1-e^{-i\epsilon_1})(1-e^{-i\epsilon_2})\!\!\left( \sum_{I<J}^n\left(e^{i\phi_I+i\phi_J}+e^{-i\phi_I+i\phi_J}+e^{i\phi_I-i\phi_J}+e^{-i\phi_I-i\phi_J}\right)
    +\chi\sum_{I=1}^n\left(e^{i\phi_I}+e^{-i\phi_I}\right)+n\right) \nn \\
    \hspace{-1cm}&& -\left(e^{-i\epsilon_1}+e^{-i\epsilon_2}\right)\left(\sum_{I=1}^n\left(e^{2i\phi_I}+e^{-2i\phi_I}\right)+\chi\right)\Bigg] \frac{\left(e^{im'}+e^{-im'}\right)}{2}\sum_{t\in Z} e^{it(1+a)}
\ee
for the hypermultiplet in the adjoint representation.
From the equivariant index, we read the contour integral formula for $O(k)_+$ dual gauge group
\be
    I^k_+\!\!&\!\!\sim\!\!&\!\!\oint\prod_{I=1}^n\frac{d\phi_I}{2\pi}\left[\frac{\sin\pi\frac{m'\pm\epsilon_-}{1+a}}{\sin\pi\frac{\epsilon_+\pm\epsilon_-}{1+a}}
    \prod_{I=1}^n\frac{\sin^2\pi\frac{\phi_I}{1+a}\sin\pi\frac{\phi_I\pm 2\epsilon_+}{1+a}\sin\pi\frac{\phi_I\pm m'\pm\epsilon_-}{1+a}}
    {\sin\pi\frac{\phi_I\pm\epsilon_+\pm\epsilon_-}{1+a}\sin\pi\frac{\phi_I\pm m'\pm\epsilon_+}{1+a}}
    \prod_{i=1}^N\frac{\sin\pi\frac{\lambda_i\pm m'}{1+a}}{\sin\pi\frac{\lambda_i\pm \epsilon_+}{1+a}}\right]^\chi \\
    &&\!\!\times\prod_{I=1}^n\left[\frac{\sin\pi \frac{2\epsilon_+}{1+a}\sin\pi\frac{m'\pm\epsilon_-}{1+a}\sin\pi\frac{2\phi_I\pm m'\pm\epsilon_-}{1+a}}{\sin\pi\frac{\epsilon_+\pm\epsilon_-}{1+a}\sin\pi\frac{m'\pm\epsilon_+}{1+a}\sin\pi\frac{2\phi_I\pm\epsilon_+\pm\epsilon_-}{1+a}}
    \prod_{i=1}^N\frac{\sin\pi\frac{\phi_I\pm\lambda_i\pm m'}{1+a}}{\sin\pi\frac{\phi_I\pm\lambda_i\pm\epsilon_+}{1+a}}\right] \nn \\
    &&\!\!\times\prod_{I<J}^n\left[\frac{\sin^2\pi\frac{\phi_I\pm\phi_J}{1+a}\sin\pi\frac{\phi_I\pm\phi_J\pm2\epsilon_+}{1+a}\sin\pi\frac{\phi_I\pm\phi_J\pm m'\pm\epsilon_-}{1+a}}
    {\sin\pi\frac{\phi_I\pm\phi_J\pm\epsilon_+\pm\epsilon_-}{1+a}\sin\pi\frac{\phi_I\pm\phi_J\pm m'\pm\epsilon_+}{1+a}}\right] \nn
\ee

Now we turn to the equivariant index for $O(k)_-$ action. Since the $O(k)_-$ action behaves in very different way for even and odd $k$, we shall deal with two cases separately.
Firstly, we write the equivariant index for odd $k$ in the adjoint representation
\be
    \hspace{-1.3cm}{\rm ind}^{k_-:{\rm odd}}_{{\rm adj}}\!\!&\!\!=\!\!&\!\!
        \Bigg[\!\!-e^{-i\epsilon_+}\sum_{I=1}^n\sum_{i=1}^N\left(e^{i\phi_I+i\lambda_i}+e^{-i\phi_I+i\lambda_i}+e^{i\phi_I-i\lambda_i}+e^{-i\phi_I-i\lambda_i}\right)
        - e^{-i\epsilon_+}e^{i\frac{\rho}{2}}\sum_{i=1}^N\left(e^{i\lambda_i}+e^{-i\lambda_i}\right) \nn \\
    \hspace{-1.3cm}&&  +(1-e^{-i\epsilon_1})(1-e^{-i\epsilon_2})\!\!\left( \sum_{I<J}^n\left(e^{i\phi_I+i\phi_J}+e^{-i\phi_I+i\phi_J}+e^{i\phi_I-i\phi_J}+e^{-i\phi_I-i\phi_J}\right)
    +e^{i\frac{\rho}{2}}\sum_{I=1}^n\left(e^{i\phi_I}+e^{-i\phi_I}\right)\!+\!n\!\right) \nn \\
    \hspace{-1.3cm}&& -\left(e^{-i\epsilon_1}+e^{-i\epsilon_2}\right)\left(\sum_{I=1}^n\left(e^{2i\phi_I}+e^{-2i\phi_I}\right)+1\right)\!\!\Bigg] \sum_{t\in Z} e^{it(1+a)}
\ee
and that of the vector multiplet and that of the hypermultiplet can be obtained from this index, which are given by
\be
    {\rm ind}^{k_-:{\rm odd}}_{{\rm vec}} = -\frac{1+e^{2i\epsilon_+}}{2}{\rm ind}^{k_-:{\rm odd}}_{{\rm adj}} \ , \quad
    {\rm ind}^{k_-:{\rm odd}}_{{\rm hyper}} = e^{i\epsilon_+}\frac{e^{im'}+e^{-im'}}{2}{\rm ind}^{k_-:{\rm odd}}_{{\rm adj}}
\ee
Then the final instanton index has the following contour integral form
\be
    I^{k:{\rm odd}}_-\!\!&\!\!\sim\!\!&\!\!\oint\prod_{I=1}^n\frac{d\phi_I}{2\pi}\left[\frac{\sin\pi\frac{m'\pm\epsilon_-}{1+a}}{\sin\pi\frac{\epsilon_+\pm\epsilon_-}{1+a}}
    \prod_{I=1}^n\frac{\cos^2\pi\frac{\phi_I}{1+a}\cos\pi\frac{\phi_I\pm 2\epsilon_+}{1+a}\cos\pi\frac{\phi_I\pm m'\pm\epsilon_-}{1+a}}
    {\cos\pi\frac{\phi_I\pm\epsilon_+\pm\epsilon_-}{1+a}\cos\pi\frac{\phi_I\pm m'\pm\epsilon_+}{1+a}}
    \prod_{i=1}^N\frac{\cos\pi\frac{\lambda_i\pm m'}{1+a}}{\cos\pi\frac{\lambda_i\pm \epsilon_+}{1+a}}\right] \nn \\
    &&\!\!\times\prod_{I=1}^n\left[\frac{\sin\pi \frac{2\epsilon_+}{1+a}\sin\pi\frac{m'\pm\epsilon_-}{1+a}\sin\pi\frac{2\phi_I\pm m'\pm\epsilon_-}{1+a}}{\sin\pi\frac{\epsilon_+\pm\epsilon_-}{1+a}\sin\pi\frac{m'\pm\epsilon_+}{1+a}\sin\pi\frac{2\phi_I\pm\epsilon_+\pm\epsilon_-}{1+a}}
    \prod_{i=1}^N\frac{\sin\pi\frac{\phi_I\pm\lambda_i\pm m'}{1+a}}{\sin\pi\frac{\phi_I\pm\lambda_i\pm\epsilon_+}{1+a}}\right] \nn \\
    &&\!\!\times\prod_{I<J}^n\left[\frac{\sin^2\pi\frac{\phi_I\pm\phi_J}{1+a}\sin\pi\frac{\phi_I\pm\phi_J\pm2\epsilon_+}{1+a}\sin\pi\frac{\phi_I\pm\phi_J\pm m'\pm\epsilon_-}{1+a}}
    {\sin\pi\frac{\phi_I\pm\phi_J\pm\epsilon_+\pm\epsilon_-}{1+a}\sin\pi\frac{\phi_I\pm\phi_J\pm m'\pm\epsilon_+}{1+a}}\right]
\ee
Here we take the periodicity for $\phi_I$ as $\rho=\frac{1+a}{2}$.

Secondly, the equivariant index for even $k$ in the adjoint representation is given by
\be
    \hspace{-1.3cm}&&{\rm ind}^{k_-:{\rm even}}_{{\rm adj}}=
        \Bigg[\!\!-e^{-i\epsilon_+}\sum_{I=1}^{n-1}\sum_{i=1}^N\left(e^{i\phi_I+i\lambda_i}+e^{-i\phi_I+i\lambda_i}+e^{i\phi_I-i\lambda_i}+e^{-i\phi_I-i\lambda_i}\right)
        - e^{-i\epsilon_+}(1+e^{i\frac{\rho}{2}})\sum_{i=1}^N\left(e^{i\lambda_i}+e^{-i\lambda_i}\right) \nn \\
    \hspace{-1.3cm}&&  +(1-e^{-i\epsilon_1})(1-e^{-i\epsilon_2})\!\!\left( \sum_{I<J}^{n-1}\left(e^{i\phi_I+i\phi_J}+e^{-i\phi_I+i\phi_J}+e^{i\phi_I-i\phi_J}+e^{-i\phi_I-i\phi_J}\right)
    +(1+e^{i\frac{\rho}{2}})\sum_{I=1}^{n-1}\left(e^{i\phi_I}+e^{-i\phi_I}\right)\!+\!n\!\right) \nn \\
    \hspace{-1.3cm}&& -\left(e^{-i\epsilon_1}+e^{-i\epsilon_2}\right)\left(\sum_{I=1}^n\left(e^{2i\phi_I}+e^{-2i\phi_I}\right)+1+e^{i\frac{\rho}{2}}\right)+(1+e^{-2i\epsilon_+})(-1+e^{i\frac{\rho}{2}})\Bigg] \sum_{t\in Z} e^{it(1+a)}
\ee
The vector multiplet index and the hypermultiplet index are then given by
\be
    {\rm ind}^{k_-:{\rm odd}}_{{\rm vec}} = -\frac{1+e^{2i\epsilon_+}}{2}{\rm ind}^{k_-:{\rm odd}}_{{\rm adj}} \ , \quad
    {\rm ind}^{k_-:{\rm odd}}_{{\rm hyper}} = e^{i\epsilon_+}\frac{e^{im'}+e^{-im'}}{2}{\rm ind}^{k_-:{\rm odd}}_{{\rm adj}}
\ee
Thus, the contour integral formula of the instanton index becomes
\be
    I^{k:{\rm even}}_-\!\!&\!\!\sim\!\!&\!\!\oint\prod_{I=1}^{n-1}\frac{d\phi_I}{2\pi}\left[\frac{\cos\pi\frac{2\epsilon_+}{1+a}\cos\pi\frac{m'\pm\epsilon_-}{1+a}\sin^2\pi\frac{m'\pm\epsilon_-}{1+a}}
    {\cos\pi\frac{m'\pm\epsilon_+}{1+a}\cos\pi\frac{\epsilon_+\pm\epsilon_-}{1+a}\sin^2\pi\frac{\epsilon_+\pm\epsilon_-}{1+a}}
    \prod_{I=1}^{n-1}\frac{\sin^22\pi\frac{\phi_I}{1+a}\sin2\pi\frac{\phi_I\pm 2\epsilon_+}{1+a}\sin2\pi\frac{\phi_I\pm m'\pm\epsilon_-}{1+a}}
    {\sin2\pi\frac{\phi_I\pm\epsilon_+\pm\epsilon_-}{1+a}\sin2\pi\frac{\phi_I\pm m'\pm\epsilon_+}{1+a}}\right] \nn \\
    &&\!\!\times\prod_{i=1}^N\frac{\sin2\pi\frac{\lambda_i\pm m'}{1+a}}{\sin2\pi\frac{\lambda_i\pm \epsilon_+}{1+a}}\times\prod_{I=1}^{n-1}\left[\frac{\sin\pi \frac{2\epsilon_+}{1+a}\sin\pi\frac{m'\pm\epsilon_-}{1+a}\sin\pi\frac{2\phi_I\pm m'\pm\epsilon_-}{1+a}}{\sin\pi\frac{\epsilon_+\pm\epsilon_-}{1+a}\sin\pi\frac{m'\pm\epsilon_+}{1+a}\sin\pi\frac{2\phi_I\pm\epsilon_+\pm\epsilon_-}{1+a}}
    \prod_{i=1}^N\frac{\sin\pi\frac{\phi_I\pm\lambda_i\pm m'}{1+a}}{\sin\pi\frac{\phi_I\pm\lambda_i\pm\epsilon_+}{1+a}}\right] \nn \\
    &&\!\!\times\prod_{I<J}^{n-1}\left[\frac{\sin^2\pi\frac{\phi_I\pm\phi_J}{1+a}\sin\pi\frac{\phi_I\pm\phi_J\pm2\epsilon_+}{1+a}\sin\pi\frac{\phi_I\pm\phi_J\pm m'\pm\epsilon_-}{1+a}}
    {\sin\pi\frac{\phi_I\pm\phi_J\pm\epsilon_+\pm\epsilon_-}{1+a}\sin\pi\frac{\phi_I\pm\phi_J\pm m'\pm\epsilon_+}{1+a}}\right]
\ee

We can evaluate the above contour integrals for some small $k$. Especially, for $k=1$ cases, there is no integral variable and the result becomes simply
\be
    I^{k=1}=\frac{1}{2}\big[I^{k=1}_+ + I^{k=1}_-\big]=\frac{1}{2}\frac{\sin\pi\frac{m'\pm\epsilon_-}{1+a}}{\sin\pi\frac{\epsilon_+\pm\epsilon_-}{1+a}}
    \left[\prod_{i=1}^N\frac{\sin\pi\frac{\lambda_i\pm m'}{1+a}}{\sin\pi\frac{\lambda_i\pm \epsilon_+}{1+a}}
    +\prod_{i=1}^N\frac{\cos\pi\frac{\lambda_i\pm m'}{1+a}}{\cos\pi\frac{\lambda_i\pm \epsilon_+}{1+a}}\right]
\ee
This becomes $1$ as we take the limit $a\rightarrow 0$. In particular, this shows
that the conjecture (\ref{instanton-conjecture}) cannot be generally correct.
It might be interesting to see if the non-simply-laced nature of $Sp(N)$ is causing
complication.

\subsection{Instanton partition functions with maximal SUSY}

The main purpose of this subsection is to derive the $SO(2N)$ instanton
partition function at the maximal SUSY point. As a simple residue formula
like (\ref{U(N)-residue}) is unavailable, one should directly work with
the contour integral expressions. As the argument for $SO(2N)$
generalizes that for $U(N)$, let us first reconsider the derivation of
the $U(N)$ $Z_{\rm inst}$ with maximal SUSY using contour integral formula.

The contour integrand for the $U(N)$ $k$ instantons
at the third fixed point is:
\begin{equation}
  \hspace*{-1cm}\prod_{I,i}\frac{\sin\pi\frac{\phi_I-\lambda_i-m-\frac{1+c}{2}}{1+c}
  \sin\pi\frac{\phi_I-\lambda_i+m+\frac{1+c}{2}}{1+c}}
  {\sin\pi\frac{\phi_I-\lambda_i+\frac{3c}{2}}{1+c}
  \sin\pi\frac{\phi_I-\lambda_i-\frac{3c}{2}}{1+c}}\cdot
  \prod_{I,J}^\prime\frac{\sin\pi\frac{\phi_{IJ}}{1+c}\sin\pi\frac{\phi_{IJ}+3c}{1+c}}
  {\sin\pi\frac{\phi_{IJ}+m+\frac{3}{2}}{1+c}
  \sin\pi\frac{\phi_{IJ}+m+\frac{3}{2}+3c}{1+c}}
  \cdot\prod_{I,J}\frac{\sin\pi\frac{\phi_{IJ}+m-\frac{1}{2}+a}{1+c}
  \sin\pi\frac{\phi_{IJ}+m-\frac{1}{2}+b}{1+c}}
  {\sin\pi\frac{\phi_{IJ}-a+c}{1+c}\sin\pi\frac{\phi_{IJ}-b+c}{1+c}}
\end{equation}
where we inserted $\epsilon_+=-\frac{3c}{2}$ and $\epsilon_-=\frac{a-b}{2}$
at this fixed point. The prime in the second product means that
$\sin\pi\frac{\phi_{IJ}}{1+c}$ factor is absent when $I=J$. The residues at a pole
will yield $Z_{Y}$, and from the first product, one $\sin$ factor in the denominator
among $\sin\pi\frac{\phi_i-\lambda_i+\epsilon_+}{1+c}=
\sin\pi\frac{\phi_i-\lambda_i-\frac{3c}{2}}{1+c}$ is absent in the residue, as this
is the location of the pole. Also, $k-1$ of the denominators in the third product are
absent as they are also the locations of the pole.

Supposing that the factor in the product is neither containing the poles nor
has an absent numerator (in the second product), one finds that most of the
$\frac{\sin\sin}{\sin\sin}$ factors simplify at $m=\frac{1}{2}$ in one of the
three limits: $a\rightarrow 0$, $b\rightarrow 0$ or $c\rightarrow 0$.
Firstly, one finds from the factor in the third product that
\begin{equation}\label{3rd-normal}
  \frac{\sin\pi\frac{\phi_{IJ}+m-\frac{1}{2}+a}{1+c}
  \sin\pi\frac{\phi_{IJ}+m-\frac{1}{2}+b}{1+c}}
  {\sin\pi\frac{\phi_{IJ}-a+c}{1+c}\sin\pi\frac{\phi_{IJ}-b+c}{1+c}}
  \rightarrow\frac{\sin\pi\frac{\phi_{IJ}+a}{1+c}\sin\pi\frac{\phi_{IJ}+b}{1+c}}
  {\sin\pi\frac{\phi_{IJ}-a+c}{1+c}\sin\pi\frac{\phi_{IJ}-b+c}{1+c}}
\end{equation}
when $m=\frac{1}{2}$. When we are taking the limit $a$ or $b\rightarrow 0$, one finds that
the corresponding residue is $0$ from one of this factor. This is because the factor with
$I=J$ (i.e. $\phi_{IJ}=0$) is included, and the numerator
$\sin\pi\frac{a}{1+c}\sin\pi\frac{b}{1+c}$ implies that this residue is zero.
To fully justify the last statement, one would also have to check that there are no
other factors of the form $\frac{\sin\sin}{\sin\sin}$ which diverges in these limits,
which turns out to be true. Exchaning the roles of $a,b,c$, this implies that
$Z_{\rm inst}^{(1)}=Z^{(2)}_{\rm inst}=0$ in the $c\rightarrow 0$ limit.

Thus, we only need to consider the limit $c\rightarrow 0$ limit of $Z_{\rm inst}^{(3)}$
given by the above contour integrand. We consider various ratios of sine functions
inside the products separately. We first consider the factors in the products which
do contain neither poles in the denominators (for the first and third products)
nor absent sine factors in the numerators (for the second product). The other cases
are considered later separately.

We first consider a factor in the first product when it does not contain a pole.
This happens when for $\phi_I$ and $\lambda_i$ which does not satisfy
$\phi_I-\lambda_i\neq\frac{3c}{2}$. At $m=\frac{1}{2}$, the $c\rightarrow 0$ limit
can be obtained by just plugging in $c=0$, as the denominator do not become zero
in the limit at the pole values of $\phi_I-\lambda_i$ in this case. One obtains
\begin{equation}
  \frac{\sin\pi(\phi_I-\lambda_i-1)\sin\pi(\phi_I-\lambda_i+1)}
  {\sin\pi(\phi_I-\lambda_i)\sin\pi(\phi_I-\lambda_i)}=1\ .
\end{equation}
We then consider the $c\rightarrow 0$ limit of the second factor at $m=\frac{1}{2}$,
supposing that $I\neq J$. Again one can just plug in $c=0$, as the denominators
remain finite in the limit at the pole values of $\phi_{IJ}$. One obtains
\begin{equation}
  \frac{\sin\pi\phi_{IJ}\sin\pi\phi_{IJ}}{\sin\pi(\phi_{IJ}+2)\sin\pi(\phi_{IJ}+2)}=1
\end{equation}
for given $I\neq J$. We also consider the $c\rightarrow 0$ limit for the
factor in the third product, namely (\ref{3rd-normal}). We first postpone considering
$k-1$ cases of $I,J$ that yields poles. We also exclude the case labeled by $J,I$
when the case with $I,J$ contains a pole, as the latter will also turn out to be
subtle in the $c\rightarrow 0$ limit. In the other cases of $I,J$, again one can naivly
plug in $c=0$ to obgain
\begin{equation}
  \frac{\sin\pi(\phi_{IJ}+a)\sin\pi(\phi_{IJ}-a)}
  {\sin\pi(\phi_{IJ}-a)\sin\pi(\phi_{IJ}+a)}=1\ .
\end{equation}
So far, all the trivial factors $1$ is heading towards the proof that $Z_{Y}=1$
at the third fixed point in this limit.

To complete the proof, we should check the $k$ factors containing the poles in the
first and third products, and also the $k$ terms with $I=J$ in the second product.
Firstly, the single factor containing the pole in the first product
has $\sin\pi(\phi-\lambda_i-\frac{3c}{2})=0$. So inserting $\phi_I-\lambda_i=\frac{3c}{2}$
in the remaining three $\sin$'s, one obtains at $m=\frac{1}{2}$
\begin{equation}
  \frac{\sin\pi\frac{c-1}{1+c}\sin\pi\frac{2c+1}{1+c}}{\sin\pi\frac{3c}{1+c}}
  =\frac{\sin\pi\frac{2c}{1+c}\sin\pi\frac{c}{1+c}}{\sin\pi\frac{3c}{1+c}}
  \ \ \stackrel{c\rightarrow 0}{\longrightarrow}\ \ \frac{2\pi c}{3}\ ,
\end{equation}
where the last expression is for small $c$.
From the second product, there are $k$ factors with $I=J$. At $m=\frac{1}{2}$
one obtains
\begin{equation}
  \left[\frac{\sin\pi\frac{3c}{1+c}}{\sin\pi\frac{2}{1+c}\sin\pi\frac{2+3c}{1+c}}\right]^k
  =\left[\frac{\sin\pi\frac{3c}{1+c}}{\sin\pi\frac{-2c}{1+c}\sin\pi\frac{c}{1+c}}\right]^k
  \ \ \stackrel{c\rightarrow 0}{\longrightarrow}\ \ \left(-\frac{3}{2\pi c}\right)^k\ .
\end{equation}
Finally, the $k-1$ cases of $I,J$ in the third product with poles, and the related
cases $k-1$ with $J,I$, are considered. Since one of the two sine factors in
the denominator yields a pole, there are two possibilities between
$\phi_{IJ}=a-c$ and $\phi_{IJ}=b-c$. In the first case with $\phi_{IJ}=a-c$,
one obtains by combining the $I,J$ factor and the $J,I$ factor the following:
\begin{equation}
  \frac{\sin\pi\frac{2a-c}{1+c}\sin\pi\frac{a-c+b}{1+a}}{\sin\pi\frac{a-b}{1+c}}\cdot
  \frac{\sin\pi\frac{c}{1+c}\sin\pi\frac{c-a+b}{1+c}}
  {\sin\pi\frac{2(c-a)}{1+c}\sin\pi\frac{2c-a-b}{1+c}}
  =\frac{\sin\pi\frac{2a-c}{1+c}\sin\pi\frac{2c}{1+c}}{\sin\pi\frac{a-b}{1+c}}
  \cdot\frac{\sin\pi\frac{c}{1+c}\sin\pi\frac{-2a}{1+c}}
  {\sin\pi\frac{2(c-a)}{1+c}\sin\pi\frac{3c}{1+c}}
\end{equation}
before taking any limit. In the second case with $\phi_{IJ}=b-c$ one similarly
obtains
\begin{equation}
  \frac{\sin\pi\frac{2b-c}{1+c}\sin\pi\frac{2c}{1+c}}{\sin\pi\frac{b-a}{1+c}}
  \cdot\frac{\sin\pi\frac{c}{1+c}\sin\pi\frac{-2b}{1+c}}
  {\sin\pi\frac{2(c-b)}{1+c}\sin\pi\frac{3c}{1+c}}\ .
\end{equation}
Their $c\rightarrow 0$ limits become
\begin{equation}
  \phi_{IJ}=a-c\ \ {\rm poles}\ :\ \frac{2\pi c}{3\cos(\pi a)}\ \ \ \ ,\ \ \ \
  \phi_{IJ}=b-c\ \ {\rm poles}\ :\ \frac{2\pi c}{3\cos(\pi b)}\nonumber
\end{equation}
For these factors, one further has to take the $a=-b\rightarrow 0$ limit
to obtain the final expression. In both cases, one obtains $\frac{\pi c}{3}$.
So combining all, one obtains
\begin{equation}
  \frac{2\pi c}{3}\cdot\left(-\frac{3}{2\pi c}\right)^k\cdot
  \left(\frac{2\pi c}{3}\right)^{k-1}=(-1)^k\ .
\end{equation}
The last $(-1)^k$ is simply canceled with similar sign factors one
has to include in the $\phi_I$ integration measure \cite{Nekrasov:2002qd},
which leads to (\ref{U(N)-residue}). This completes the proof of
$Z_{\rm inst}=\eta^{-N}$ without using the handy residue formula (\ref{U(N)-residue}).

For $SO(2N)$, apart from the measure of the above form, resembling that for $U(2N)$
with restricted VEV $\lambda_i$ in $SO(2N)$ Cartan, one finds the following extra
measure for $SO(2N)$ at the third fixed point:
\begin{eqnarray}\label{SO(2N)-maximal-extra}
  &&\prod_{I<J}\frac{\sin\pi\frac{\phi_I+\phi_J}{1+c}
  \sin\pi\frac{\phi_I+\phi_J+3c}{1+c}}{\sin\pi\frac{\phi_I+\phi_J+m+\frac{3}{2}}{1+c}
  \sin\frac{\phi_I+\phi_J+m+\frac{3}{2}+3c}{1+c}}\times
  \frac{\sin\pi\frac{\phi_I+\phi_J}{1+c}
  \sin\pi\frac{\phi_I+\phi_J-3c}{1+c}}{\sin\pi\frac{\phi_I+\phi_J-m-\frac{3}{2}}{1+c}
  \sin\frac{\phi_I+\phi_J-m-\frac{3}{2}-3c}{1+c}}\nonumber\\
  &&\times\frac{\sin\pi\frac{\phi_I+\phi_J+m-\frac{1}{2}+a}{1+c}
  \sin\pi\frac{\phi_I+\phi_J+m-\frac{1}{2}+b}{1+c}}
  {\sin\pi\frac{\phi_I+\phi_J-a+c}{1+c}\sin\pi\frac{\phi_I+\phi_J-b+c}{1+c}}\times
  \frac{\sin\pi\frac{\phi_I+\phi_J-m+\frac{1}{2}-a}{1+c}
  \sin\pi\frac{\phi_I+\phi_J-m+\frac{1}{2}-b}{1+c}}
  {\sin\pi\frac{\phi_I+\phi_J+a-c}{1+c}\sin\pi\frac{\phi_I+\phi_J+b-c}{1+c}}\nonumber\\
  &&\times\prod_I\frac{\sin\pi\frac{2\phi_I}{1+c}\sin\pi\frac{2\phi_I+3c}{1+c}}
  {\sin\pi\frac{2\phi_I+m+\frac{3}{2}}{1+c}\sin\pi\frac{2\phi+m+\frac{3}{2}+3c}{1+c}}
  \times\frac{\sin\pi\frac{2\phi_I}{1+c}\sin\pi\frac{2\phi_I-3c}{1+c}}
  {\sin\pi\frac{2\phi_I-m-\frac{3}{2}}{1+c}\sin\pi\frac{2\phi-m-\frac{3}{2}-3c}{1+c}}\ .
\end{eqnarray}
The $c\rightarrow 0$ limits of six $\frac{\sin\sin}{\sin\sin}$ factors
at $m=\frac{1}{2}$ can be attained by simply plugging in $c=0$, which yields
\begin{equation}
  \frac{\sin\pi\phi\sin\pi\phi}{\sin\pi(\phi+2)\sin\pi(\phi+2)}=1\ ,\ \
  \frac{\sin\pi\phi\sin\pi\phi}{\sin\pi(\phi-2)\sin\pi(\phi-2)}=1
\end{equation}
for the two factors on the first line,
\begin{equation}
  \frac{\sin\pi(\phi+b)\sin\pi(\phi+c)}{\sin\pi(\phi-b)\sin\pi(\phi-c)}
  \times\frac{\sin\pi(\phi-b)\sin\pi(\phi-c)}{\sin\pi(\phi+b)\sin\pi(\phi+c)}=1
\end{equation}
for the two on the second line, and
\begin{equation}
  \frac{\sin(2\pi\phi)\sin(2\pi\phi)}{\sin\pi(2\phi+2)\sin\pi(2\phi+2)}=1\ ,\ \
  \frac{\sin(2\pi\phi)\sin(2\pi\phi)}{\sin\pi(2\phi-2)\sin\pi(2\phi-2)}=1
\end{equation}
for the two on the last line.

The above proves that $Z_{Y_+,Y_-}=1$ for the third fixed point
with the pair of $N$-colored Young diagrams labeling the poles,
when we take the go to the maximal SUSY point by taking $c\rightarrow 0$
first. To finally understand $Z_{\rm inst}$ with maximal SUSY, one has to
understand the allowed pair $(Y_+,Y_-)$ of colored Young diagrams which
generally provide nonzero residues. One can first check that the pair
$Y_\pm=(Y_{1\pm},Y_{2\pm},\cdots,Y_{N\pm})$ has zero residue when
both $Y_{i+}$, $Y_{i-}$ are nonempty with a given $i,1,2,\cdots,N$.
Let us check this by considering the case with both $Y_{i\pm}$ being nonempty.
In this case, the $\phi$ values for the first boxes of the two
Young diagrams $Y_{i+}$, $Y_{i-}$ are $\phi_I=\lambda_i+\frac{3c}{2}$ and
$\phi_J=-\lambda_i+\frac{3c}{2}$, respectively. Then, consider the second
factor on the first line of (\ref{SO(2N)-maximal-extra}). This factor contains
$\sin\pi\frac{\phi_I+\phi_J-3c}{1+c}=0$ in its numerator. This proves that
there are no poles when both $Y_{i+}$, $Y_{i-}$ are nonempty.

So $Z_{\rm inst}=Z_{\rm inst}^{(3)}$ with maximal SUSY is the generating
function for the pair of $N$-colored Young diagrams satisfying the above constraint,
which counts the diagrams taking the gauge symmetries of this system into account.
The relevant gauge symmetry here is the Weyl group of $O(2N)$, when its $N$ Cartans
$\lambda_i$ are nonzero and all different generically.\footnote{Note
that it is more natural to regard the gauge group as $O(2N)$ rather than $SO(2N)$, as
one motivation for the existence of this type of 6d $(2,0)$ theory comes from
M5-branes at an orbifold. Brane pictures are giving $O$ type rather than $SO$
type gauge groups.} The Weyl group of $O(2N)$ has $2^NN!$ elements.
As $N!$ of them simply permutes the $N$ eigenvalues $\lambda_i$, this part of
the symmetry is broken with all $\lambda_i$'s being different. The remainig
$2^N$ elements with a given set of $\lambda_i$'s act by flipping signs of these
eigenvalues in all possible ways. Note that under the change
$\lambda_i\rightarrow-\lambda_i$, the two diagrams $Y_{i\pm}$ are exchanged.
Since we just checked that one of $y_{i\pm}$ is always empty for any $i$, we can
use these $2^N$ Weyl transformation to move all nonempty Young diagrams to $Y_+$.
So the problem boils down to the counting of one set of $N$-colored Young diagrams.
Just like the $U(N)$ case, its generating function is given by
\begin{equation}
  Z_{\rm inst}^{SO(2N)}=\eta(e^{-\frac{4\pi^2}{\beta}})^{-N}\ ,
\end{equation}
which proves the claim made in section 3.2.

\section{Off-shell action from supergravity}

In this section, we derive the classical action (\ref{gaussian2}) at the saddle
point with a constant scalar in the vector multiplet. We need some information
about the off-shell action on squashed $S^5$. The simplest way is to use the
off-shell supergravity method \cite{Festuccia:2011ws}. Off-shell
formalism of 5d supergravity is known in \cite{Kugo:2000af,Hanaki:2006pj}.

Let us first summarize our strategy. From the
dimensional reduction of twisted product space $S^5\times S^1$, we identified
the squashed $S^5$ geometry as well as other background scalar and vector fields.
They are to be regarded as nonzero background fields in the 5d supergravity coupled
to our matter vector multiplet, containing $\phi$ and the gauge field $A_\mu$.
We can also reduce the canonical Kiling spinor equation on twisted $S^5\times S^1$
to 5d. This has to be identified with the vanishing condition of gravitino SUSY variation.
By comparing the two, one can decide the values for various auxiliary fields in
the off-shell supergravity. With these fields determined, one studies SUSY
variations for other fermions to fix the values of more bosonic auxiliary fields
that we need to compute our off-shell field theory action. Finally, plugging in
the SUSY configuration of our matter vector multiplet (corresponding to our
saddle points in localization calculus), we shall obtain (\ref{gaussian2}).

In 6d, the chemical potentials in the index leads to the following shifts
of the Euclidean time translation on $S^5\times\mathbb{R}$ (we set the $S^5$ radius
$r=1$ in this section, for simplicity):
\be
    \nabla_\tau \rightarrow \nabla_\tau +aj_1 + bj_2+ cj_3 -\frac{R_1+R_2}{2} -m(R_1-R_2)\ .
\ee
The Killing spinors surviving the compactification of time direction, and thus the
5d reduction, are constant in $\tau$. In the time component of Killing spinor equation,
\be
    \left(\partial_\tau+aj_1 + bj_2+ cj_3 -\frac{R_1+R_2}{2} -m(R_1-R_2)\right)\epsilon_\pm = \mp\frac{1}{2}\epsilon_\pm\ ,
\ee
one seeks for constant solutions with generic values of $a,b,c\ (a+b+c=0)$ and $m$.
This leads us to 2 real components obeying the following conditions:
\be
    j_1\epsilon_\pm = j_2\epsilon_\pm=j_3\epsilon_\pm = -\frac{R_1+R_2}{2}\epsilon_\pm = \mp\frac{1}{2}\epsilon_\pm\ .
\ee
They are $Q$, $S$ which annihilate our BPS states in the index.

The metric on $S^5\times S^1$ with twist (\ref{squash-S5}) can be written as
\be
    ds_6^2=ds_5^2+\alpha^{-2}\left(d\tau+C\right)^2
\ee
with
\be
    \alpha^{-2} = (1-n_i^2a_i^2) \ , \quad C = \alpha^2irn_i^2a_id\phi_i\ .
\ee
$C_\mu$ and $\alpha$ should be identified as the vector and scalar in
a background vector multiplet on the squashed $S^5$ in off-shell supergravity.
As the chemical potentials for the $S^5$ angular momenta are explicitly encoded
in the geometry, the twistings in the time derivative are now restricted
only to the internal part:
\be
    \left(\partial_\tau+aj_1 + bj_2+ cj_3 -\frac{R_1+R_2}{2} -m(R_1-R_2)\right) \rightarrow\left(\partial_\tau-\frac{R_1+R_2}{2} -m(R_1-R_2)\right)
\ee
It is the last expression which replaces all time derivatives in the SUSY
and Abelian equation of motions in 6d. When we reduce 6d Abelian fields to
squashed $S^5$, we only keep the $\tau$ independent modes.
Thus, all twisted time derivatives are essentially replaced as
\be
    \partial_\tau\rightarrow -\left(\frac{R_1+R_2}{2} +m(R_1-R_2)\right)\ .
\ee

Let us define $V=dC$ to be the field strength of $C$. We also choose the vielbein as
\be
    e^a\, , \quad e^6 =\alpha^{-1}\left(d\tau+C\right)
\ee
where $a=1,2,3,4,5$ and $6$ are the 6d indices in the local Lorentz frame.
Then starting from the Killing spinor equation of twisted $S^5\times S^1$ and
reducing to 5d following the rules above, the 5d part of the Killing spinor
equation can be rewritten as
\be\label{Killing4}
    \left(D_a-\frac{i}{8\alpha}V^{bc}\gamma_{abc}\right)\epsilon_\pm=\gamma_a\eta_\pm \ , \quad \eta_\pm =i\left(-\alpha\frac{R_1+R_2}{2}-\frac{1}{4\alpha^2}V_{ab}\gamma^{ab}+\frac{i}{2\alpha}
    \partial_a\alpha\gamma^a\right)\epsilon_\pm
\ee
This can be compared with the vanishing of the gravitino SUSY variation
\be
    \delta \psi_\mu = \mathcal{D}_\mu\epsilon + \frac{1}{2}v^{ab}\gamma_{\mu ab}\epsilon - \gamma_\mu\eta=0
\ee
in 5d, where $\mathcal{D}_\mu\epsilon^i =\left(\partial_\mu +\frac{1}{4}\omega_{\mu ab}\gamma^{ab}+\frac{1}{2}b_\mu\right)\epsilon^i - \tilde{V}_{\mu\  j}^{\ i}\epsilon^j$.
For the two equations to be compatible, we set some 5d supergravity fields as
\be
    b_\mu = 0 \ , \quad \tilde{V}_\mu = -C_\mu\left(\frac{R_1+R_2}{2}\right) \ , \quad v = -\frac{i}{4\alpha}V \ , \quad \epsilon = \epsilon_\pm , \quad  \eta = \eta_\pm
\ee
The first equation for $b_\mu$ is a gauge-fixing.
The vanishing of the gaugino SUSY variation in the Weyl multiplet, together with
the $\mu=6$ component of the $S^5\times S^1$ Killing spinor equation,
also determines an auxiliary field ${\rm D}$. It appears that the solution is
as simple as ${\rm D} = -\frac{3}{8\alpha^2}V^2$.

One can also consider the SUSY of off-shell vector multiplets.
Let $I$ label the vector multiplets. In the bosonic sector, each vector
multiplet consists of a gauge field $A^I_\mu$, a real scalar $\phi^I$, and
an $SU(2)$ triplet of auxiliary fields $(D^I)^i_{\ j}$ with $i,j=1,2$.
In our problem, we need two sets of vector multiplets. One is the vector
multiplet fields which are dynamical in our QFT, which can be labeled
by many $I$ components if we are considering non-Abelian generalization
(with gaugings in supergravity). We need an extra background vector multiplet,
which we label by $I=0$, whose vector and scalar components are the background
`RR-field' $C_\mu$ and `dilaton' $\alpha$ that we identified after the
reduction of twisted $S^5\times S^1$ on the circle. In fact, the above
background values are invariant under the off-shell SUSY transformation
for the vector multiplet, provided that we suitably turn on the D-term
auxiliary fields:
\begin{equation}\label{background-vector}
  \mathcal{V}^0=(A^0_\mu,\phi^0,D^0)=(C_\mu,\alpha,-i\alpha^2\sigma^3)\ .
\end{equation}
Nonzero value of $D^0$, with a choice $\sigma^3$, breaks $SU(2)$ R-symmetry
in the curved background \cite{Hosomichi:2012ek,Imamura:2012xg}.


Now we consider the bosonic part of the off-shell action which contain
our dynamical vector multiplet fields. As the main purpose of this appendix
is to identify the correct saddle point action, we do not need to consider our
hypermultiplet here. We take the cubic polynomial in the supergravity to be
\begin{equation}
  \mathcal{N}=C_{IJK}\mathcal{V}^I\mathcal{V}^J\mathcal{V}^K=
  \frac{1}{2}\alpha\phi^2+\cdots\ ,
\end{equation}
where $\cdots$ denote the possible terms which do not contain $\phi$.
As these do not affect our QFT Lagrangian on curved space, we do not pay
attention to them. The first term is normalized to have the canonical kinetic
term for the vector multiplet fields on $S^5$ with $\alpha=1$.
Then the vector multiplet part of the bosonic action can be read off from
the off-shell supergravity action with above $\mathcal{N}$, which is
\be
    g_{YM}^2e^{-1}\mathcal{L}&=&\frac{1}{2}\left(\frac{3}{16\alpha^2}V^2+\frac{1}{4} R
    +\frac{3}{16\alpha^2}V^2\right)\alpha\phi^2 -\frac{1}{4\alpha}\phi^2V^2
    -\frac{1}{2}\phi V^{ab}F_{ab} \nn \\
    &&-2\phi\left(-\frac{1}{4}V^{ab}F_{ab}-\frac{1}{2}\partial^a\alpha D_a\phi+\frac{i}{4}\alpha^2(\sigma^3)_{ij}D^{ij}\right)\\
    && -\alpha\left(-\frac{1}{4}F_{ab}F^{ab} -\frac{1}{2} D^a\phi D_a\phi-\frac{1}{4}D_{ij} D^{ij}\right) +e^{-1}\frac{i}{8}\epsilon^{\mu\nu\lambda\rho\sigma}C_\mu F_{\nu\lambda}F_{\rho\sigma}\ .\nn
\ee
The auxiliary fields remaining in the above expression is determined above.
So this yields the vector multiplet part of the bosonic action with our dynamical
field $A_\mu,\phi$ and the triplet auxiliary field $D^i_{\ j}$.

Now we finally come to the classical action at the saddle point. Note first that
that the off-shell SUSY transformation for all vector multiplets take the same form.
So even our dynamical vector multiplet fields can have SUSY configuration proportional
to the value (\ref{background-vector}) for the background vector multiplet.
Multiplying a constant $\phi_0$ valued in the gauge algebra, one obtains
\begin{equation}
  \mathcal{V}^1=(A_\mu,\phi, D)=\phi_0\mathcal{V}^0
  =\left(\phi_0C_\mu,\phi_0\alpha,\phi_0(-i\alpha^2\sigma^3)\right)\ .
\end{equation}
$\phi_0$ labels our off-shell saddle point on the squashed $S^5$, which also
correctly reduces to the known saaddle point values of the round $S^5$ studied
previously \cite{Hosomichi:2012ek,Kallen:2012va,Kim:2012av}, up to different
definitions of $D$.
Note that similar configuration was studied in \cite{Imamura:2012xg} for a different
class of squashed $S^5$. Plugging in all auxiliary fields and our saddle point
configuration, one obtains
\be
    2g_{YM}^2e^{-1}\mathcal{L}_0=\left[\alpha^{3}\left(\frac{1}{4}R +\frac{6}{16\alpha^2} V^2\right) + \alpha^{3}\partial_a\alpha\partial^a\alpha+3\alpha^5+\frac{i}{4}e^{-1}\epsilon^{\mu\nu\lambda\rho\sigma}C_\mu V_{\nu\lambda}V_{\rho\sigma}\right]\phi_0^2
\ee
After evaluating the integral over the functions appearing in the square bracket,
one obtains
\be
    S_0=\int \sqrt{g}d^5x\mathcal{L}_0 = \frac{4\pi^3}{g_{YM}^2}\frac{\phi_0^2}{(1+a_1)(1+a_2)(1+a_3)}=
    \frac{2\pi^2\lambda^2}{\beta(1+a)(1+b)(1+c)}\ ,
\ee
where $\phi_0=\lambda$ in our setting with $r=1$. This leads to (\ref{gaussian2}).

\end{document}